\documentstyle[a4,12pt,rotating,epsfig,subfigure]{article}

\newcommand{\mycaption}[1]{\begin{center}
                             \parbox{0.85\textwidth}{
                               \caption{\small #1}
                             }
                           \end{center}}

\def\strutl{\protect\rule{0ex}{1.4em}}
\def\strutd{\protect\rule[-0.6em]{0ex}{1.9em}}

\newcommand{\GeVc}  {GeV/$c$}
\newcommand{\GeVcc} {GeV/$c^2$}
\newcommand{\ggqq}  {\mbox{$\gamma\gamma\rightarrow\mathrm q\bar q$}}
\newcommand{\ggee}  {\mbox{$\gamma\gamma\rightarrow\mathrm e^+e^-$}}
\newcommand{\ggmm}  {\mbox{$\gamma\gamma\rightarrow\mu^+\mu^-$}}
\newcommand{\ggtt}  {\mbox{$\gamma\gamma\rightarrow\tau^+\tau^-$}}

\newcommand {\ff}            {\mbox{${\mathrm f} \bar{\mathrm f}$}}
\newcommand {\ee}            {\mbox{${\mathrm e}^+{\mathrm e}^-$}}
\newcommand {\myll}          {\mbox{${\mathrm l}^+{\mathrm l}^-$}}

\newcommand {\ZZ}            {${\mathrm Z} {\mathrm Z}$}
\newcommand {\WW}            {$\mathrm W^+ W^-$}

\newcommand {\Mvis}          {$M_{\mathrm vis}$}

\newcommand {\qq}            {\mbox{${\mathrm q} \bar{\mathrm q}$}}
\newcommand {\bb}            {\mbox{${\mathrm b} \bar{\mathrm b}$}}
\newcommand {\cc}            {\mbox{${\mathrm c} \bar{\mathrm c}$}}
\newcommand {\myss}          {\mbox{${\mathrm s} \bar{\mathrm s}$}}
\newcommand {\dd}            {\mbox{${\mathrm d} \bar{\mathrm d}$}}
\newcommand {\uu}            {\mbox{${\mathrm u} \bar{\mathrm u}$}}

\newcommand {\mumu}          {\mbox{${\mu^+\mu^-}$}}

\newcommand {\tautau}        {\mbox{${\tau^+\tau^-}$}}

\newcommand {\Afb}[1]        {\mbox{$A_{\mathrm FB}^{\mathrm #1}$}}
\newcommand {\Qfb}           {\mbox{$\langle Q_{\mathrm FB}\rangle$}} 

\newcommand{\lam}  {\mbox{$\Lambda$}}
\newcommand{\eps}  {\mbox{$\epsilon$}}

\newcommand{\MLQ}    {\mbox{$M_{\mathrm LQ}$}}
\newcommand{\half}   {\mbox{$\scriptscriptstyle\frac{1}{2}$}}
\newcommand{\SL}[1]  {$\mathrm S_{#1}~(L)$}
\newcommand{\SR}[1]  {$\mathrm S_{#1}~(R)$}
\newcommand{\SBR}[1] {$\mathrm\tilde S_{#1}~(R)$}
\newcommand{\SBL}[1] {$\mathrm\tilde S_{#1}~(L)$}
\newcommand{\VR}[1]  {$\mathrm V_{#1}~(R)$}
\newcommand{\VL}[1]  {$\mathrm V_{#1}~(L)$}
\newcommand{\VBL}[1] {$\mathrm\tilde V_{#1}~(L)$}
\newcommand{\VBR}[1] {$\mathrm\tilde V_{#1}~(R)$}

\newcommand{\MZP}     {\mbox{$m_{\mathrm Z'}$}}

\newcommand{\ZP}     {\mbox{$\mathrm Z'$}}

\newcommand{\ZZERO}     {\mbox{$\mathrm Z^{0}$}}
\newcommand{\ZZEROP}     {\mbox{$\mathrm Z'^{0}$}}

\newcommand{\ES}      {$E_{6}$}

\newcommand{\ESCHI}      {$E_{6}(\chi)$}
\newcommand{\ESPSI}      {$E_{6}(\psi)$}
\newcommand{\ESETA}      {$E_{6}(\eta)$}
\newcommand{\ESI}      {$E_{6}(I)$}
\newcommand{\LRS}      {LRS}
\newcommand{\TES}     {$\theta_{E_{6}}$}
\newcommand{\TMIX}     {\mbox{$\theta_{\mathrm mix}$}}

\newcommand{\ALR}     {\mbox{$\alpha_{LR}$}}
\newcommand{\CHI}     {$\chi^{2}$}
\newcommand{\CHISM}     {$\chi^{2}_{\mathrm SM}$}

\newcommand{\Rb} {$R_{\mathrm b}$}

\newcommand{\Rc} {$R_{\mathrm c}$}

\newcommand {\mr}[1]   {\mbox {${\mathrm {#1} }$}}

\begin{document}

%%%%%%%%
%--- titlepage, abstract
%%%%%%%%
%\input ew2_conf_vanc_title.tex
%%%%%%%%%%%%%%%%%%%%%%%%%%%%%%%%%%%%%%%%%%%%%%%%%%%%%%%%
%
% title + abstract
%
%%%%%%%%%%%%%%%%%%%%%%%%%%%%%%%%%%%%%%%%%%%%%%%%%%%%%%%%

\begin{titlepage}

\vspace{3truecm}
\centerline{\Large EUROPEAN ORGANIZATION FOR PARTICLE PHYSICS}
\vspace{1truecm}

\begin{flushright}
CERN--EP/99--042 \\
% \today \\
3 March 1999 \\
%{\bf Draft 2.1} \\
\end{flushright}

%TRACKING SYSTEMATIC ???
%RADIATVE SYSTEMATIC (TAU,MU) ??? for inc cross sec
%BHABHA BACKGROUND/SYSTEMATIC to DITAU ??? none

\vspace{2cm}

\begin{center}

{\LARGE\bf
Study of Fermion Pair Production\\ 
in {\boldmath $\mathrm e^+e^-$} Collisions at 130--183~GeV}

\vspace{3cm}

{\Large The ALEPH Collaboration}

\vspace{2cm}

{\large  Abstract}

\end{center}

\noindent
The cross sections and forward-backward asymmetries of 
hadronic and leptonic events produced in \ee\ collisions at 
centre-of-mass energies from 130 to 183~GeV are presented. 
Results for \ee, \mumu, \tautau, \qq, \bb\ and \cc\ production
show no significant deviation from the Standard Model predictions.
This enables constraints to be set upon physics beyond the Standard Model
such as four-fermion contact interactions, leptoquarks, \ZP~bosons and 
R-parity violating squarks and sneutrinos.
Limits on the energy scale $\Lambda$ of $\ee\ff$ contact interactions 
are typically in the range from 2 to 10~TeV. Limits on R-parity violating 
sneutrinos reach masses of a few hundred \GeVcc\ for large values of 
their Yukawa couplings. 

\vfil

\centerline{\it To be submitted to Euro. Phys. J. C}

\vfil\noindent
%{\bf 
%Please send comments Marie-Noelle.Minard@cern.ch and Ian.Tomalin@cern.ch
%on behalf of the BEW group, and to the referees Monica.Pepe.Altarelli@cern.ch
%and duflot@lal.in2p3.fr~.
%}

\end{titlepage}

%
%--- Author list
%
\newpage
%------------------------------------------------------------------------
% authob12pt.tex
% authors' list for papers at LEP 1.5 and 2 energies
%-----------------------------------------------------------------------
\pagestyle{empty}
\newpage
\small
%
% remember the old settings
%
\newlength{\saveparskip}
\newlength{\savetextheight}
\newlength{\savetopmargin}
\newlength{\savetextwidth}
\newlength{\saveoddsidemargin}
\newlength{\savetopsep}
\setlength{\saveparskip}{\parskip}
\setlength{\savetextheight}{\textheight}
\setlength{\savetopmargin}{\topmargin}
\setlength{\savetextwidth}{\textwidth}
\setlength{\saveoddsidemargin}{\oddsidemargin}
\setlength{\savetopsep}{\topsep}
%
% text dimensions for the author list
%
\setlength{\parskip}{0.0cm}
\setlength{\topsep}{1mm}
\pretolerance=10000
%%\begin{document}
%\centerline{EUROPEAN ORGANIZATION FOR NUCLEAR RESEARCH}
%\centerline{EUROPEAN LABORATORY FOR PARTICLE PHYSICS (CERN)}
%\vspace{1cm}
%\begin{flushright}CERN-PPE/96-   \\
%22 February 1999 - last update
%\end{flushright}
\centerline{\large\bf The ALEPH Collaboration}
\footnotesize
\vspace{0.5cm}
{\raggedbottom
\begin{sloppypar}
\samepage\noindent
R.~Barate,
D.~Decamp,
P.~Ghez,
C.~Goy,
S.~Jezequel,
J.-P.~Lees,
F.~Martin,
E.~Merle,
\mbox{M.-N.~Minard},
B.~Pietrzyk
\nopagebreak
\begin{center}
\parbox{15.5cm}{\sl\samepage
Laboratoire de Physique des Particules (LAPP), IN$^{2}$P$^{3}$-CNRS,
F-74019 Annecy-le-Vieux Cedex, France}
\end{center}\end{sloppypar}
\vspace{2mm}
\begin{sloppypar}
\noindent
R.~Alemany,
M.P.~Casado,
M.~Chmeissani,
J.M.~Crespo,
E.~Fernandez,
M.~Fernandez-Bosman,
Ll.~Garrido,$^{15}$
E.~Graug\`{e}s,
A.~Juste,
M.~Martinez,
G.~Merino,
R.~Miquel,
Ll.M.~Mir,
P.~Morawitz,
A.~Pacheco,
I.C.~Park,
I.~Riu
\nopagebreak
\begin{center}
\parbox{15.5cm}{\sl\samepage
Institut de F\'{i}sica d'Altes Energies, Universitat Aut\`{o}noma
de Barcelona, 08193 Bellaterra (Barcelona), E-Spain$^{7}$}
\end{center}\end{sloppypar}
\vspace{2mm}
\begin{sloppypar}
\noindent
A.~Colaleo,
D.~Creanza,
M.~de~Palma,
G.~Gelao,
G.~Iaselli,
G.~Maggi,
M.~Maggi,
S.~Nuzzo,
A.~Ranieri,
G.~Raso,
F.~Ruggieri,
G.~Selvaggi,
L.~Silvestris,
P.~Tempesta,
A.~Tricomi,$^{3}$
G.~Zito
\nopagebreak
\begin{center}
\parbox{15.5cm}{\sl\samepage
Dipartimento di Fisica, INFN Sezione di Bari, I-70126 Bari, Italy}
\end{center}\end{sloppypar}
\vspace{2mm}
\begin{sloppypar}
\noindent
X.~Huang,
J.~Lin,
Q. Ouyang,
T.~Wang,
Y.~Xie,
R.~Xu,
S.~Xue,
J.~Zhang,
L.~Zhang,
W.~Zhao
\nopagebreak
\begin{center}
\parbox{15.5cm}{\sl\samepage
Institute of High-Energy Physics, Academia Sinica, Beijing, The People's
Republic of China$^{8}$}
\end{center}\end{sloppypar}
\vspace{2mm}
\begin{sloppypar}
\noindent
D.~Abbaneo,
U.~Becker,$^{19}$
G.~Boix,$^{6}$
M.~Cattaneo,
V.~Ciulli,
G.~Dissertori,
H.~Drevermann,
R.W.~Forty,
M.~Frank,
F.~Gianotti,
A.W.~Halley,
J.B.~Hansen,
J.~Harvey,
P.~Janot,
B.~Jost,
I.~Lehraus,
O.~Leroy,
C.~Loomis,
P.~Maley,
P.~Mato,
A.~Minten,
A.~Moutoussi,
F.~Ranjard,
L.~Rolandi,
D.~Rousseau,
D.~Schlatter,
M.~Schmitt,$^{20}$
O.~Schneider,$^{23}$
W.~Tejessy,
F.~Teubert,
I.R.~Tomalin,
E.~Tournefier,
M.~Vreeswijk,
A.E.~Wright
\nopagebreak
\begin{center}
\parbox{15.5cm}{\sl\samepage
European Laboratory for Particle Physics (CERN), CH-1211 Geneva 23,
Switzerland}
\end{center}\end{sloppypar}
\vspace{2mm}
\begin{sloppypar}
\noindent
Z.~Ajaltouni,
F.~Badaud
G.~Chazelle,
O.~Deschamps,
S.~Dessagne,
A.~Falvard,
C.~Ferdi,
P.~Gay,
C.~Guicheney,
P.~Henrard,
J.~Jousset,
B.~Michel,
S.~Monteil,
\mbox{J-C.~Montret},
D.~Pallin,
P.~Perret,
F.~Podlyski
\nopagebreak
\begin{center}
\parbox{15.5cm}{\sl\samepage
Laboratoire de Physique Corpusculaire, Universit\'e Blaise Pascal,
IN$^{2}$P$^{3}$-CNRS, Clermont-Ferrand, F-63177 Aubi\`{e}re, France}
\end{center}\end{sloppypar}
\vspace{2mm}
\begin{sloppypar}
\noindent
J.D.~Hansen,
J.R.~Hansen,
P.H.~Hansen,
B.S.~Nilsson,
B.~Rensch,
A.~W\"a\"an\"anen
\nopagebreak
\begin{center}
\parbox{15.5cm}{\sl\samepage
Niels Bohr Institute, 2100 Copenhagen, DK-Denmark$^{9}$}
\end{center}\end{sloppypar}
\vspace{2mm}
\begin{sloppypar}
\noindent
G.~Daskalakis,
A.~Kyriakis,
C.~Markou,
E.~Simopoulou,
A.~Vayaki
\nopagebreak
\begin{center}
\parbox{15.5cm}{\sl\samepage
Nuclear Research Center Demokritos (NRCD), GR-15310 Attiki, Greece}
\end{center}\end{sloppypar}
\vspace{2mm}
\begin{sloppypar}
\noindent
A.~Blondel,
\mbox{J.-C.~Brient},
F.~Machefert,
A.~Roug\'{e},
M.~Swynghedauw,
R.~Tanaka,
A.~Valassi,$^{12}$
H.~Videau
\nopagebreak
\begin{center}
\parbox{15.5cm}{\sl\samepage
Laboratoire de Physique Nucl\'eaire et des Hautes Energies, Ecole
Polytechnique, IN$^{2}$P$^{3}$-CNRS, \mbox{F-91128} Palaiseau Cedex, France}
\end{center}\end{sloppypar}
\vspace{2mm}
\begin{sloppypar}
\noindent
E.~Focardi,
G.~Parrini,
K.~Zachariadou
\nopagebreak
\begin{center}
\parbox{15.5cm}{\sl\samepage
Dipartimento di Fisica, Universit\`a di Firenze, INFN Sezione di Firenze,
I-50125 Firenze, Italy}
\end{center}\end{sloppypar}
\vspace{2mm}
\begin{sloppypar}
\noindent
R.~Cavanaugh,
M.~Corden,
C.~Georgiopoulos
\nopagebreak
\begin{center}
\parbox{15.5cm}{\sl\samepage
Supercomputer Computations Research Institute,
Florida State University,
Tallahassee, FL 32306-4052, USA $^{13,14}$}
\end{center}\end{sloppypar}
\vspace{2mm}
\begin{sloppypar}
\noindent
A.~Antonelli,
G.~Bencivenni,
G.~Bologna,$^{4}$
F.~Bossi,
P.~Campana,
G.~Capon,
F.~Cerutti,
V.~Chiarella,
P.~Laurelli,
G.~Mannocchi,$^{5}$
F.~Murtas,
G.P.~Murtas,
L.~Passalacqua,
M.~Pepe-Altarelli$^{1}$
\nopagebreak
\begin{center}
\parbox{15.5cm}{\sl\samepage
Laboratori Nazionali dell'INFN (LNF-INFN), I-00044 Frascati, Italy}
\end{center}\end{sloppypar}
\vspace{2mm}
%\pagebreak
\begin{sloppypar}
\noindent
M.~Chalmers,
L.~Curtis,
J.G.~Lynch,
P.~Negus,
V.~O'Shea,
B.~Raeven,
C.~Raine,
D.~Smith,
P.~Teixeira-Dias,
A.S.~Thompson,
J.J.~Ward
\nopagebreak
\begin{center}
\parbox{15.5cm}{\sl\samepage
Department of Physics and Astronomy, University of Glasgow, Glasgow G12
8QQ,United Kingdom$^{10}$}
\end{center}\end{sloppypar}
%\vspace{2mm}
%\pagebreak
\begin{sloppypar}
\noindent
O.~Buchm\"uller,
S.~Dhamotharan,
C.~Geweniger,
P.~Hanke,
G.~Hansper,
V.~Hepp,
E.E.~Kluge,
A.~Putzer,
J.~Sommer,
K.~Tittel,
S.~Werner,$^{19}$
M.~Wunsch
\nopagebreak
\begin{center}
\parbox{15.5cm}{\sl\samepage
Institut f\"ur Hochenergiephysik, Universit\"at Heidelberg, D-69120
Heidelberg, Germany$^{16}$}
\end{center}\end{sloppypar}
\vspace{2mm}
\begin{sloppypar}
\noindent
R.~Beuselinck,
D.M.~Binnie,
W.~Cameron,
P.J.~Dornan,$^{1}$
M.~Girone,
S.~Goodsir,
N.~Marinelli,
E.B.~Martin,
J.~Nash,
J.~Nowell,
A.~Sciab\`a,
J.K.~Sedgbeer,
P.~Spagnolo,
E.~Thomson,
M.D.~Williams
\nopagebreak
\begin{center}
\parbox{15.5cm}{\sl\samepage
Department of Physics, Imperial College, London SW7 2BZ,
United Kingdom$^{10}$}
\end{center}\end{sloppypar}
\vspace{2mm}
\begin{sloppypar}
\noindent
V.M.~Ghete,
P.~Girtler,
E.~Kneringer,
D.~Kuhn,
G.~Rudolph
\nopagebreak
\begin{center}
\parbox{15.5cm}{\sl\samepage
Institut f\"ur Experimentalphysik, Universit\"at Innsbruck, A-6020
Innsbruck, Austria$^{18}$}
\end{center}\end{sloppypar}
\vspace{2mm}
\begin{sloppypar}
\noindent
A.P.~Betteridge,
C.K.~Bowdery,
P.G.~Buck,
P.~Colrain,
G.~Crawford,
G.~Ellis,
A.J.~Finch,
F.~Foster,
G.~Hughes,
R.W.L.~Jones,
N.A.~Robertson,
M.I.~Williams
\nopagebreak
\begin{center}
\parbox{15.5cm}{\sl\samepage
Department of Physics, University of Lancaster, Lancaster LA1 4YB,
United Kingdom$^{10}$}
\end{center}\end{sloppypar}
\vspace{2mm}
\begin{sloppypar}
\noindent
P.~van~Gemmeren,
I.~Giehl,
F.~H\"olldorfer,
C.~Hoffmann,
K.~Jakobs,
K.~Kleinknecht,
M.~Kr\"ocker,
H.-A.~N\"urnberger,
G.~Quast,
B.~Renk,
E.~Rohne,
H.-G.~Sander,
S.~Schmeling,
H.~Wachsmuth
C.~Zeitnitz,
T.~Ziegler
\nopagebreak
\begin{center}
\parbox{15.5cm}{\sl\samepage
Institut f\"ur Physik, Universit\"at Mainz, D-55099 Mainz, Germany$^{16}$}
\end{center}\end{sloppypar}
\vspace{2mm}
\begin{sloppypar}
\noindent
J.J.~Aubert,
C.~Benchouk,
A.~Bonissent,
J.~Carr,$^{1}$
P.~Coyle,
A.~Ealet,
D.~Fouchez,
F.~Motsch,
P.~Payre,
M.~Talby,
M.~Thulasidas,
A.~Tilquin
\nopagebreak
\begin{center}
\parbox{15.5cm}{\sl\samepage
Centre de Physique des Particules, Facult\'e des Sciences de Luminy,
IN$^{2}$P$^{3}$-CNRS, F-13288 Marseille, France}
\end{center}\end{sloppypar}
\vspace{2mm}
\begin{sloppypar}
\noindent
M.~Aleppo,
M.~Antonelli,
F.~Ragusa
\nopagebreak
\begin{center}
\parbox{15.5cm}{\sl\samepage
Dipartimento di Fisica, Universit\`a di Milano e INFN Sezione di
Milano, I-20133 Milano, Italy.}
\end{center}\end{sloppypar}
\vspace{2mm}
\begin{sloppypar}
\noindent
R.~Berlich,
V.~B\"uscher,
H.~Dietl,
G.~Ganis,
K.~H\"uttmann,
G.~L\"utjens,
C.~Mannert,
W.~M\"anner,
\mbox{H.-G.~Moser},
S.~Schael,
R.~Settles,
H.~Seywerd,
H.~Stenzel,
W.~Wiedenmann,
G.~Wolf
\nopagebreak
\begin{center}
\parbox{15.5cm}{\sl\samepage
Max-Planck-Institut f\"ur Physik, Werner-Heisenberg-Institut,
D-80805 M\"unchen, Germany\footnotemark[16]}
\end{center}\end{sloppypar}
\vspace{2mm}
\begin{sloppypar}
\noindent
P.~Azzurri,
J.~Boucrot,
O.~Callot,
S.~Chen,
M.~Davier,
L.~Duflot,
\mbox{J.-F.~Grivaz},
Ph.~Heusse,
A.~Jacholkowska,
M.~Kado,
J.~Lefran\c{c}ois,
L.~Serin,
\mbox{J.-J.~Veillet},
I.~Videau,$^{1}$
J.-B.~de~Vivie~de~R\'egie,
D.~Zerwas
\nopagebreak
\begin{center}
\parbox{15.5cm}{\sl\samepage
Laboratoire de l'Acc\'el\'erateur Lin\'eaire, Universit\'e de Paris-Sud,
IN$^{2}$P$^{3}$-CNRS, F-91898 Orsay Cedex, France}
\end{center}\end{sloppypar}
\vspace{2mm}
\begin{sloppypar}
\noindent
%\samepage
G.~Bagliesi,
S.~Bettarini,
T.~Boccali,
C.~Bozzi,$^{24}$
G.~Calderini,
R.~Dell'Orso,
I.~Ferrante,
A.~Giassi,
A.~Gregorio,
F.~Ligabue,
A.~Lusiani,
P.S.~Marrocchesi,
A.~Messineo,
F.~Palla,
G.~Rizzo,
G.~Sanguinetti,
G.~Sguazzoni,
R.~Tenchini,
C.~Vannini,
A.~Venturi,
P.G.~Verdini
\samepage
\begin{center}
\parbox{15.5cm}{\sl\samepage
Dipartimento di Fisica dell'Universit\`a, INFN Sezione di Pisa,
e Scuola Normale Superiore, I-56010 Pisa, Italy}
\end{center}\end{sloppypar}
\vspace{2mm}
\begin{sloppypar}
\noindent
G.A.~Blair,
J.~Coles,
G.~Cowan,
M.G.~Green,
D.E.~Hutchcroft,
L.T.~Jones,
T.~Medcalf,
J.A.~Strong,
J.H.~von~Wimmersperg-Toeller
\nopagebreak
\begin{center}
\parbox{15.5cm}{\sl\samepage
Department of Physics, Royal Holloway \& Bedford New College,
University of London, Surrey TW20 OEX, United Kingdom$^{10}$}
\end{center}\end{sloppypar}
\vspace{2mm}
\begin{sloppypar}
\noindent
D.R.~Botterill,
R.W.~Clifft,
T.R.~Edgecock,
P.R.~Norton,
J.C.~Thompson
\nopagebreak
\begin{center}
\parbox{15.5cm}{\sl\samepage
Particle Physics Dept., Rutherford Appleton Laboratory,
Chilton, Didcot, Oxon OX11 OQX, United Kingdom$^{10}$}
\end{center}\end{sloppypar}
\vspace{2mm}
%\pagebreak
\begin{sloppypar}
\noindent
\mbox{B.~Bloch-Devaux},
P.~Colas,
B.~Fabbro,
G.~Fa\"if,
E.~Lan\c{c}on,
\mbox{M.-C.~Lemaire},
E.~Locci,
P.~Perez,
H.~Przysiezniak,
J.~Rander,
\mbox{J.-F.~Renardy},
A.~Rosowsky,
A.~Trabelsi,$^{21}$
B.~Tuchming,
B.~Vallage
\nopagebreak
\begin{center}
\parbox{15.5cm}{\sl\samepage
CEA, DAPNIA/Service de Physique des Particules,
CE-Saclay, F-91191 Gif-sur-Yvette Cedex, France$^{17}$}
\end{center}\end{sloppypar}
\pagebreak
\vspace{2mm}
\begin{sloppypar}
\noindent
S.N.~Black,
J.H.~Dann,
H.Y.~Kim,
N.~Konstantinidis,
A.M.~Litke,
M.A. McNeil,
G.~Taylor
\nopagebreak
\begin{center}
\parbox{15.5cm}{\sl\samepage
Institute for Particle Physics, University of California at
Santa Cruz, Santa Cruz, CA 95064, USA$^{22}$}
\end{center}\end{sloppypar}
%\pagebreak
\vspace{2mm}
\begin{sloppypar}
\noindent
C.N.~Booth,
S.~Cartwright,
F.~Combley,
P.N.~Hodgson,
M.S.~Kelly,
M.~Lehto,
L.F.~Thompson
\nopagebreak
\begin{center}
\parbox{15.5cm}{\sl\samepage
Department of Physics, University of Sheffield, Sheffield S3 7RH,
United Kingdom$^{10}$}
\end{center}\end{sloppypar}
\vspace{2mm}
\begin{sloppypar}
\noindent
K.~Affholderbach,
A.~B\"ohrer,
S.~Brandt,
C.~Grupen,
A.~Misiejuk,
G.~Prange,
U.~Sieler
\nopagebreak
\begin{center}
\parbox{15.5cm}{\sl\samepage
Fachbereich Physik, Universit\"at Siegen, D-57068 Siegen, Germany$^{16}$}
\end{center}\end{sloppypar}
\vspace{2mm}
\begin{sloppypar}
\noindent
G.~Giannini,
B.~Gobbo
\nopagebreak
\begin{center}
\parbox{15.5cm}{\sl\samepage
Dipartimento di Fisica, Universit\`a di Trieste e INFN Sezione di Trieste,
I-34127 Trieste, Italy}
\end{center}\end{sloppypar}
\vspace{2mm}
\begin{sloppypar}
\noindent
J.~Putz,
J.~Rothberg,
S.~Wasserbaech,
R.W.~Williams
\nopagebreak
\begin{center}
\parbox{15.5cm}{\sl\samepage
Experimental Elementary Particle Physics, University of Washington, WA 98195
Seattle, U.S.A.}
\end{center}\end{sloppypar}
\vspace{2mm}
\begin{sloppypar}
\noindent
S.R.~Armstrong,
E.~Charles,
P.~Elmer,
D.P.S.~Ferguson,
Y.~Gao,
S.~Gonz\'{a}lez,
T.C.~Greening,
O.J.~Hayes,
H.~Hu,
S.~Jin,
P.A.~McNamara III,
J.M.~Nachtman,$^{2}$
J.~Nielsen,
W.~Orejudos,
Y.B.~Pan,
Y.~Saadi,
I.J.~Scott,
J.~Walsh,
Sau~Lan~Wu,
X.~Wu,
G.~Zobernig
\nopagebreak
\begin{center}
\parbox{15.5cm}{\sl\samepage
Department of Physics, University of Wisconsin, Madison, WI 53706,
USA$^{11}$}
\end{center}\end{sloppypar}
}
\footnotetext[1]{Also at CERN, 1211 Geneva 23, Switzerland.}
\footnotetext[2]{Now at University of California at Los Angeles (UCLA),
Los Angeles, CA 90024, U.S.A.}
\footnotetext[3]{Also at Dipartimento di Fisica, INFN Sezione di Catania,
95129 Catania, Italy.}
\footnotetext[4]{Also Istituto di Fisica Generale, Universit\`{a} di
Torino, 10125 Torino, Italy.}
\footnotetext[5]{Also Istituto di Cosmo-Geofisica del C.N.R., Torino,
Italy.}
\footnotetext[6]{Supported by the Commission of the European Communities,
contract ERBFMBICT982894.}
\footnotetext[7]{Supported by CICYT, Spain.}
\footnotetext[8]{Supported by the National Science Foundation of China.}
\footnotetext[9]{Supported by the Danish Natural Science Research Council.}
\footnotetext[10]{Supported by the UK Particle Physics and Astronomy Research
Council.}
\footnotetext[11]{Supported by the US Department of Energy, grant
DE-FG0295-ER40896.}
\footnotetext[12]{Now at LAL, 91898 Orsay, France.}
\footnotetext[13]{Supported by the US Department of Energy, contract
DE-FG05-92ER40742.}
\footnotetext[14]{Supported by the US Department of Energy, contract
DE-FC05-85ER250000.}
\footnotetext[15]{Permanent address: Universitat de Barcelona, 08208 Barcelona,
Spain.}
\footnotetext[16]{Supported by the Bundesministerium f\"ur Bildung,
Wissenschaft, Forschung und Technologie, Germany.}
\footnotetext[17]{Supported by the Direction des Sciences de la
Mati\`ere, C.E.A.}
\footnotetext[18]{Supported by Fonds zur F\"orderung der wissenschaftlichen
Forschung, Austria.}
\footnotetext[19]{Now at SAP AG, 69185 Walldorf, Germany}
\footnotetext[20]{Now at Harvard University, Cambridge, MA 02138, U.S.A.}
\footnotetext[21]{Now at D\'epartement de Physique, Facult\'e des Sciences de Tunis, 1060 Le Belv\'ed\`ere, Tunisia.}
\footnotetext[22]{Supported by the US Department of Energy,
grant DE-FG03-92ER40689.}
\footnotetext[23]{Now at Universit\'e de Lausanne, 1015 Lausanne, Switzerland.}
\footnotetext[24]{Now at INFN Sezione di Ferrara, 44100 Ferrara, Italy.}
%
% restore the previous settings
%
\setlength{\parskip}{\saveparskip}
\setlength{\textheight}{\savetextheight}
\setlength{\topmargin}{\savetopmargin}
\setlength{\textwidth}{\savetextwidth}
\setlength{\oddsidemargin}{\saveoddsidemargin}
\setlength{\topsep}{\savetopsep}
%%%%%%%%%%%%%%%%%%%%%%%%%%%%%%%%%%%%%%%%%
\normalsize
\newpage
\pagestyle{plain}
\setcounter{page}{1}

\clearpage

\pagestyle{plain}
\setcounter{page}{1}

%%%%%%%%
%--- section 1 : introduction + def + hadrons
%%%%%%%%
%\input ew2_conf_vanc_had.tex
%%%%%%%%%%%%%%%%%%%%%%%%%%%%%%%%%%%%%%%%%%%%%%%%%%%%%%%%
%
% Marie-Noelle's stuff: intro, definitions, hadrons etc
%
%%%%%%%%%%%%%%%%%%%%%%%%%%%%%%%%%%%%%%%%%%%%%%%%%%%%%%%%
 
\section{Introduction}
 \label{intro}

In the period 1995-97, the centre-of-mass energy of the LEP \ee\ collider was 
increased in five energy steps from 130 to 183~GeV, so opening a new energy 
regime for electroweak cross section and asymmetry measurements. Such 
measurements provide a test of the Standard Model (SM) and allow one to 
place limits on possible extensions to it.

This paper begins by providing in Section~\ref{def} the definitions of
cross section and asymmetry used here. A brief description of the ALEPH 
detector is given in Section~\ref{detector} and details of the luminosity 
measurement and data/Monte Carlo samples in Section~\ref{lumi}.

Section~\ref{sec:had} describes the measurement of the \qq\ cross section.
It also explores heavy quark production, providing measurements of 
\Rb\ (\Rc) which are here defined as the ratio of the \bb\ (\cc) cross
section to the total \qq\ cross section. Constraints on \qq\
forward-backward asymmetries are obtained using jet charge measurements.
In Section~\ref{sec:leptons}, cross section and asymmetry results are
reported for the three lepton species.

Based on these results, Section~\ref{interpretations} gives limits
on extensions to the SM involving contact interactions,
R-parity violating sneutrinos, leptoquarks, \ZP~bosons and R-parity violating 
squarks.

%%%%%%%%%%%%%%%%%%%%%%%%%%%%%%%%%%%%%%%%%%%%%%%%%%%%%%%%
% def
%%%%%%%%%%%%%%%%%%%%%%%%%%%%%%%%%%%%%%%%%%%%%%%%%%%%%%%%
\section{Definition of Cross Section and Asymmetry}
\label{def}

Cross section results for all fermion species are provided for

\begin{enumerate}
  \item
   the {\it inclusive} process, comprising all events with
   $\sqrt{s^{\prime}/s}>0.1$, so including events having hard initial 
   state radiation (ISR).
  \item
   the {\it exclusive} process, comprising all events with 
   $\sqrt{s^{\prime}/s}>0.9$, so excluding radiative events, such 
   as those in which a return to the Z~resonance occurs.
\end{enumerate}    
Here, the variable $s$ is the square of the centre-of-mass 
energy. For leptonic final states the variable $s^{\prime}$ is 
defined as the square of the mass of the outgoing lepton pair.
For hadronic final states $s^{\prime}$ is defined as the mass squared of 
the Z/$\gamma^*$ propagator. This latter choice is necessary, because as 
a result of gluon radiation, (which may occur before or after final state 
photon radiation), the mass of the outgoing quark pair is not well defined.

Interference effects between ISR and final state radiative (FSR) photons
affect the exclusive cross sections at the level of a few percent
and are not accurately described by existing Monte Carlo (MC) generators.
They are particularly prominent when the outgoing fermions make a small
angle to the incoming \ee\ beams. To reduce uncertainties related to this,
the exclusive cross section and asymmetry results presented here are
defined so as to include only the polar angle region $|{\cos\theta}| < 0.95$,
where $\theta$ is the polar angle of the outgoing fermion. This is 
unnecessary for the inclusive cross sections, since they are relatively
insensitive to radiative photons. The inclusive results are therefore
defined to include the full angular acceptance.

When selecting events experimentally, the variable $s^{\prime}_m$ 
is used, which provides a good approximation to $s^\prime$ when only
one ISR photon is present:
\begin{equation}
 s^{\prime}_m\;=\;\frac{\sin\theta_1\;+\;\sin\theta_2\;-\;
|{\sin(\theta_1+\theta_2)}|}{\sin\theta_1\;+\;\sin\theta_2\;+\;
|{\sin(\theta_1+\theta_2)}|}\;\times\;s~.
\end{equation}
Here $\theta_1$ and $\theta_2$ are the angles of the final state 
fermions f and $\bar{\mathrm f}$ measured with respect to the direction of the 
incoming e$^-$ beam or with respect to the direction of a photon 
seen in the apparatus and consistent with ISR. If two or more such photons
are found, the angles are measured with respect to the sum of their 
three-momenta. The fermion flight directions 
are determined in electron and muon pair events simply from the 
directions of the reconstructed tracks, in tau pair events from the jets 
reconstructed from the visible tau decay products,
and in hadronic events from the jets formed when forcing the event into 
two jets  after removing isolated, high energy photons detected in the 
apparatus.

If an event contains two or more ISR photons, then unless these photons all
go in the same direction, the variable $s^{\prime}_m$ ceases to be a good 
approximation to $s^{\prime}$. Such events can pass the exclusive selection 
by being reconstructed with $\sqrt{s^{\prime}_m/s}>0.9$, but nonetheless 
have $\sqrt{s^{\prime}/s}<0.9$. They are called `radiative background'.

For dilepton events, the differential cross section is measured
as a function of the angle $\theta^*$, defined by
\begin{equation}
\cos\theta^* = \frac{\sin\frac{1}{2}(\theta_+ - \theta_-)}
                    {\sin\frac{1}{2}(\theta_+ + \theta_-)}~,
\end{equation}
where $\theta_-$ and $\theta_+$ are the angles of the negatively
and positively charged leptons, respectively, with respect to the incoming 
e$^-$ beam. The angle $\theta^*$ corresponds to the scattering angle between
the incoming e$^-$ and the outgoing l$^-$, measured in the \myll\ 
rest frame, provided that no large-angle ISR photons are present.

The forward-backward asymmetries are determined from the formula
\begin{equation}
    \Afb{} = \frac{\sigma_F - \sigma_B}{\sigma_F + \sigma_B}~,
\end{equation}
where $\sigma_F$ and $\sigma_B$ are the cross sections to produce events with
the negatively charged lepton in the forward ($\theta_- < 90^\circ$) and
backward ($\theta_- > 90^\circ$) hemispheres, respectively, defined in the
same limited angular acceptance as given above. Determining the forward-backward
asymmetries from this formula avoids any specific assumption about 
the angular dependence of the differential cross section.

%%%%%%%%%%%%%%%%%%%%%%%%%%%%%%%%%%%%%%%%%%%%%%%%%%%%%%%%
% detector
%%%%%%%%%%%%%%%%%%%%%%%%%%%%%%%%%%%%%%%%%%%%%%%%%%%%%%%%
\section{The ALEPH Detector}
 \label{detector}

The ALEPH detector and performance are fully described in \cite{ALDET}
and \cite{ALPERFORM}. In October 1995, the silicon vertex detector (VDET)
described in these papers was replaced by an improved detector \cite{VDET}, 
which is used for the analyses presented here. A brief description of the
ALEPH detector follows.

The main tracking detector is a time projection chamber (TPC) lying between
radii of 30 and 180~cm from the beam axis. It provides up to 21 
three-dimensional coordinates per track. Inside the TPC is a small drift 
chamber (ITC) and within this, the new VDET. The latter has two layers
of silicon, each providing three-dimensional coordinates. All three
tracking detectors contribute coordinates to tracks for polar angles
to the beam axis up to $|{\cos\theta}| < 0.95$. They are immersed in a 1.5~T
axial magnetic field provided by a superconducting solenoid. This allows
the momentum $p$ of charged tracks to be measured with a resolution of 
$\sigma(p)/p = 6\times 10^{-4} p_T\oplus 0.005$ (where $p_T$ is the momentum
component perpendicular to the beam axis in \GeVc). The three-dimensional
impact parameter resolution is measured with an accuracy of 
$(34 + 70/p)\times (1 + 1.6\cos^4\theta)$~$\mu$m (where $p$ is measured 
in \GeVc).

Between the tracking detectors and the solenoid is an electromagnetic 
calorimeter (ECAL), which 
provides identification of electrons and photons, and measures their
energies $E$ with a resolution of 
$\sigma(E)/E = 0.18/\sqrt{E({\mathrm GeV})} + 0.009$. Outside the solenoid, 
is a hadron calorimeter (HCAL), which, combined with the ECAL, 
measures the energy of hadrons with a resolution of 
$\sigma(E)/E = 0.85/\sqrt{E({\mathrm GeV})}$. The HCAL is also used for
muon identification, together with muon chambers lying outside it.
The ECAL and HCAL acceptances extend down to polar angles of 190 and 100 mrad
to the beam axis, respectively.

The luminosity is measured with a lead/proportional-chamber electromagnetic 
calorimeter (LCAL) covering the small angle region between 46 and 122 mrad
from the beam axis. A tungsten/silicon electromagnetic calorimeter (SICAL)
covering the angular range from 24 to 58 mrad is used to provide a cross-check.

In general, charged tracks are considered {\it good} for the analyses presented
here if they originate within a cylinder of radius 2~cm and length 10~cm, 
centered at the interaction point, and whose axis is parallel to the beam 
axis. They must also have at least 
four TPC hits, a momentum larger than 0.1~\GeVc\ and a polar angle to the
beam axis satisfying $|{\cos\theta}| < 0.95$.

By relating charged tracks to energy deposits found in the 
calorimeters and using photon, electron and muon identification information,
a list of charged and neutral {\it energy flow particles} \cite{ALPERFORM} is 
created for each event and used in the following analyses.

%%%%%%%%%%%%%%%%%%%%%%%%%%%%%%%%%%%%%%%%%%%%%%%%%%%%%%%%
% lumi
%%%%%%%%%%%%%%%%%%%%%%%%%%%%%%%%%%%%%%%%%%%%%%%%%%%%%%%%
\section{Data and Monte Carlo Samples and the Luminosity Measurement}
 \label{lumi}

The data used were taken at five centre-of-mass energies 
which are given in Table~\ref{tab:lumi}. This table also shows the
integrated luminosity recorded at each energy point, together with its 
statistical and systematic uncertainties.
\begin{table}[htbp]
  \mycaption{Centre-of-mass energies and integrated luminosities of the 
           high energy data samples. The two uncertainties quoted on each 
           integrated luminosity correspond to its statistical and systematic
           uncertainty respectively. \label{tab:lumi}}

  \begin{center}
    \begin{tabular}{|c|c|}
\hline
Energy (GeV) & Luminosity (pb$^{-1}$) \\
\hline
 130.2 & $~6.03 \pm 0.03 \pm 0.05$ \\
 136.2 & $~6.10 \pm 0.03 \pm 0.05$ \\
 161.3 & $11.08 \pm 0.04 \pm 0.07$ \\
 172.1 & $10.65 \pm 0.05 \pm 0.06$ \\
 182.7 & $56.78 \pm 0.11 \pm 0.29$ \\
\hline
    \end{tabular}
  \end{center}
\end{table}
For analyses relying heavily on the VDET, such as the study of \bb\ production,
data collected at 130 and 136~GeV in 1995 are discarded, 
since the new VDET was not fully installed. However, data
taken at these two energy points in 1997, corresponding to integrated 
luminosities of 3.30 and 3.51~$\mathrm{pb}^{-1}$, respectively, are used.

The luminosity is measured using the LCAL calorimeter following the analysis 
procedure described in Ref.~\cite{LCAL}, with a slightly reduced 
acceptance due to the shadowing of the LCAL detector by the SICAL 
below 59~mrad. The systematic uncertainty on the luminosity, which is assumed to be 
fully correlated between the different energy points, includes a 
theoretical uncertainty, estimated to be 0.25\%~\cite{LUMI_ERR} from the 
BHLUMI~\cite{BHLUMI} Monte Carlo generator. The luminosity measurement is 
checked by comparison with the luminosity measured independently using the 
SICAL detector. The two are consistent within the estimated uncertainties. 

Samples of Monte Carlo events were produced as follows. The generator 
BHWIDE~v1.01 \cite{BHWIDE} is used for the electron pair channel and 
KORALZ~v4.2 \cite{KORALZ} for the muon and tau pair channels. 
Simulation of diquark events relies on production of the initial \qq\ system 
and accompanying ISR photons with either KORALZ or PYTHIA~v5.7 
\cite{PYTHIA}. The simulation of FSR and fragmentation is then 
carried out with the program JETSET~v7.4 \cite{PYTHIA}. 
The PYTHIA generator is also used for four-fermion 
processes such as the Z pair and Ze$^+$e$^-$ channels. The programs 
PHOT02~\cite{PHOJET}, HERWIG~v5.9~\cite{HERWIG} and PYTHIA are used to generate 
the two-photon events. Finally, backgrounds from W~pair production are studied 
using the generators KORALW~v1.21~\cite{KORALW} and EXCALIBUR~\cite{EXCALI}.

%%%%%%%%%%%%%%%%%%%%%%%%%%%%%%%%%%%%%%%%%%%%%%%%%%%%%%%%
% had
%%%%%%%%%%%%%%%%%%%%%%%%%%%%%%%%%%%%%%%%%%%%%%%%%%%%%%%%
\section{Hadronic Final States}
\label{sec:had}

Section~\ref{subhad} describes the measurement of the \qq\ cross section.
Sections~\ref{rb} and \ref{rc} study heavy quark production, providing 
measurements of \Rb\ (\Rc) which are here defined as the ratio of the 
\bb\ (\cc) cross section to the total \qq\ cross section. The measured
values of \Rb\ (\Rc) are statistically independent of the \qq\ cross 
section measurement. Furthermore, by measuring these ratios, rather than
the \bb\ and \cc\ cross sections, one benefits from the cancellation of 
some systematic uncertainties.
Section~\ref{jet_charge} places constraints upon \qq\ forward-backward
asymmetries using a jet charge technique applied to b-enriched and b-depleted
event samples.

%%%%%%%%%%%%%%%%%%%%%%%%%%%%%%%%%%%%%%%%%%%%%%%%%%%%%%%%
% subhad
\
\subsection {The Hadronic Cross Section }
\label{subhad}

\begin{figure}[htbp]
\mbox{\epsfig{file=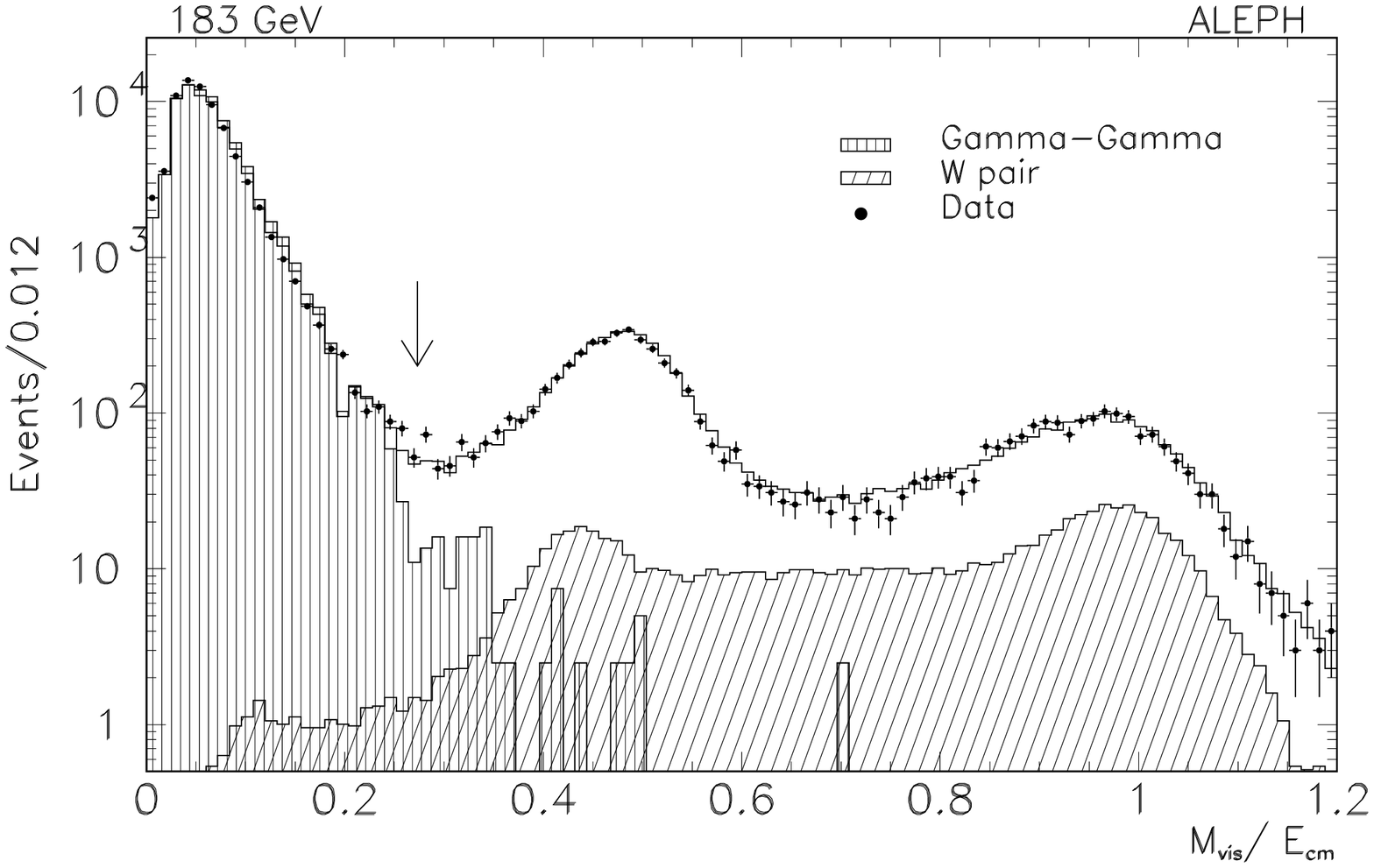,width=0.9\textwidth}} 
\mycaption{\label{evis_183}
         Visible mass distribution at a centre-of-mass energy of 183~GeV for 
         events having at least seven tracks. The white histogram shows
         the expected signal, whilst the hashed histograms give the expected
         contributions of the $\gamma\gamma$ and W~pair backgrounds.
         The arrow indicates the cut used in the inclusive selection.}
\end{figure}

The hadronic event selection begins by requiring events to have at least 
seven good charged tracks. The energy flow particles are then clustered into 
jets using the JADE algorithm~\cite{JADE} with a clustering 
parameter $y_{\mathrm cut}$ of 0.008~. Thin, low multiplicity jets 
with an electromagnetic energy content 
of at least 90\% and an energy of more than 10~GeV are considered to be ISR 
photon candidates. 
The visible mass \Mvis\ of the event is then measured using charged and 
neutral energy flow particles, but excluding these photons and 
energy flow particles which make an angle of less than $2^\circ$ to the beam 
axis. The distribution of \Mvis\ is shown in Fig.~\ref{evis_183} for 
data taken at 183~GeV. It is required to be more than 50~\GeVcc.

The inclusive selection makes the additional requirement that 
$\sqrt{s^{\prime}_m/s}>0.1$. There is actually negligible acceptance for 
events with $0.1 < \sqrt{s^{\prime}/s} < 0.3$ as a result of the cut on 
\Mvis. The inclusive cross sections are therefore extrapolated down to 
$\sqrt{s^{\prime}/s} = 0.1$ using the KORALZ generator.

\begin{figure}[p]
\mbox{\epsfig{file=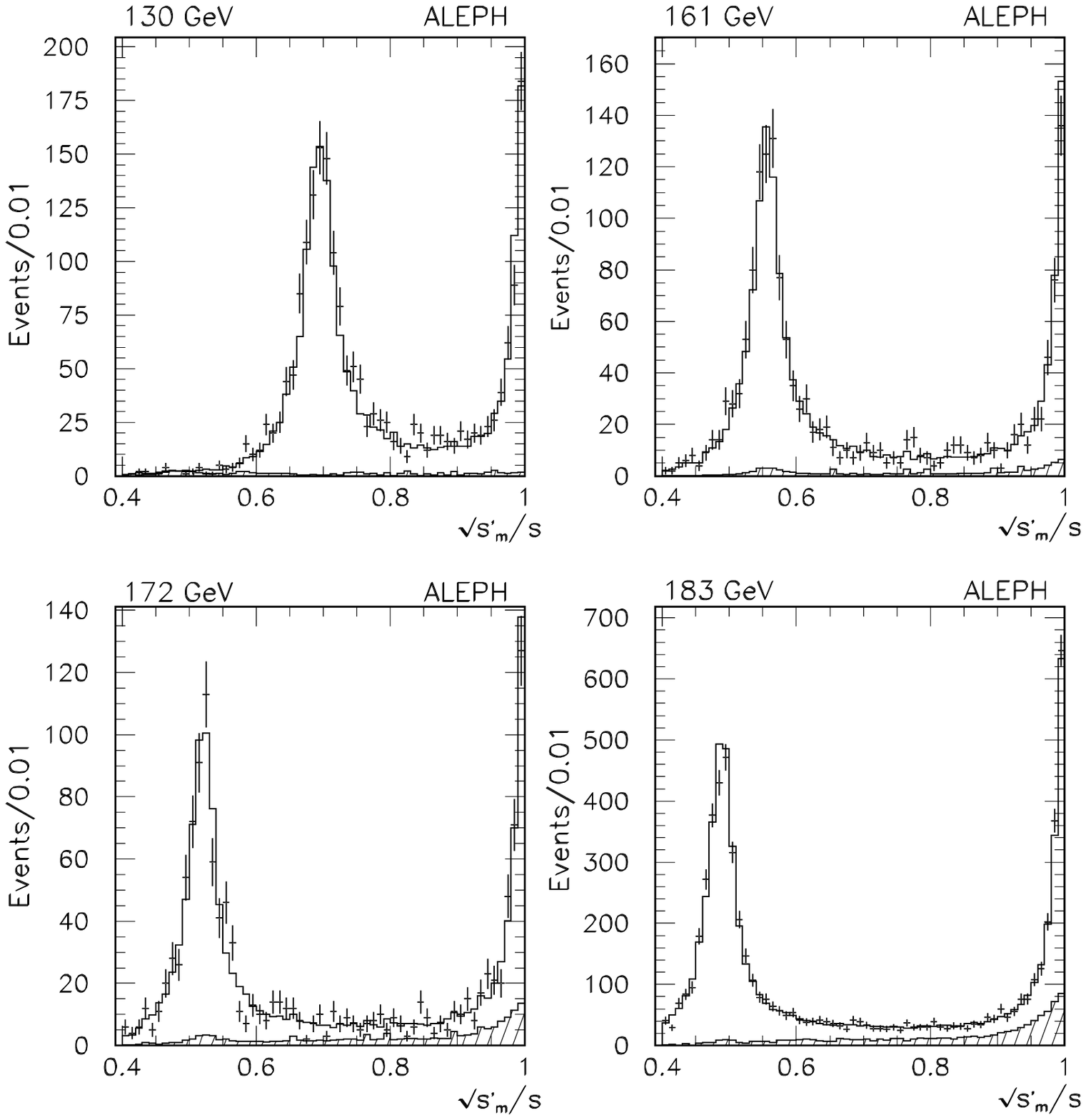,width=0.9\textwidth}} 
\mycaption{\label{sprim_all}
         $\sqrt{s^{\prime}_m/s}$ distribution for hadronic
         events at centre-of-mass energies from 130 to 183 GeV.
         This is compared to the Monte Carlo expectations shown by the white 
         histograms. The hashed areas correspond to background
         contributions and are dominated by W pair production. The
         distribution at 136~GeV is omitted because it closely resembles
         that at 130~GeV.}
\end{figure}

The exclusive selection requires $\sqrt{s^{\prime}_m/s}>0.9$. Here $s^{\prime}_m$ is determined from the
reconstructed jet directions, when the event is clustered into two jets,
after first removing reconstructed ISR photons. 
For the exclusive selection, these jets are required to have 
$|{\cos\theta}| < 0.95$~.
The $\sqrt{s^{\prime}_{m}/s}$ distributions for centre-of-mass energies of 
130, 161, 172 and 183~GeV are displayed in Fig.~\ref{sprim_all}, together 
with the expected background.

\begin{figure}[htbp]
\mbox{\epsfig{file=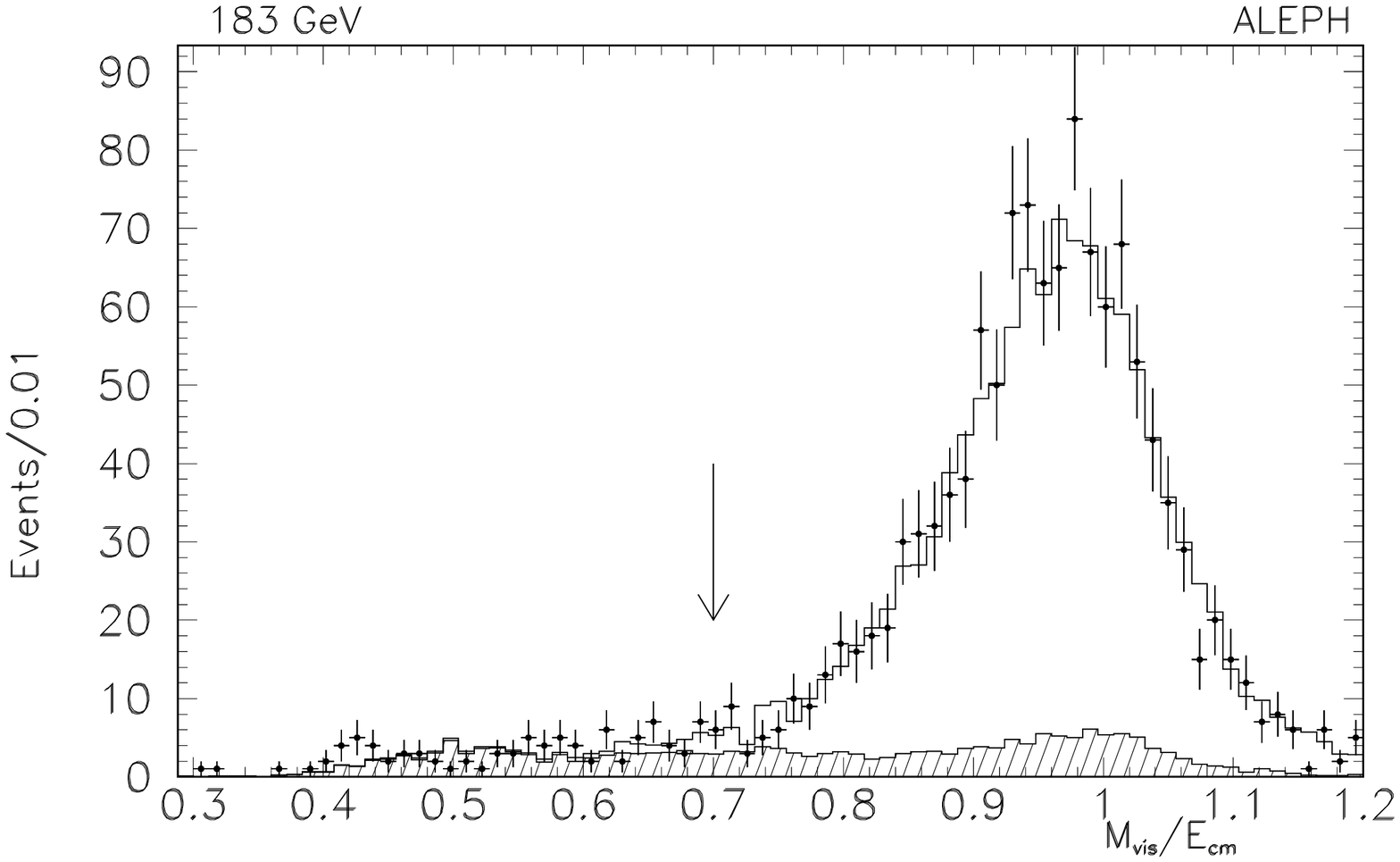,width=0.9\textwidth}} 
\mycaption{\label{mvis_183}
         Visible mass distribution at 183~GeV centre-of-mass energy for 
         hadronic events with $\sqrt{s^{\prime}_m/s}>0.9$. This is compared 
         with Monte Carlo expectations shown by the white histogram.
         The hashed area represents the expected contribution from events 
         with generated $\sqrt{s^{\prime}/s}<0.9$. The arrow indicates the
         cut used in the exclusive selection.}
\end{figure}

\begin{figure}[htbp]
\mbox{\epsfig{file=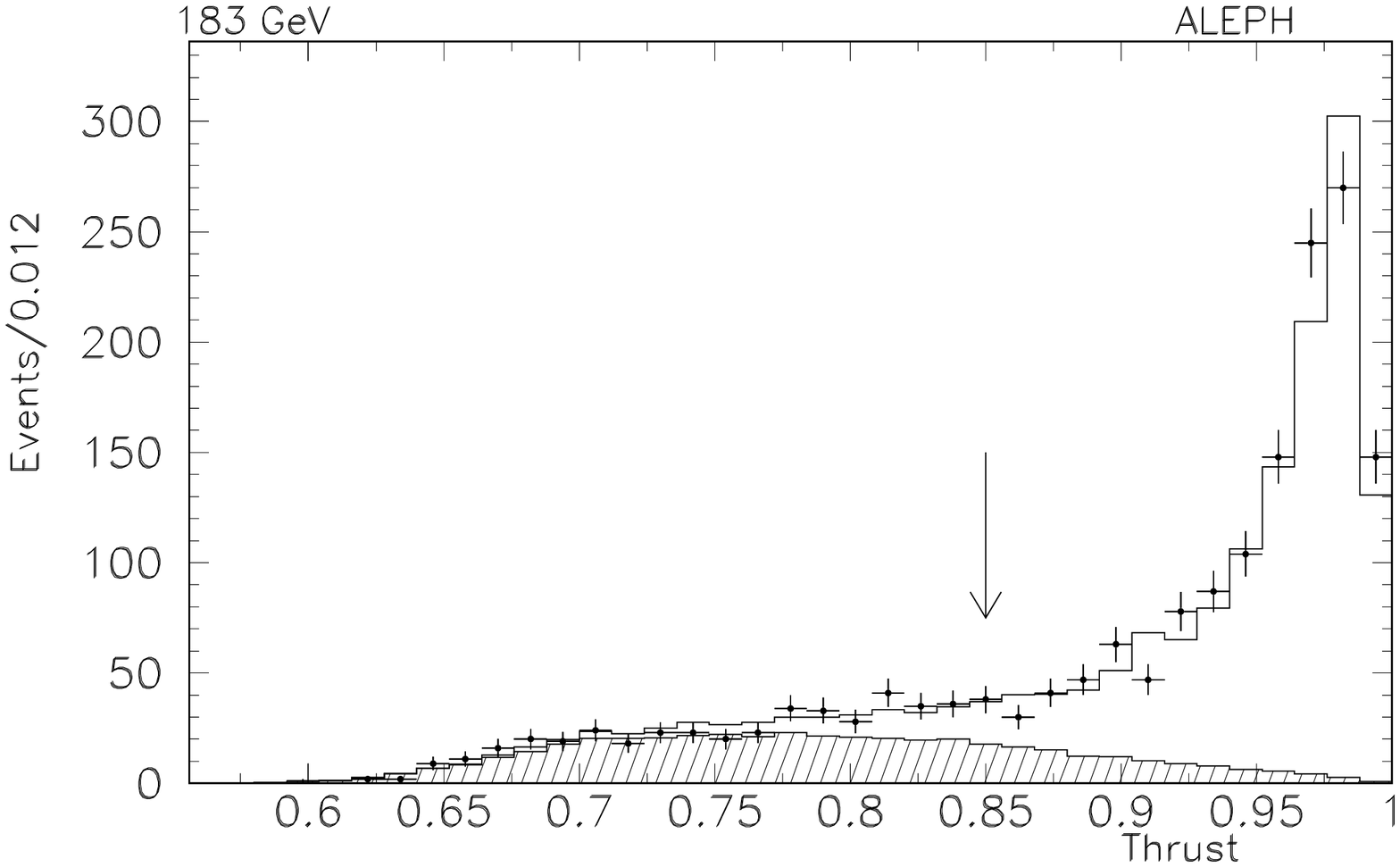,width=0.9\textwidth}} 
\mycaption{\label{thrust_183}
         Thrust distribution at 183~GeV centre-of-mass energy for 
         hadronic events with $\sqrt{s^{\prime}_m/s}>0.9$. The white histogram
         shows the expected signal and the hashed one corresponds to the 
         W pair background. The arrow indicates the cut used in the exclusive 
         selection.}
\end{figure}

For the exclusive process, two additional cuts are then applied. Firstly,
\Mvis\  is required to exceed 70\% of the centre-of-mass energy. 
This suppresses residual events with a radiative return to the Z. Such events can
have $\sqrt{s^{\prime}_m/s}>0.9$ if they emit two or more ISR photons.
Figure~\ref{mvis_183} shows \Mvis\ for events with 
$\sqrt{s^{\prime}_m/s} > 0.9$. The contribution from doubly radiative 
events is clearly seen at low masses.
Secondly, when above the \WW\ threshold, (i.e. $\sqrt{s}\geq 161$~GeV), about 
80\% of \WW\ background is eliminated by requiring that the thrust of the 
event exceeds 0.85~. The thrust distribution is shown in Fig.~\ref{thrust_183}.

%\FIGS{sprim_all}{0.9\textwidth}
%        {$\sqrt{s^{\prime}_m/s}$ distribution for hadronic
%         events at centre-of-mass energies from 130 to 183 GeV.
%         This is compared to the Monte Carlo expectations shown by the white 
%         histograms. The hashed areas correspond to background
%         contributions and are dominated by W pair production. The
%         distribution at 136~GeV is omitted since it closely resembles
%         the one at 130~GeV.}

The selection efficiencies are estimated using the KORALZ Monte Carlo 
samples. The efficiencies are given for all centre-of-mass 
energies in Table~\ref{EFF_ALL}. 
%
%%%%%%%%
%--- table : efficiencies + backgrounds
%%%%%%%%
%\input ew2_conf_vanc_tabeff.tex
%%%%%%%%%%%%%%%%%%%%%%%%%%%%%%%%%%%%%%%%%%%%%%%%%%%%%%%%
%
% table: efficiency + backgrounds
%
%%%%%%%%%%%%%%%%%%%%%%%%%%%%%%%%%%%%%%%%%%%%%%%%%%%%%%%%

\begin{table}[p]
\mycaption{\label{EFF_ALL} 
   Selection efficiencies and background fractions. For \ee\ production, 
   numbers are given for (1) $-0.9<\cos\theta^*<0.9$ and 
   (2) $-0.9<\cos\theta^*<0.7$.}
\begin{center}
\begin{tabular}{| c | c | c || c | c |}
\hline\strutl
 $\sqrt{s^{\prime}/s}$ & $\mathrm E_{cm}$  & Event & Efficiency & Background \\
        cut            &       (GeV)       &  type &    (\%)    &    (\%)    \\
\hline\hline
  0.1  &  130  &     \qq          &  $89.6\pm 0.9$  & $~2.4\pm 0.3$  \\
       &       &  $\mu^+\mu^-$    &  $78.8\pm 1.6$  & $~0.4\pm 0.1$  \\
       &       &  $\tau^+\tau^-$  &  $55.3\pm 1.1$  & $~5.7\pm 2.4$  \\
\cline{2-5}
       &  136  &     \qq          &  $89.5\pm 0.7$  & $~1.6\pm 0.2$  \\
       &       &  $\mu^+\mu^-$    &  $78.8\pm 1.6$  & $~0.5\pm 0.1$  \\
       &       &  $\tau^+\tau^-$  &  $53.9\pm 1.3$  & $~8.9\pm 4.0$  \\
\cline{2-5}
       &  161  &     \qq          &  $88.4\pm 0.6$  & $~4.0\pm 0.1$  \\
       &       &  $\mu^+\mu^-$    &  $76.5\pm 1.5$  & $~3.3\pm 1.2$  \\
       &       &  $\tau^+\tau^-$  &  $44.2\pm 1.4$  & $~8.3\pm 4.0$  \\
\cline{2-5}
       &  172  &     \qq          &  $87.3\pm 0.6$  & $10.2\pm 0.1$  \\
       &       &  $\mu^+\mu^-$    &  $75.6\pm 1.6$  & $~4.1\pm 1.3$  \\
       &       &  $\tau^+\tau^-$  &  $44.8\pm 1.0$  & $11.3\pm 3.7$  \\
\cline{2-5}
       &  183  &     \qq          &  $83.9\pm 0.7$  & $17.8\pm 0.2$  \\
       &       &  $\mu^+\mu^-$    &  $75.6\pm 1.6$  & $~5.4\pm 1.2$  \\
       &       &  $\tau^+\tau^-$  &  $47.4\pm 0.9$  & $12.9\pm 2.6$  \\
\hline\hline
  0.9  &  130  &     \qq          &  $92.2\pm 1.0$  & $~9.5\pm 0.3$  \\
       &       &  \ee (1)         &  $90.0\pm 1.6$  & $10.4\pm 0.9$  \\
       &       &  \ee (2)         &  $95.5\pm 1.4$  & $11.5\pm 1.0$  \\
       &       &  $\mu^+\mu^-$    &  $95.1\pm 1.8$  & $~3.3\pm 0.7$  \\
       &       &  $\tau^+\tau^-$  &  $63.9\pm 1.6$  & $11.9\pm 3.2$  \\
\cline{2-5}
       &  136  &     \qq          &  $89.2\pm 0.8$  & $~8.8\pm 0.4$  \\
       &       &  \ee (1)         &  $90.6\pm 1.5$  & $10.8\pm 1.0$  \\
       &       &  \ee (2)         &  $96.9\pm 1.4$  & $14.1\pm 1.6$  \\
       &       &  $\mu^+\mu^-$    &  $95.1\pm 1.8$  & $~3.3\pm 0.7$  \\
       &       &  $\tau^+\tau^-$  &  $63.6\pm 1.5$  & $15.9\pm 3.1$  \\
\cline{2-5}
       &  161  &     \qq          &  $89.3\pm 0.7$  & $~7.7\pm 0.3$  \\
       &       &  \ee (1)         &  $89.1\pm 1.5$  & $10.0\pm 0.9$  \\
       &       &  \ee (2)         &  $94.8\pm 1.3$  & $11.0\pm 1.3$  \\
       &       &  $\mu^+\mu^-$    &  $96.0\pm 1.8$  & $~4.5\pm 0.9$  \\
       &       &  $\tau^+\tau^-$  &  $62.3\pm 1.7$  & $~9.5\pm 2.7$  \\
\cline{2-5}
       &  172  &     \qq          &  $90.2\pm 0.6$  & $~6.4\pm 0.3$  \\
       &       &  \ee (1)         &  $90.7\pm 1.5$  & $10.7\pm 1.0$  \\
       &       &  \ee (2)         &  $96.3\pm 1.2$  & $11.9\pm 1.4$  \\
       &       &  $\mu^+\mu^-$    &  $96.8\pm 1.8$  & $~6.0\pm 1.3$  \\
       &       &  $\tau^+\tau^-$  &  $64.6\pm 1.7$  & $13.9\pm 3.8$  \\
\cline{2-5}
       &  183  &     \qq          &  $85.9\pm 0.8$  & $~9.2\pm 0.1$  \\
       &       &  \ee (1)         &  $87.5\pm 1.4$  & $10.9\pm 0.7$  \\
       &       &  \ee (2)         &  $95.2\pm 1.2$  & $12.4\pm 0.9$  \\
       &       &  $\mu^+\mu^-$    &  $95.9\pm 1.8$  & $~6.5\pm 1.3$  \\
       &       &  $\tau^+\tau^-$  &  $67.9\pm 1.8$  & $13.3\pm 3.0$  \\
\hline
\end{tabular}
\end{center}
\end{table}
The uncertainties in the efficiencies arise 
from the statistical uncertainties on the Monte Carlo and also from a number 
of systematic effects, which are assessed as described below.

\begin{enumerate}

\item Uncertainties related to the simulation of ISR are estimated from
      the difference in the efficiencies determined from KORALZ and PYTHIA.
      Only the generator KORALZ simulates ISR using YFS exponentiation 
      \cite{KORALZ}.

\item The overall ECAL energy calibration is determined each year, from 
      Bhabha events recorded when running at the Z peak. Its uncertainty is
      estimated to be $\pm 0.9\%$, based upon a comparison of the detector 
      response to low energy ($\approx 1$~GeV) electrons in \ggee\ events, 
      with that to high energy electrons in Bhabha events.
      The calibration of the HCAL energy scale uses minimum ionizing 
      particles. The statistical uncertainty on this calibration is $\pm 2\%$. 
      These uncertainties on the calorimeter energy scales are considered 
      to be uncorrelated from year to year. 

\item The energy response to 45~GeV jets, in data and Monte Carlo events 
      produced at the Z resonance, is compared and the difference parametrized as 
      a function of polar angle to the beam axis. The selection efficiency is
      corrected using this parametrization. The systematic uncertainty due
      to the energy response is derived from the change in the measured
      cross section when this correction is applied.
      This uncertainty, which is the largest of the detector 
      related ones, is considered to be correlated from year to year.

\item The measured polar angles of jets with respect to the beam axis can 
      suffer from small systematic biases, particularly at low polar angles. 
      These biases are studied using events produced at the Z~resonance. 
      Differences between the data and Monte Carlo are taken into account via
      their effect on $s^{\prime}_m$. The associated systematic uncertainty
      is given by the change in the measured cross section when applying these
      corrections. It is considered to be correlated from year to year.
\end{enumerate}  

The background fractions remaining after the selection are also given
in Table~\ref{EFF_ALL}. The uncertainty in these backgrounds receives
contributions from the detector calibration and energy response, estimated
as above. Additional uncertainties are studied as follows:

\begin{enumerate}

\item The two-photon background is simulated with the PYTHIA, PHOT02 and 
      HERWIG generators. It is normalized to the data in the region 
      $M_{\mathrm vis} < 50$~\GeVcc\ (Fig.~\ref{evis_183}), and the difference 
      between the expected cross section from PHOT02 and the normalized one 
      is used to calculate the uncertainty on the process. This background 
      dominates at centre-of-mass energies of 130 and 136~GeV. 

\item Above the W pair production threshold, the \WW\ background is estimated 
      using KORALW. At 161~GeV the input cross section is taken
      from the theoretical expectation, computed with the GENTLE \cite{GENTLE}
      program for a W mass of $80.39\pm 0.06$~\GeVcc \cite{WW98}.
      At 172 and 183~GeV the cross sections are taken from the average
      measurement of the four LEP experiments~\cite{WW98}.
      Uncertainties in the W pair cross sections are propagated to 
      estimate the associated systematic uncertainties on the \qq\ cross sections.

\item Other four-fermion backgrounds such as ZZ and Ze$^+$e$^-$ are 
      estimated using PYTHIA and EXCALIBUR. They introduce only a $\pm 0.1$\%
      systematic uncertainty on the exclusive \qq\ cross section measurement.

\item Events where ISR occurs from the incoming electron and positron remain
      an important background for the exclusive selection. Systematic 
      uncertainties on this background arise from the simulation of ISR
      photons and from the energy scale and response of the detector. These
      are already taken into account as described above.
\end{enumerate} 

The cross section measurements at each centre-of-mass energy are listed in 
Table~\ref{CROSSALL}, where they are compared with the SM
predictions. These predictions are discussed in Section~\ref{smpred}.
For the exclusive results, the measurements and predictions 
refer to \qq\ final states with $|{\cos\theta}|<0.95$,
where $\theta$ is the polar angle of the quark with respect to the 
beam axis. A breakdown of contributions to the systematic uncertainties on 
the cross section measurements is given in Table~\ref{HADSYS}.

%%%%%%%%
%--- table : xsections
%%%%%%%%
%\input ew2_conf_vanc_tabxsec.tex
%%%%%%%%%%%%%%%%%%%%%%%%%%%%%%%%%%%%%%%%%%%%%%%%%%%%%%%%
%
% table: xsections
%
%%%%%%%%%%%%%%%%%%%%%%%%%%%%%%%%%%%%%%%%%%%%%%%%%%%%%%%%

\begin{table}[p]
\vspace{-1cm}
\mycaption{\label{CROSSALL}
   Measured cross sections with statistical and systematic uncertainties for 
   different channels at centre-of-mass energies from 130 to 183~GeV.
   The SM predictions are also given, together with 
   the number of selected events (before background subtraction). 
   The exclusive cross sections correspond to the restricted angular 
   range $|\cos \theta| < 0.95$. For the Bhabha process, results are given 
   for (1) $-0.9 < \cos\theta^* < 0.9$ and (2) $-0.9 < \cos\theta^* < 0.7$.}
\begin{center}
\begin{tabular}{| c | c | c || r | c | c |}
\hline\strutl
 $\sqrt{s^{\prime}/s}$ & $\mathrm E_{cm}$  & Event &  No.~~ & $\sigma$ & SM prediction \\
         cut           &      (GeV)        &  type & Events &   (pb)   &      (pb)      \\
\hline\hline
 0.1 & 130 &    \qq         & 1858 & $335.6~\pm 7.9~\pm 4.5~$ & 327.0~ \\
     &     & $\mu^+\mu^-$   &  110 & $~22.5~\pm 2.2~\pm 0.5~$ & ~21.9~ \\
     &     & $\tau^+\tau^-$ &   94 & $~25.9~\pm 2.9~\pm 0.6~$ & ~21.9~ \\
\cline{2-6}
     & 136 &    \qq         & 1558 & $280.8~\pm 7.2~\pm 3.7~$ & 269.8~ \\
     &     & $\mu^+\mu^-$   &  103 & $~20.4~\pm 2.1~\pm 0.5~$ & ~18.7~ \\
     &     & $\tau^+\tau^-$ &   67 & $~17.8~\pm 2.5~\pm 0.5~$ & ~18.6~ \\
\cline{2-6}
     & 161 &    \qq         & 1520 & $149.0~\pm 4.0~\pm 1.7~$ & 147.6~ \\
     &     & $\mu^+\mu^-$   &  107 & $~12.2~\pm 1.2~\pm 0.3~$ & ~11.2~ \\
     &     & $\tau^+\tau^-$ &   78 & $~14.6~\pm 1.8~\pm 0.3~$ & ~11.2~ \\
\cline{2-6}
     & 172 &    \qq         & 1270 & $122.6~\pm 3.8~\pm 1.2~$ & 122.8~ \\
     &     & $\mu^+\mu^-$   &   74 & $~~8.8~\pm 1.1~\pm 0.2~$ & ~~9.5~ \\
     &     & $\tau^+\tau^-$ &   51 & $~~9.5~\pm 1.5~\pm 0.3~$ & ~~9.5~ \\
\cline{2-6}
     & 183 &    \qq         & 6072 & $104.8~\pm 1.6~\pm 0.9~$ & 104.4~ \\
     &     & $\mu^+\mu^-$   &  406 & $~~8.84\pm 0.47\pm 0.19$ & ~~8.22 \\
     &     & $\tau^+\tau^-$ &  241 & $~~7.80\pm 0.58\pm 0.17$ & ~~8.22 \\
\hline\hline
 0.9 & 130 &    \qq         &  440 & $~71.6~\pm 3.8~\pm 1.1~$ & ~70.7~ \\
     &     & \ee (1)        & 1186 & $191.3~\pm 6.2~\pm 3.5~$ & 186.7~ \\
     &     & \ee (2)        &  274 & $~41.1~\pm 2.8~\pm 0.9~$ & ~39.0~ \\
     &     & $\mu^+\mu^-$   &   48 & $~~7.9~\pm 1.2~\pm 0.2~$ & ~~7.0~ \\
     &     & $\tau^+\tau^-$ &   49 & $~10.9~\pm 1.8~\pm 0.4~$ & ~~7.3~ \\
\cline{2-6}
     & 136 &    \qq         &  351 & $~58.8~\pm 3.5~\pm 0.9~$ & ~57.3~ \\
     &     & \ee (1)        & 1051 & $162.2~\pm 5.6~\pm 3.5~$ & 167.3~ \\
     &     & \ee (2)        &  212 & $~29.5~\pm 2.4~\pm 0.7~$ & ~33.9~ \\
     &     & $\mu^+\mu^-$   &   43 & $~~6.9~\pm 1.1~\pm 0.2~$ & ~~6.1~ \\
     &     & $\tau^+\tau^-$ &   27 & $~~5.6~\pm 1.3~\pm 0.2~$ & ~~6.3~ \\
\cline{2-6}
     & 161 &    \qq         &  321 & $~29.94\pm 1.8~\pm 0.4~$ & ~30.7~ \\
     &     & \ee (1)        & 1393 & $119.7~\pm 3.7~\pm 2.3~$ & 119.0~ \\
     &     & \ee (2)        &  302 & $~25.6~\pm 1.7~\pm 0.5~$ & ~24.8~ \\
     &     & $\mu^+\mu^-$   &   50 & $~~4.49\pm 0.69\pm 0.09$ & ~~3.88 \\
     &     & $\tau^+\tau^-$ &   44 & $~~5.75\pm 0.96\pm 0.23$ & ~~4.01 \\
\cline{2-6}
     & 172 &    \qq         &  271 & $~26.4~\pm 1.7~\pm 0.4~$ & ~25.1~ \\
     &     & \ee (1)        & 1166 & $107.8~\pm 3.5~\pm 2.1~$ & 102.5~ \\
     &     & \ee (2)        &  268 & $~23.0~\pm 1.6~\pm 0.5~$ & ~21.5~ \\
     &     & $\mu^+\mu^-$   &   29 & $~~2.64\pm 0.53\pm 0.06$ & ~~3.32 \\
     &     & $\tau^+\tau^-$ &   26 & $~~3.26\pm 0.74\pm 0.09$ & ~~3.43 \\
\cline{2-6}
     & 183 &    \qq         & 1165 & $~21.71\pm 0.70\pm 0.23$ & ~21.1~ \\
     &     & \ee (1)        & 5063 & $~90.9~\pm 1.4~\pm 1.7~$ & ~90.9~ \\
     &     & \ee (2)        & 1171 & $~18.99\pm 0.63\pm 0.36$ & ~19.11 \\
     &     & $\mu^+\mu^-$   &  175 & $~~2.98\pm 0.24\pm 0.06$ & ~~2.89 \\
     &     & $\tau^+\tau^-$ &  129 & $~~2.90\pm 0.29\pm 0.09$ & ~~2.98 \\
\hline
\end{tabular} 
\end{center}
\end{table}

%%%%%%%%%%%%%%%%%%%%%%%%%%%%%%%%%%%%%%%%%%%%%%%%%%%%%%%%
%
% table: systematic uncertainty breakdown
%
%%%%%%%%%%%%%%%%%%%%%%%%%%%%%%%%%%%%%%%%%%%%%%%%%%%%%%%%
\begin{table}[htbp]
\mycaption{\label{HADSYS}{Contributions to the systematic uncertainties on the 
\qq\ cross section measurements, for all energies and for both inclusive and 
exclusive processes. All quoted values are in percent.}}
\begin{center}
\begin{tabular}{| c | l || l | l | l | l | l |}
\hline\strutl
 $\sqrt{s^{\prime}/s}$ & \multicolumn{1}{c||}{Description} & 
                         \multicolumn{5}{c|}{$\mathrm E_{cm}$ (GeV)} \\
\cline{3-7}
          cut          &             & 130 & 136 & 161 & 172 & 183 \\ 
\hline\hline
  0.1 & MC statistics                & 0.3 & 0.3 & 0.3  & 0.3  & 0.1  \\
      & ISR simulation               & 0.3 & 0.3 & 0.3  & 0.3  & 0.3  \\
      & Energy scale                 & 0.4 & 0.5 & 0.4  & 0.4  & 0.3  \\
      & Detector response            & 0.6 & 0.5 & 0.5  & 0.5  & 0.6  \\ 
      & \ggqq                        & 0.3 & 0.2 & 0.05 & 0.05 & 0.04 \\
      & \WW                          & --- & --- & 0.05 & 0.05 & 0.04 \\
      & Luminosity                   & 1.0 & 1.0 & 0.7  & 0.7  & 0.5  \\
\hline
  0.9 & MC statistics                & 0.4 & 0.4  & 0.4  & 0.4  & 0.2  \\
      & ISR simulation               & 0.3 & 0.4  & 0.4  & 0.7  & 0.4  \\
      & Energy scale                 & 0.4 & 0.3  & 0.3  & 0.3  & 0.3  \\ 
      & Detector response            & 0.9 & 0.9  & 0.7  & 0.7  & 0.7  \\
      & \WW                          & --- & ---  & 0.02 & 0.02 & 0.01 \\
      & \ZZ                          & --- & ---  & 0.01 & 0.01 & 0.01 \\
      & Other four-fermion           & --- & ---  & 0.03 & 0.03 & 0.03 \\
%      & Detector response            & 0.9 & 0.8  & 0.6  & 0.6  & 0.7  \\ 
%      & Radiative background         & 0.3 & 0.4  & 0.3  & 0.3  & 0.1  \\
      & Luminosity                   & 1.0 & 1.0  & 0.7  & 0.7  & 0.5  \\
\hline
\end{tabular}
\end{center}
\end{table}

%%%%%%%%
%--- subsection  : bbbar
%%%%%%%%
%\input ew2_conf_vanc_bbar.tex
%%%%%%%%%%%%%%%%%%%%%%%%%%%%%%%%%%%%%%%%%%%%%%%%%%%%%%%%
%
% bbbar (R_b)
%
%%%%%%%%%%%%%%%%%%%%%%%%%%%%%%%%%%%%%%%%%%%%%%%%%%%%%%%%
\subsection{Measurement of the \boldmath$\mathrm b\bar b$ Production Fraction 
$R_{\mathrm b}$}
\label{rb} 

The ratio \Rb\ of the \bb\ to \qq\ production cross sections is 
measured at centre-of-mass energies in the range 130--183~GeV. 
Hadronic events are chosen following the exclusive selection described in 
Section~\ref{subhad}. The $\rm{b\bar{b}}$ events are separated from the 
hadronic ones using the relatively long lifetime of b~hadrons.
From the measured impact parameters of the tracks in an event, the confidence
level that all these tracks originate from the primary vertex is 
calculated~\cite{rblep1} and required to be less than a certain cut, which is 
chosen to minimize the total uncertainty on $\sigma_{\rm{b}\bar{\rm b}}$. 

The estimated selection efficiency and background fraction at each 
centre-of-mass energy are given in Table~\ref{rbeff}.
At low centre-of-mass energies about two thirds of the background is due to 
radiative events, with the remainder being dominated by \cc\ events. 
At higher energies, radiative background, \cc\ events and background from 
four-fermion events contribute in roughly equal proportions.

\begin{table}[htbp]
\mycaption{\label{rbeff} 
    Selection efficiencies and background fractions for the measurement 
    of \Rb.}
\begin{center}
\begin{tabular}{| c || c | c |}
\hline\strutl
 $\mathrm E_{cm}$  &   Efficiency  &   Background  \\
      (GeV)        &      (\%)     &      (\%)     \\
\hline
        130        & $40.3\pm 1.0$ &  $12.1\pm 1.6$ \\
        136        & $40.8\pm 1.0$ &  $10.0\pm 1.5$ \\
        161        & $40.7\pm 1.5$ &  $10.8\pm 1.2$ \\
        172        & $38.0\pm 1.4$ &  $10.6\pm 1.3$ \\
        183        & $37.8\pm 0.9$ &  $~9.6\pm 1.0$ \\
\hline
\end{tabular}
\end{center}
\end{table}

The largest contribution to the systematic uncertainty arises from the 
b~tagging efficiency. This is evaluated by using the same technique as 
above to measure the fraction \Rb\ of \bb\ events in Z data 
taken in the same year. The difference between this measurement and the very 
precisely known world average measurement of \Rb\ at the Z peak is then taken
as the systematic uncertainty. The b~tagging efficiency is
almost independent of the centre-of-mass energy, because the impact 
parameters of tracks from a decaying b~hadron depend little on its energy. 
The systematic uncertainties evaluated at the Z peak can therefore be directly
translated to higher centre-of-mass energies. 
The small decrease in tagging efficiency with centre-of-mass energy seen in 
Table~\ref{rbeff} is due to a decay length cut, which the b~tag uses to reject
tracks from ${\mathrm K}^0$ decay. This introduces no significant additional 
uncertainty on the tagging efficiency.

Systematic uncertainties due to the charm and light quark backgrounds are 
smaller. If one neglects the dependence of the quark production fractions on 
centre-of-mass energy, they are already taken into account by the study 
using Z peak data mentioned above. The impact parameter 
resolution is monitored using tracks with negative impact parameters.

\begin{figure}[htbp]
\mbox{\epsfig{file=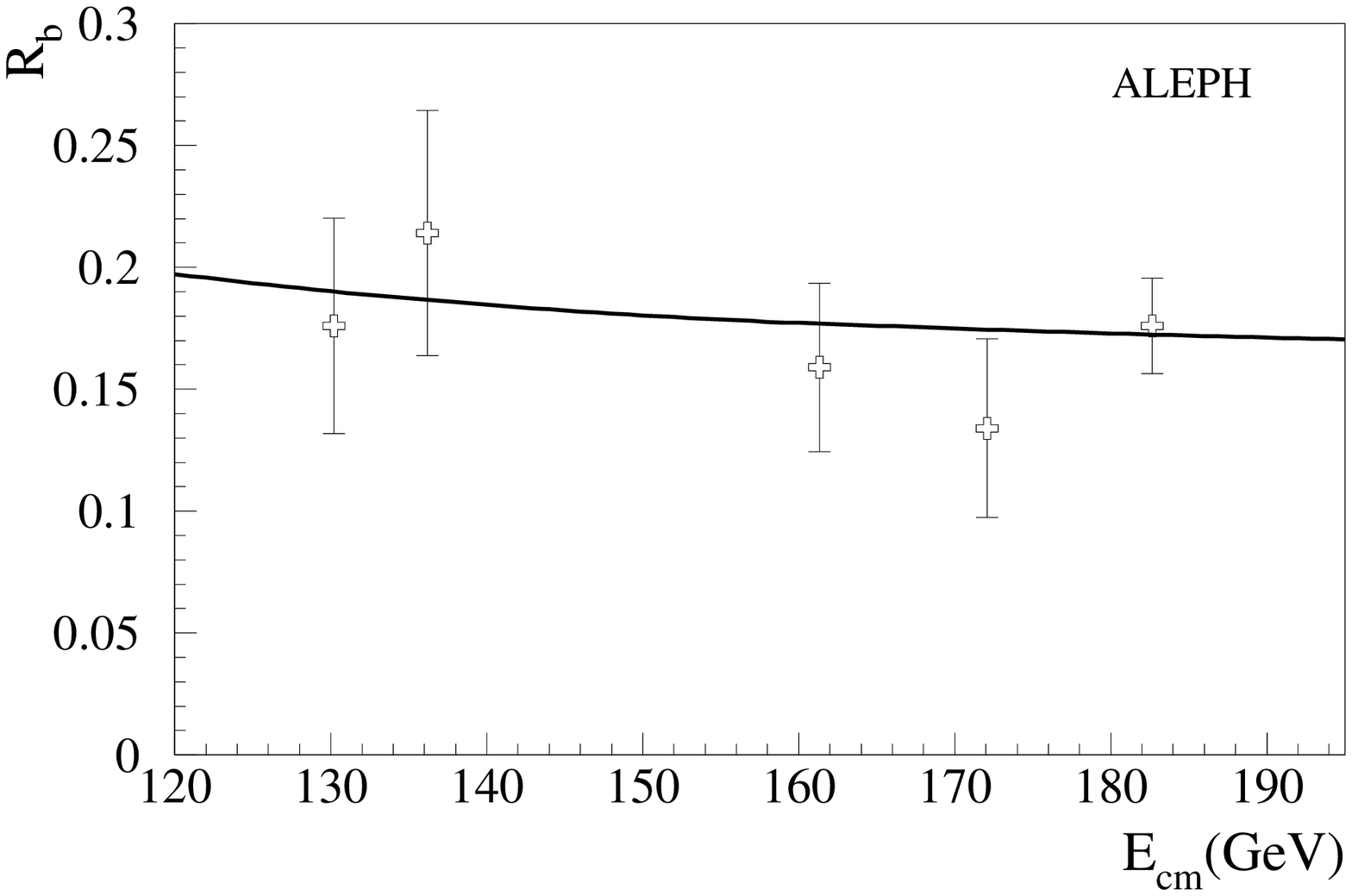,width=0.9\textwidth}} 
\mycaption{Measured values of the ratio \Rb\ at various centre-of-mass energies, 
         compared with the SM expectations.
\label{bbsum}}
\end{figure}

The number of observed events, together with the measured and predicted 
values of \Rb\ are given in Table~\ref{rbcross}. Figure~\ref{bbsum} summarizes 
the measurements of \Rb\ as a function of $\sqrt{s}$.

\begin{table}[htbp]
\mycaption{\label{rbcross}
    Measured values of \Rb\ with statistical and systematic uncertainties for 
    \bb\  production with $\sqrt{s^\prime/s} > 0.9$ and 
    $|{\cos\theta}| < 0.95$, where $\theta$ is the polar angle of the b~quark
    with respect to the beam axis.
    The SM predictions are also given, together with 
    the number of selected events (before background subtraction).}
\begin{center}
\begin{tabular}{| c || c | c | c |}
\hline\strutl
 $\mathrm E_{cm}$  &   No.  &         \Rb          & SM prediction \\
      (GeV)        & Events &                      &                 \\
\hline
       130         &   19   & $0.176\pm 0.044\pm 0.004$ &      0.190   \\
       136         &   20   & $0.214\pm 0.050\pm 0.005$ &      0.186   \\
       161         &   24   & $0.159\pm 0.034\pm 0.006$ &      0.175   \\
       172         &   16   & $0.134\pm 0.036\pm 0.007$ &      0.173   \\
       183         &   91   & $0.176\pm 0.019\pm 0.005$ &      0.171   \\
\hline
\end{tabular} 
\end{center}
\end{table}

%%%%%%%%
%--- subsection  : ccbar
%%%%%%%%
%\input ew2_conf_vanc_ccbar.tex
%%%%%%%%%%%%%%%%%%%%%%%%%%%%%%%%%%%%%%%%%%%%%%%%%%%%%%%%
%
% ccbar (R_c)
%
%%%%%%%%%%%%%%%%%%%%%%%%%%%%%%%%%%%%%%%%%%%%%%%%%%%%%%%%

%\bibliographystyle{unsrt}   % for BibTeX - sorted numerical labels by order of
                             % first citation.

\subsection{Measurement of the \boldmath$\mathrm c\bar c$ Production Fraction 
$R_{\mathrm c}$}
\label{rc}

The ratio \Rc\ of the \cc\ to \qq\ production cross sections is 
measured at a centre-of-mass energy of 183~GeV.
Hadronic events are chosen using the exclusive selection described in 
Section~\ref{subhad}, and clustered into two jets using the JADE algorithm. 
An additional acceptance cut requiring that both jets 
have $|{\cos\theta}| < 0.9$ is applied to ensure that the event is well 
contained inside the VDET. 

Background from \bb\  events is suppressed by taking advantage of the 
relatively long lifetime and high mass of b~hadrons. Three algorithms are used 
sequentially. The first rejects events on the basis of track impact parameter
significances \cite{rblep1}, the second using the decay length significance
of reconstructed secondary vertices \cite{Bs_oscillation} and the third 
relying on a comparison of the total invariant mass of high impact 
parameter significance tracks with the charm hadron mass \cite{rblep1}.
Together the hadronic selection and \bb\  rejection have a 60\% efficiency
for \cc\  events with $\sqrt{s^\prime/s} > 0.9$. The residual fraction of 
\bb\ events in the resulting sample is 4\%.

\begin{figure}[htbp]
\begin{center}
\mbox{\epsfig{figure=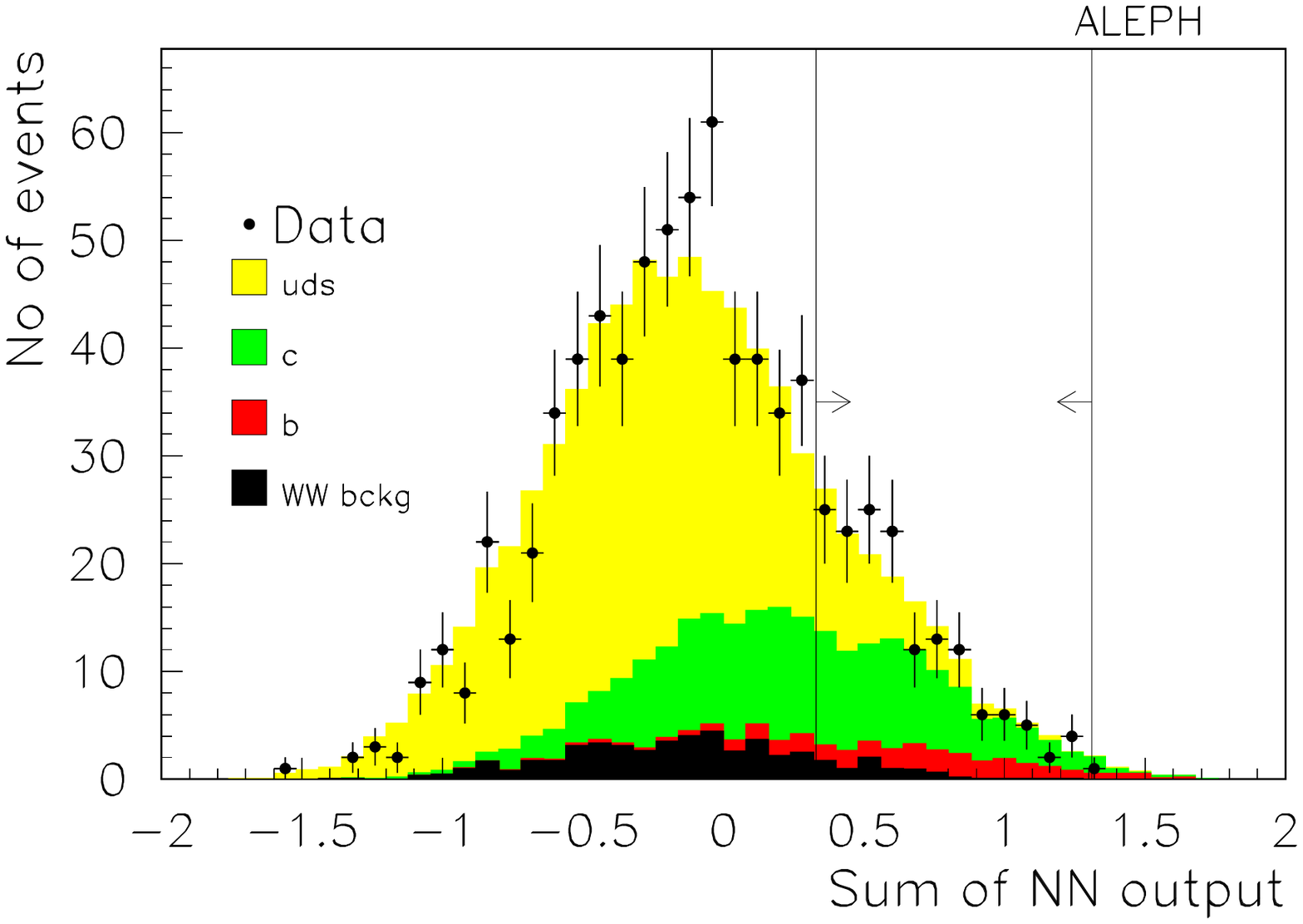,width=0.9\textwidth,height=0.7\textwidth}} 
\mycaption{Distribution of the sum of the neural network outputs for hadronic
events at 183~GeV centre-of-mass energy after hadronic selection and 
\bb\ ~rejection. The lines indicate the chosen cuts.\label{nnet}}
\end{center}
\end{figure}

The final \cc\ selection uses a neural network. This was trained to separate
\cc\ jets (giving a neural network output close to one) from light quark 
jets (giving an output close to minus one). This network uses twelve variables 
per jet. These are listed below, ordered according to decreasing weight.
\begin{itemize}
\setlength{\itemsep}{0mm}
\item The sum of the rapidities with respect to the jet axis of energy flow 
      particles within $40^\circ$ of this axis.
\item The sphericity of the four most energetic energy flow particles in the 
      jet, calculated in their rest~frame.
\item The total energy of the four most energetic energy flow objects in the 
      jet.
\item The number of identified leptons (electrons or muons) in the jet with 
      momentum larger than 1.5 \GeVc.
\item The transverse momentum squared $p_\perp^2$ with respect to the
      jet axis of the $\pi_{\mathrm soft}$ candidate from 
      ${\mathrm D}^*\rightarrow\pi_{\mathrm soft}\mathrm X$, defined as the charged 
      track in the jet with the smallest value of $p_\perp$ and an energy 
      between 1 and 4~GeV.
\item The confidence level that all charged tracks in the jet originate 
      from the primary vertex.
\item The energy of a subjet of mass 2.1~\GeVcc\ built around the leading 
      energy flow particle in the jet.
\item The momentum of the leading energy flow particle in the jet.
\item The number of energy flow objects in a cone of half-angle 
      $40^\circ$ around the jet axis.
\item The confidence level that all charged tracks in the jet having a
      rapidity with respect to the jet axis exceeding 4.9, originate 
      from the primary vertex.
\item The decay length significance of a reconstructed secondary vertex.
\item The energy of a reconstructed D~meson (if any). D~meson reconstruction
      is attempted in the charged decay channels \mr{K\pi}, \mr{K\pi\pi} 
      and \mr{K\pi\pi\pi}.

\end{itemize}
The distribution of the sum of the neural network outputs for the two jets
in each event is shown in Fig.~\ref{nnet}. A lower cut is placed on this
to reject light quark events. It was also found that an upper cut served to
reject a tail of remaining \bb\  events. The final selection efficiencies and event
fractions for each flavour are summarized in Table~\ref{flavour}.

To evaluate the systematic uncertainty introduced by a given selection 
variable, the ratio of distributions of this variable in data to Monte Carlo 
is determined. After smoothing to reduce statistical fluctuations, this ratio is
then used to reweight individually \bb, \cc\  and light quark event
flavours in the Monte Carlo. The relative change in the resulting efficiency 
for each flavour is taken as its systematic uncertainty. A cross-check is 
performed by measuring \Rc\ on the Z data taken in the same year; the 
measured value is consistent with the SM prediction. All contributions to 
the systematic uncertainty are given in Table~\ref{syst}.

\begin{table}[htbp]
\begin{center}
\begin{minipage}{.45\linewidth}
\mycaption{\label{flavour}
Selection efficiency for each flavour, and its expected fraction in the final 
event sample, assuming SM cross sections.}
\begin{center}
\begin{tabular}{|l|r|r|}
\hline
 & \multicolumn{1}{c|}{Efficiency (\%)} & \multicolumn{1}{c|}{Fraction (\%)} \\
\hline
 \uu   &  $6.18 \pm 0.25$~ & $14.8 \pm 0.6$~ \\
 \dd   &  $6.66 \pm 0.33$~ & $10.3 \pm 0.5$~ \\
 \myss &  $6.09 \pm 0.31$~ &  $9.7 \pm 0.5$~ \\
 \cc   & $22.07 \pm 0.48$~ & $51.8 \pm 0.8$~ \\
 \bb   &  $8.61 \pm 0.37$~ & $13.4 \pm 0.6$~ \\
\hline 
\end{tabular}
\end{center}
\end{minipage}
\hfil
\begin{minipage}{.50\linewidth}
\mycaption{\label{syst} 
Contributions to the systematic uncertainty on the measured value of \Rc.}
\begin{center}
\begin{tabular}{|l|c|}
\hline
 \multicolumn{1}{|c|}{Description} & Uncertainty \\
\hline
Luminosity                        & $\pm 0.001$ \\
Efficiency of \qq\ selection      & $\pm 0.001$ \\
Background in \qq\ selection      & $\pm 0.001$ \\
Performance of \bb\  rejection tag  & $\mbox{}^{+0.012}_{-0.003}$  \\
Performance of neural network     & $\mbox{}^{+0.006}_{-0.010}$  \\
MC statistics                     & $\pm 0.007$  \\
\hline\strutl
Total                             & $\mbox{}^{+0.016}_{-0.013}$ \\ 
\hline 
\end{tabular}
\end{center}
\end{minipage}
\end{center}
\end{table}

In the data, 153 events are selected, of which 74.8 are estimated from the
Monte Carlo simulation to be background. The value of \Rc\ at 
$\sqrt{s}=183$~GeV, for $\sqrt{s^{\prime}/s}>0.9$ and $|{\cos\theta}| < 0.95$,
where $\theta$ is the polar angle of the c~quark with respect to the 
beam axis, is found to be
\begin{equation}
R_{\mathrm c} = 0.276\pm 0.041({\mathrm stat})\ 
^{+0.016}_{-0.013}({\mathrm syst}) - 0.284\times (R _{\mathrm uds} - 0.584) 
                         - 0.390\times (R_{\mathrm b} - 0.171)~,
\end{equation}
where the dependence on a possible departure of the light quark or \bb\ 
fractions from their SM expectations is explicitly given. The measured 
value of \Rc\ is consistent with its SM expectation of 0.244~. 

%%%%%%%%
%--- subsection  : jet charge
%%%%%%%%
%\input ew2_conf_vanc_jet_charge.tex
%%%%%%%%%%%%%%%%%%%%%%%%%%%%%%%%%%%%%%%%%%%%%%%%%%%%%%%%
%
% jet charge
%
%%%%%%%%%%%%%%%%%%%%%%%%%%%%%%%%%%%%%%%%%%%%%%%%%%%%%%%%
%
\subsection{\boldmath Measurement of $A_{\mathrm FB}^{\mathrm q}$ using 
                      Jet Charge}
\label{jet_charge}
Constraints upon the forward-backward asymmetries \Afb{q}\ of \qq\
events at $\sqrt{s}=183$~GeV are obtained using a jet charge technique. 
This is performed separately for b-enriched and b-depleted events, which allows
the asymmetry in \bb\ events to be well determined.

The analysis uses the hadronic events with $\sqrt{s'/s}>0.9$, selected as 
described in Section~\ref{subhad}. To ensure that the events are well 
contained in the tracking chambers, they are also required to satisfy 
$|{\cos\theta^*}|<0.9$. 
The events are then divided into two samples, according to whether they pass 
or fail the b~lifetime tag of Section~\ref{rb} (here used with looser cuts). 
This gives a 91\% pure sample of \bb\  events plus a sample of
predominantly light/charm quark events.

After clustering all the events into two jets, the jet charge 
$Q_{\mathrm jet}$ of each jet is determined, where

\begin{equation}
Q_{\mathrm jet} = 
\left.
{\displaystyle\sum_{i=1}^{\mathrm N_{track}} p_{\parallel i}^{\kappa} Q_i}
\right/
{\displaystyle\sum_{i=1}^{\mathrm N_{track}} p_{\parallel i}^{\kappa}}
\end{equation}
and the sums extend over the charged tracks in the jet. The track momentum
component parallel to the jet and its charge are $p_{\parallel i}$ and $Q_i$, 
respectively. The parameter $\kappa$ is set to 0.3, because this minimizes
the uncertainty on the final result. 
The mean charge difference between the forward and backward jets 
$\Qfb = \langle Q_{\mathrm jet}^{\mathrm F}\rangle - 
\langle Q_{\mathrm jet}^{\mathrm B}\rangle$ is then formed. The first row of 
Table~\ref{Qfb} shows the observed value of \Qfb\ in the data.

\begin{table}[htbp]
\mycaption{\label{Qfb} Comparison of \Qfb\ in data 
with the prediction of Equation~\ref{Qfbeqn} using SM values of
$\sigma_{\mathrm q}$ and \Afb{q}. Also shown is the contribution to this 
prediction from signal and background events.}
\begin{center}
\begin{tabular}{|c||c|c|}
\hline
         \Qfb                   & b tagged events   & anti-b tagged events  \\
\hline
          Data                  & $-0.029\pm 0.018$ &    $0.022\pm 0.007$   \\
     Expectation in SM          & $-0.052$          &    ~$0.014$           \\
 Contribution from signal       & $-0.048$          &    ~$0.018$           \\
 Contribution from background   & $-0.004$          &    $-0.003$           \\
\hline
\end{tabular}
\end{center}
\end{table}

The expected value of \Qfb\ depends on the 
\qq\ cross sections $\sigma_{\mathrm q}$ and asymmetries \Afb{q}. 
Defining $\epsilon_i$ as the selection efficiency for event type $i$, 
the prediction is

\begin{equation}
\label{Qfbeqn}
 \Qfb = \frac
 {\displaystyle \sum_{\mathrm q} \sigma_{\mathrm q} \epsilon_{\mathrm q} 
                \Afb{q} \delta_{\mathrm q} D_{\mathrm q} +
                \sum_{\mathrm x} \sigma_{\mathrm x} \epsilon_{\mathrm x} 
                {\Qfb}_x
 }
 {\displaystyle \sum_{\mathrm q} \sigma_{\mathrm q} \epsilon_{\mathrm q} +
                \sum_{\mathrm x} \sigma_{\mathrm x} \epsilon_{\mathrm x}
 }~,
\end{equation}
where the sums extend over the quark flavours~q and the various background 
types~x. The parameters $\delta_{\mathrm q} = 
\langle {Q_{\mathrm jet}^{\mathrm q}}\rangle - 
\langle {Q_{\mathrm jet}^{\mathrm\bar q}}\rangle$ give the mean charge 
separation between the jet containing the quark and that containing the antiquark.
Table~\ref{qsep} shows the parameters $\delta_{\mathrm q}$ for each flavour. 
They are obtained from Monte Carlo simulation, but with additional corrections 
based upon a precise analysis at the Z peak \cite{Halley}. The parameters
$\delta_q$ vary only slowly with centre-of-mass energy so these corrections
are still applicable at LEP2. The most important correction is that 
to $\delta_c$, which amounts to 26\%. This is due to an inadequate simulation 
of charm hadron decay modes. However, even this only alters the predicted 
value of \Qfb\ by about one half of the statistical error in the data.
Systematic uncertainties on the $\delta_q$ can therefore at present be 
neglected.
The parameters $D_{\mathrm q}$ ($\approx 0.98$) give the dilution in 
${\Qfb}_{\mathrm q} = \Afb{q} \delta_{\mathrm q} D_{\mathrm q}$ caused by the 
small angular dependence of the acceptance efficiency. 

\begin{table}[htbp]
\mycaption{\label{qsep} Mean jet charge separations for each quark flavour.}
\begin{center}
\begin{tabular}{|c|c|c|c|c|}
\hline
 $\delta_{\mathrm u}$ & $\delta_{\mathrm d}$ & $\delta_{\mathrm s}$ & $\delta_{\mathrm c}$ & $\delta_{\mathrm b}$ \\
\hline
   $0.205$    &   $-0.130$   &  $-0.153$    &   $0.155$    &   $-0.108$   \\
\hline
\end{tabular}
\end{center}
\end{table}
Assuming SM values of $\sigma_{\mathrm q}$ and \Afb{q}, the predicted values
of \Qfb\ are given in the second row of Table~\ref{Qfb}. They agree with those 
measured in the data, both for the b-enriched and the b-depleted samples. 
By comparing the measured value of \Qfb\ in each of these two samples with
the predictions of Equation~\ref{Qfbeqn}, two independent constraint equations
are obtained for the allowed values of $\sigma_{\mathrm q}$ and \Afb{q}. 

Providing that deviations from the SM are small, it is convenient to 
approximate each of these two constraints by a linear equation. 
These two linear equations will be used when placing limits on physics beyond 
the SM in Section~\ref{interpretations}.

If $\sigma_{\mathrm q}$ and \Afb{q} differ from their SM predictions by 
$\Delta\sigma_{\mathrm q}$ and $\Delta\Afb{q}$, respectively, then a Taylor 
expansion of Equation~\ref{Qfbeqn} yields
\begin{equation}
\label{Taylor}
\Qfb - {\Qfb}_{\mathrm SM} =
\sum_{\mathrm q} \frac{\partial\Qfb}{\partial\sigma_{\mathrm q}}\Delta\sigma_{\mathrm q} +
       \frac{\partial\Qfb}{\partial\Afb{q}}\Delta\Afb{q}~.
\end{equation}
Dividing this equation throughout by the estimated uncertainty on the measured
value of \Qfb\ in the data gives an equation of the form
\begin{equation}
\label{Afbeqn}
  \gamma = \sum_{\mathrm q} \alpha_{\mathrm q} \Delta\sigma_{\mathrm q} + \beta_{\mathrm q}\Delta\Afb{q}~.
\end{equation}
Here $\gamma$ is the difference between the measured value of 
\Qfb\ and that expected in the SM, divided by 
the measurement uncertainty (such that $\gamma$ has an uncertainty
of $\pm 1$). The coefficients $\alpha_{\mathrm q}$ and $\beta_{\mathrm q}$,
defined by this equation, can be expressed in terms of $\epsilon_i$, 
$\delta_{\mathrm q}$, $D_{\mathrm q}$ and ${\Qfb}_x$. The values of 
$\gamma$, $\alpha_{\mathrm q}$ and $\beta_{\mathrm q}$ for the b-enriched 
and b-depleted event samples are given in Table~\ref{Afb}. The uncertainties
in $\alpha_{\mathrm q}$ and $\beta_{\mathrm q}$ can be neglected.
The equation obtained from the b-enriched sample can be interpreted as a 
measurement of \Afb{b}, since $\beta_b$ dominates the other coefficients. 
It gives 
%$1.3\pm 1.0 = -5.4 \Delta\Afb{b}$,
%which using the SM prediction for \Afb{b}\ of 0.57, gives 
a measured value of $\Afb{b} = 0.33 \pm 0.19$, compared with the SM 
prediction of 0.57~.

\begin{table}[htbp]
\mycaption{\label{Afb} Coefficients of the linear constraint 
Equation~\ref{Afbeqn} derived for the b-enriched and b-depleted data samples.
The units of $\alpha$ are pb$^{-1}$.}
\begin{center}
\begin{tabular}{|l||r||c|c|c|c|c||c|c|c|c|c|}
\hline
 Event Sample         & \multicolumn{1}{c||}{$\gamma$}   &
           $\alpha_u$ & $\alpha_d$ & $\alpha_s$ & $\alpha_c$ & $\alpha_b$ &
           $\beta_u$  & $\beta_d$  & $\beta_s$  & $\beta_c$  & $\beta_b$ \\
\hline
$b$ tagged              & $1.3\pm 1.0$&
                          $0.0$ & $ 0.0$ & $ 0.0$ & $0.0$ & $-0.1$ &
                          $0.0$ & $ 0.0$ & $ 0.0$ & $0.4$ & $-5.4$ \\
anti-$b$ tagged         & $1.2\pm 1.0$ &
                          $0.7$ & $-0.6$ & $-0.7$ & $0.5$ & $-0.2$ &
                          $6.6$ & $-2.7$ & $-3.4$ & $5.1$ & $-1.1$ \\
\hline
\end{tabular}
\end{center}
%\vspace{2cm}
\end{table}

%%%%%%%%
%--- section : leptons
%%%%%%%%
%\input ew2_conf_vanc_leptons.tex
%%%%%%%%%%%%%%%%%%%%%%%%%%%%%%%%%%%%%%%%%%%%%%%%%%%%%%%%
%
% leptons
%
%%%%%%%%%%%%%%%%%%%%%%%%%%%%%%%%%%%%%%%%%%%%%%%%%%%%%%%%

\section{Leptonic Final States }
\label{sec:leptons}

%%%%%%%%%%%%%%%%%%%%%%%%%%%%%%%%%%%%%%%%%%%%%%%%%%%%%%%%
% bhabhas
%%%%%%%%%%%%%%%%%%%%%%%%%%%%%%%%%%%%%%%%%%%%%%%%%%%%%%%%
\subsection{The  \boldmath$\mathrm e^{+}e^{-}$ Channel}
\label{sec:bhabha}

Electron pair events are selected by requiring the presence of two good tracks of 
opposite charge with a polar angle to the beam axis of $|{\cos\theta}| < 0.9$.
The sum of the momenta of the two tracks must exceed
30\% of the centre-of-mass energy. The total energy associated with them in 
the ECAL must be at least 40\% of the centre-of-mass energy. When calculating 
this energy, if one of the tracks passes near a crack in the ECAL, 
then the associated HCAL energy is included if it matches the track 
extrapolation within 25~mrad. Furthermore, the energy of bremsstrahlung 
photons is included when within cones of $20^\circ$ around each 
track.

\begin{figure}[htbp]
\begin{center}
\mbox{\epsfig{figure=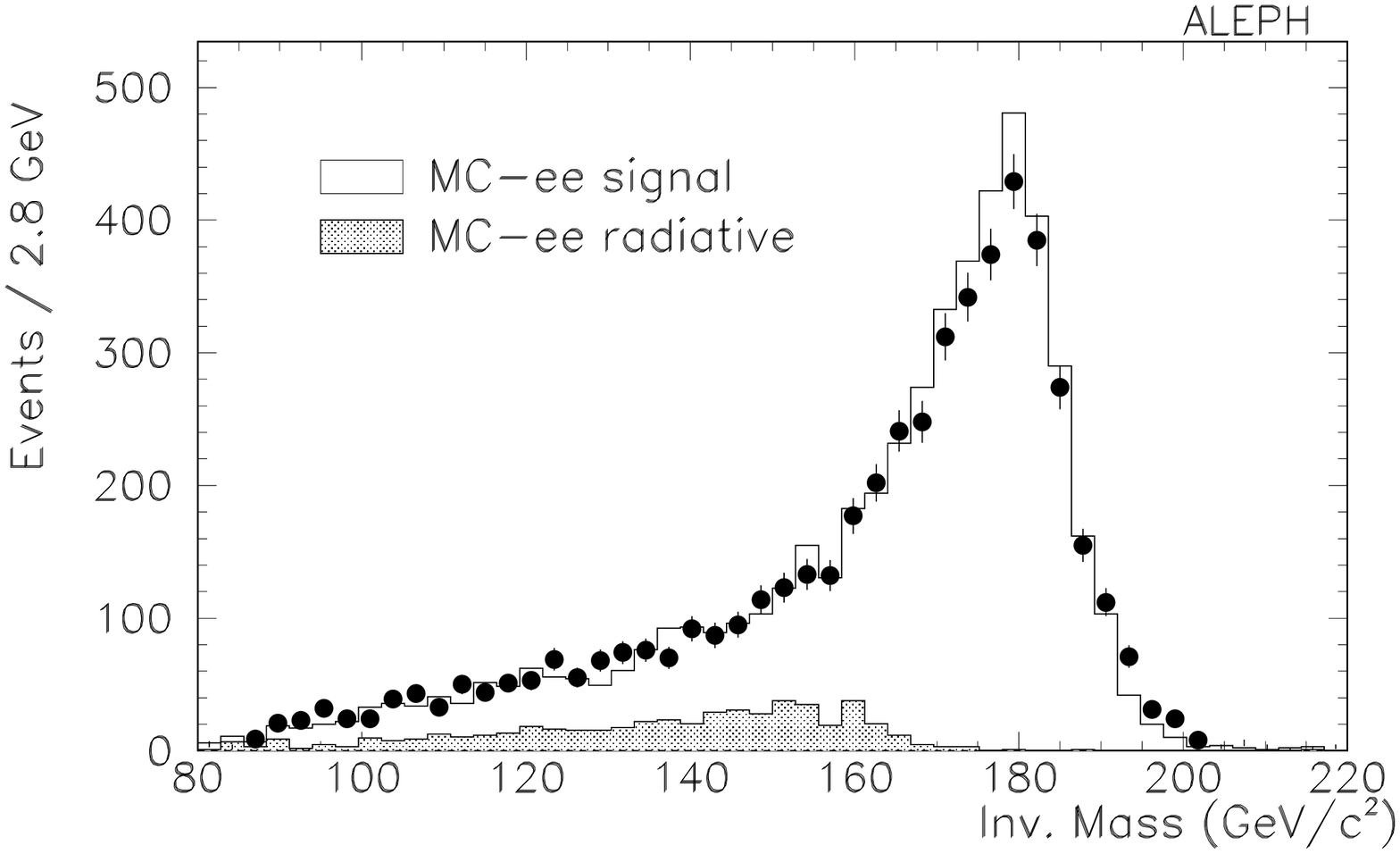,width=0.9\textwidth}} 
\mycaption{Electron pair invariant mass distribution at 183~GeV 
centre-of-mass energy. The data, which is shown by points, is compared
with the Monte Carlo expectations for the signal and the radiative background.
\label{dielmass}}
\end{center}
\end{figure}

Only the exclusive cross section is measured, so the requirement 
$\sqrt{s^\prime_m/s}> 0.9$ is applied. 
%Pairs which are not back-to-back w.r.t.\ the polar angle 
%are removed by demanding that the pseudo-rapidity $Y$ defined as
%$  Y = 0.5\,\ln(\,(\sin \theta_+ (1 + \cos \theta_-) + \sin \theta_- (1 + \cos \theta_+))\,/\, 
%                  (\sin \theta_+ (1 - \cos \theta_-) + \sin \theta_- (1 - \cos \theta_+))\, ) $
%be less than 0.105. 
%$\theta_- (\theta_+)$ refers to the polar angle of the electron (positron) in the laboratory frame.  
The main background is due to ISR and this is reduced to a level of 10 -- 12\%, 
by requiring that the invariant mass of the \ee\ final state, determined from 
the measured track momenta, exceeds
80~\GeVcc. The resulting invariant mass distribution is shown in 
Fig.~\ref{dielmass}. The discrepancy between data and Monte Carlo seen in
this figure arises from an inaccurate simulation of the momentum
resolution of tracks at small angles to the beam axis.
As the cut applied on the electron pair invariant mass is very loose, this
does not lead to a significant systematic uncertainty.
The normalization of the radiative background is determined from a fit to 
Fig.~\ref{dielmass}, using the expected shapes of the signal and background. 
The statistical uncertainty on the fit result leads to the systematic 
uncertainty due to radiative background quoted in Table~\ref{ELSYS}.     

The selection efficiencies are determined from BHWIDE Monte Carlo samples
at each energy point. They are given, together with the estimated background,
in Table~\ref{EFF_ALL}. The uncertainties on the efficiencies are estimated 
by varying the calorimeter energy scales as in Section~\ref{subhad}. 

The exclusive cross section is determined in two polar
angle ranges: $-0.9<\cos\theta^*<0.9$ and $-0.9<\cos\theta^*<0.7$. It is
dominated by $t$~channel photon exchange, particularly in the forward region.
The cross section measurements are given in Table~\ref{CROSSALL}.
The contributions to the systematic uncertainties on these measurements are 
given in Table~\ref{ELSYS}. Uncertainties from possible bias in the polar
angle measurement of tracks are negligible. They have been assessed 
by redetermining the cross section without using tracks having
$|\cos\theta| > 0.9$.

\begin{table}[htbp]
\mycaption{\label{ELSYS}{Contributions to the systematic uncertainties on the 
\ee\ exclusive cross section measurements, for both $\cos\theta^*$ 
intervals. All quoted values are in percent.}}
\begin{center}
\begin{tabular}{| c | c | l || l | l | l | l | l |}
\hline\strutl
 $\sqrt{s^{\prime}/s}$ & $\cos\theta^*$ range & \multicolumn{1}{c||}{Description} & 
                               \multicolumn{5}{c|}{$\mathrm E_{cm}$ (GeV)} \\
\cline{4-8}
          cut          &       &       & 130  & 136  & 161  & 172  & 183  \\ 
\hline\hline
  0.9 & $-0.9<\cos\theta^*<0.9$ 
                & MC statistics        & 0.4  & 0.4  & 0.4  & 0.4  & 0.4  \\
      &         & Energy scale         & 1.7  & 1.6  & 1.4  & 1.6  & 1.6  \\
      &         & \tautau              & 0.01 & 0.01 & 0.01 & 0.01 & 0.01 \\
      &         & Radiative background & 0.9  & 1.0  & 0.9  & 1.0  & 0.9  \\
      &         & Luminosity           & 1.0  & 1.0  & 0.7  & 0.7  & 0.5  \\
\hline
  0.9 & $-0.9<\cos\theta^*<0.7$ 
                & MC statistics        & 0.6  & 0.5  & 0.6  & 0.6  & 0.6  \\
      &         & Energy scale         & 1.4  & 1.3  & 1.2  & 1.2  & 1.1  \\
      &         & \tautau              & 0.2  & 0.04 & 0.03 & 0.03 & 0.03 \\
      &         & Radiative background & 1.3  & 1.6  & 1.3  & 1.4  & 1.3  \\
      &         & Luminosity           & 1.0  & 1.0  & 0.7  & 0.7  & 0.5  \\
\hline
\end{tabular}
\end{center}
\end{table}

%%%%%%%%%%%%%%%%%%%%%%%%%%%%%%%%%%%%%%%%%%%%%%%%%%%%%%%%
% dimu
%%%%%%%%%%%%%%%%%%%%%%%%%%%%%%%%%%%%%%%%%%%%%%%%%%%%%%%%
\subsection{The \boldmath $\mu^+\mu^-$\unboldmath\ Channel}
\label{dimu}

The muon pair selection requires events to contain two good, oppositely charged 
tracks with momenta exceeding 6~\GeVc\ and angles to the beam
axis of $|{\cos\theta}|<0.95$. The scalar sum of the momenta of the two
tracks must exceed 60~\GeVc. The total number of good charged tracks in
the events must be no more than eight. 
To limit the background from cosmic ray events, 
both tracks are required to originate near the primary vertex, and to have at 
least four associated ITC hits, which confirms that they were produced within 
a few nanoseconds of the LEP beam crossing. 

Both tracks must be identified as muons, where muon identification is based 
either on the digital hit pattern associated with a track in the HCAL or 
on its energy deposition in the calorimeters:

\begin{itemize}
\item
  The track should fire at least 10 of the 23 drift tube planes in the HCAL. 
  It should also fire at least half of the HCAL planes which the track is 
  expected to cross (taking into account HCAL cracks), and furthermore 
  should fire 3 or more of the outermost 10 planes of the HCAL. 
  Alternatively, the track should have at least one associated hit in the 
  muon chambers.
\item
  The energy deposition must be consistent with that of a minimum ionizing 
  particle,
  i.e., the sum of the energies associated with the track in the ECAL and HCAL 
  should not exceed $60\%$ of the track momentum. Moreover, the sum of these 
  energies and the track momentum should be smaller than $60\%$ of the 
  centre-of-mass energy. To control a small misidentification background 
  related to calorimeter cracks, tracks are also required to have at least 
  one associated hit in the 10 outermost layers of the HCAL.
\end{itemize}

\begin{figure}[htbp]
\begin{center}
\mbox{\epsfig{figure=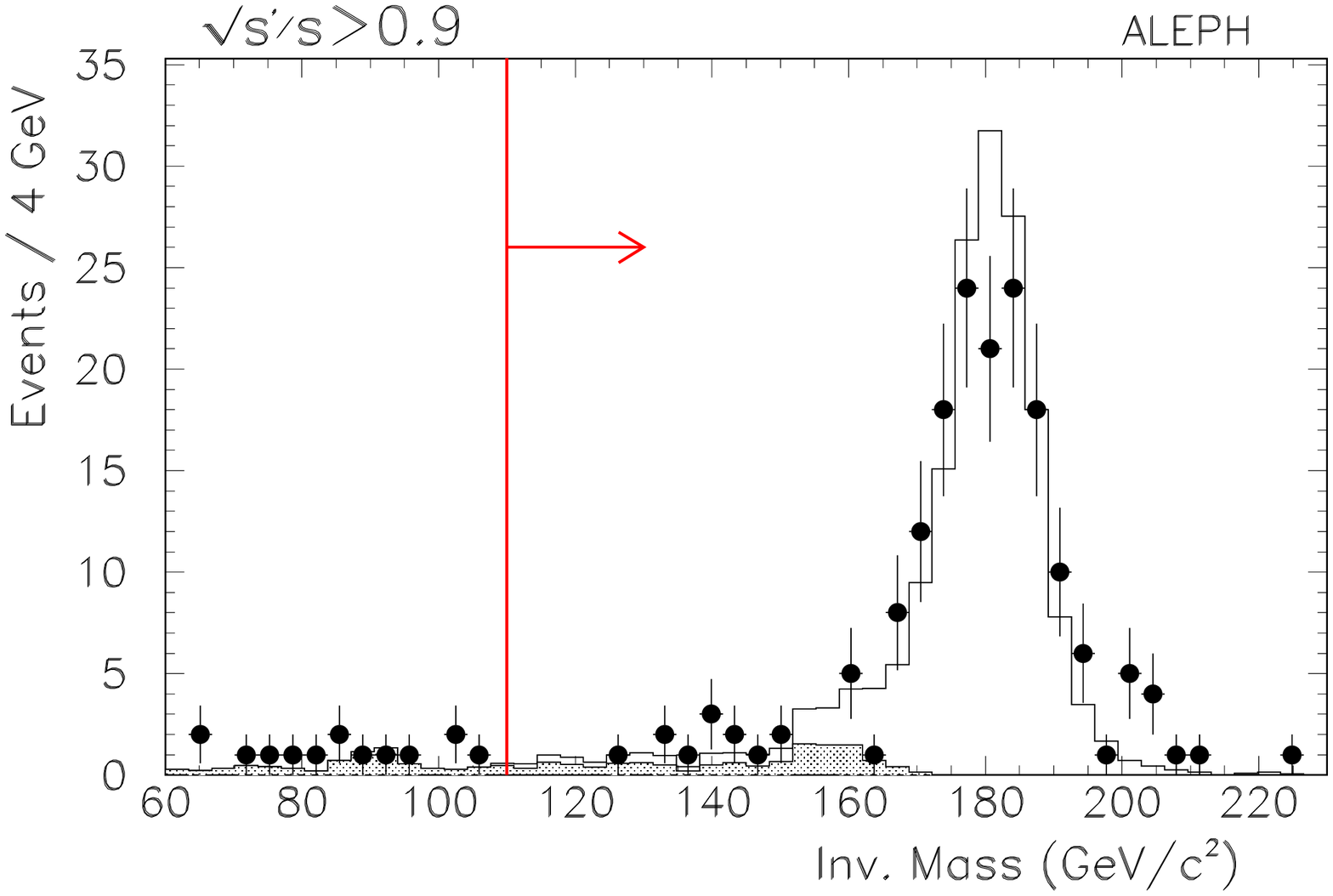,width=0.9\textwidth}} 
\mycaption{Distribution of the invariant mass of \mumu\ events at 
$\sqrt{s}=183$~GeV, with $\sqrt{s^{\prime}_{m}/s}>$ 0.9. The data points are 
compared to the white histogram of Monte Carlo \mumu\ events, normalized to the same 
integrated luminosity. The multi-radiative background is shown as the shaded 
histogram. The vertical line shows the cut used for the exclusive selection.
\label{dimuon_mass}}
\end{center}
\end{figure}

In the case of the exclusive selection, $\sqrt{s^{\prime}_{m}/s}>$0.9, it is 
also required that the muon pair invariant mass exceed 110~\GeVcc, reducing
background from radiative events by about 40\%. The invariant mass 
distribution is shown in Fig.~\ref{dimuon_mass}. 

The efficiency of the kinematic selection cuts is estimated using KORALZ 
Monte Carlo events, whilst the muon identification efficiency is measured 
using muon pair events in data recorded at the Z~peak in the same year.
This efficiency is typically uncertain by about $\pm 1.6$\% as a result of 
the limited number of Z~events. 

\begin{figure}[htbp]
\begin{center}
\mbox{\epsfig{figure=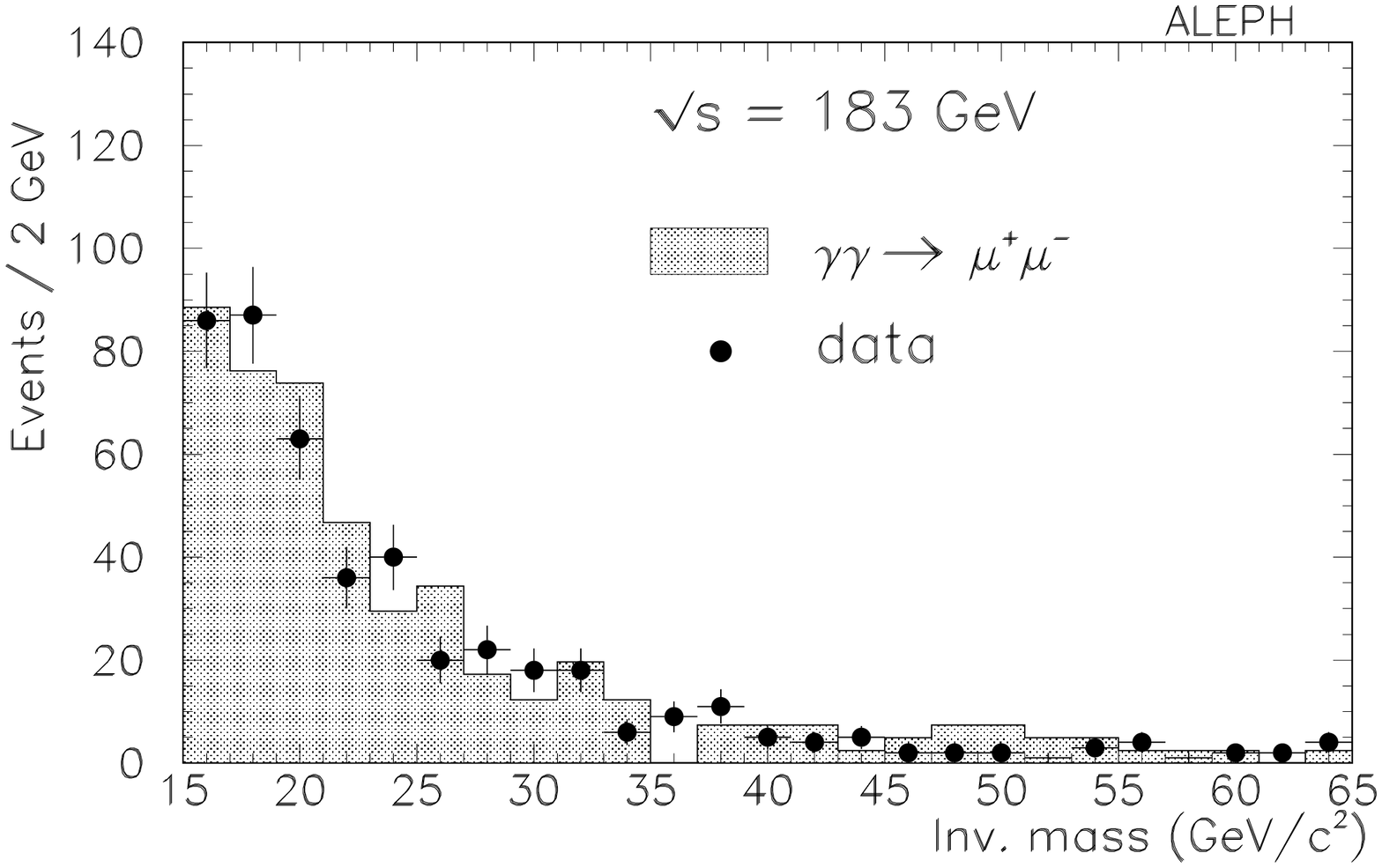,width=0.9\textwidth}} 
\mycaption{Distribution of the invariant mass of \mumu\ events at 183~GeV 
centre-of-mass energy in the low mass region where the process \ggmm\ is 
dominant. The Monte Carlo prediction is indicated by the shaded histogram.
\label{ggmu}}
\end{center}
\end{figure}

For the inclusive process, the main background contamination stems
from \ggmm. The systematic uncertainty associated with the normalization 
of this background is estimated by comparing data and Monte Carlo in the 
region of \mumu\ mass below 50~\GeVcc, which is shown in Fig.~\ref{ggmu}.

For the exclusive process, the main background comes from radiative events
and is assessed using the mass region below $0.9\sqrt{s}$ of 
Fig.~\ref{dimuon_mass}. 
%Residual cosmic ray contamination is estimated by 
%relaxing the cuts upon the distance between the two tracks in $r\phi$ and $z$.
Residual cosmic ray contamination is estimated by 
relaxing cuts on track impact parameters, and is found to be less than 0.5\%.

Efficiencies and background levels are summarized in Table~\ref{EFF_ALL}, and 
the cross section measurements are given in Table~\ref{CROSSALL}. 
The contributions to the systematic uncertainties on the cross sections are 
given in Table~\ref{MUSYS}.

\begin{table}[htbp]
\mycaption{\label{MUSYS}{Contributions to the systematic uncertainties on the 
muon pair cross section measurements, for all energies and for both inclusive and 
exclusive processes. All quoted values are in percent.}}
\begin{center}
\begin{tabular}{| c | l || l | l | l | l | l |}
\hline\strutl
 $\sqrt{s^{\prime}/s}$ & \multicolumn{1}{c||}{Description} & 
                         \multicolumn{5}{c|}{$\mathrm E_{cm}$ (GeV)} \\
\cline{3-7}
          cut          &             & 130  & 136  & 161  & 172  & 183  \\ 
\hline\hline
  0.1 & MC statistics                & 0.4  & 0.4  & 0.4  & 0.4  & 0.4  \\
      & $\mu$ identification         & 2.0  & 2.0  & 1.9  & 2.0  & 1.9  \\
      & \ggmm                        & 0.0  & 0.0  & 1.0  & 1.0  & 0.7  \\
      & \tautau                      & 0.04 & 0.05 & 0.1  & 0.1  & 0.1  \\
      & \WW                          & ---  & ---  & 0.05 & 0.2  & 0.3  \\
      & Luminosity                   & 1.0  & 1.0  & 0.7  & 0.7  & 0.5  \\
\hline
  0.9 & MC statistics                & 0.4  & 0.4  & 0.4  & 0.3  & 0.4  \\
      & $\mu$ identification         & 1.9  & 1.9  & 1.8  & 1.8  & 1.9  \\
      & \tautau                      & 0.00 & 0.00 & 0.02 & 0.1  & 0.1  \\
      & \WW                          & ---  & ---  & 0.0  & 0.03 & 0.2  \\
      & Radiative background         & 0.5  & 0.5  & 0.6  & 0.8  & 0.6  \\
      & Luminosity                   & 1.0  & 1.0  & 0.7  & 0.7  & 0.5  \\
\hline
\end{tabular}
\end{center}
\end{table}
    
%%%%%%%%%%%%%%%%%%%%%%%%%%%%%%%%%%%%%%%%%%%%%%%%%%%%%%%%
% ditau
%%%%%%%%%%%%%%%%%%%%%%%%%%%%%%%%%%%%%%%%%%%%%%%%%%%%%%%%
\subsection{The \boldmath$\tau^+\tau^-$\unboldmath\ Channel} 
\label{sec:ditau}
The tau pair selection begins by clustering events into jets,
using the JADE algorithm with the clustering parameter $y_{\mathrm cut}$ equal 
to 0.008~. 
Tau jet candidates must contain between one and eight charged tracks. Events 
with two tau jet candidates are selected, providing that the invariant mass 
of the two jets exceeds 25~\GeVcc. This requirement removes a large part of 
the $\gamma\gamma$ background. 

Following an approach already used for tau identification at 
LEP1~\cite{TAULEP}, each tau jet candidate is analysed and classified 
as a tau lepton decay into an electron, a muon, charged hadrons or 
charged hadrons plus one or more $\pi^{0}$. Both tau jets must be 
classified in this way. 
Events are required to have at least one tau jet candidate identified
as a decay into a muon or into charged hadrons or charged hadrons plus $\pi^0$.
To suppress background from $\gamma\gamma\rightarrow\mu^+\mu^-$ or \mumu,
events with two jets classified as muonic tau decay are discarded.

%The tau selection explicitly identifies tau leptons decaying into 
%muons, charged hadrons or charged hadrons plus one or more $\pi^{0}$. 
%Events are kept if at least one $\tau$ lepton is identified in this way.
%To reduce contamination from Bhabha events and muon pairs, events are 
%rejected if they are compatible with electron pairs, or if both hemispheres 
%are identified as muons. 

For the exclusive selection, W pair events in which both W decay
leptonically represent an important background. However, most of this
background is rejected by requiring that the acoplanarity angle between
the two taus be less than 250~mrad.
 
Estimated selection efficiencies and background levels are given in 
Table~\ref{EFF_ALL}. A more detailed breakdown of the selection efficiencies
for the various tau decay channels is given in Table~\ref{EFF_TAU}.
The main uncertainty on the selection efficiency arises 
from the energy scale of the calorimeters and is estimated as in 
Section~\ref{subhad}. For the inclusive cross section the dominant background 
comes from \ggtt. The systematic uncertainty associated with the normalization
of this background is estimated from a comparison of data and Monte Carlo in 
the low visible mass range of selected tau pair events 
(15~\GeVcc~$< M_{\mathrm vis} <$~50~\GeVcc). 
In the exclusive selection, the dominant background is Bhabha events, which
sometimes pass the selection criteria if they enter cracks in the ECAL
acceptance.

The cross section measurements are listed in Table~\ref{CROSSALL}. 
The contributions to the systematic uncertainties are given in 
Table~\ref{TAUSYS}. 

\begin{table}[htbp]
\mycaption{\label{EFF_TAU} Percentage efficiency of the \tautau\ selection for 
the various tau decay channels at a centre-of-mass energy of 183~GeV.}
\begin{center}
\begin{tabular}{| c c c | r | r |}
\hline\strutl
 $\tau_1$ decay & / & $\tau_2$ decay & $\sqrt{{\rm s}^{\prime}{\rm /s}}\,>\,$0.1 \hspace*{3mm} & $\sqrt{{\rm s}^{\prime}{\rm /s}}\,>\,$0.9 \hspace*{3mm} \\

\hline
 $\mu\nu_{\mu}\nu_{\tau}$ & / & $\mu\nu_{\mu}\nu_{\tau}$ & 
 0.8 $\pm$ 0.5 \hspace*{3mm} & 1.1 $\pm$ 1.1 \hspace*{3mm} \\

 $\mu\nu_{\mu}\nu_{\tau}$ & / & hadrons$\,\nu_{\tau}$ & 
 56.3 $\pm$ 1.8 \hspace*{3mm} & 77.1 $\pm$ 2.5 \hspace*{3mm} \\

 $\mu\nu_{\mu}\nu_{\tau}$ & / & hadrons$\,\pi^0\,\nu_{\tau}$ & 
 61.7 $\pm$ 1.3 \hspace*{3mm} & 87.3 $\pm$ 1.5 \hspace*{3mm} \\

 $\mu\nu_{\mu}\nu_{\tau}$ & / & e$\nu_{\mathrm e}\nu_{\tau}$ & 
 49.2 $\pm$ 2.1 \hspace*{3mm} & 74.9 $\pm$ 2.9 \hspace*{3mm} \\

 hadrons$\,\nu_{\tau}$ & / & hadrons$\,\nu_{\tau}$ & 
 48.0 $\pm$ 2.2 \hspace*{3mm} & 67.5 $\pm$ 3.3 \hspace*{3mm} \\

 hadrons$\,\nu_{\tau}$ & / & hadrons$\,\pi^0\,\nu_{\tau}$ & 
 53.6 $\pm$ 1.2 \hspace*{3mm} & 77.9 $\pm$ 1.6 \hspace*{3mm} \\

 hadrons$\,\nu_{\tau}$ & / & e$\nu_{\mathrm e}\nu_{\tau}$ & 
 31.6 $\pm$ 1.7 \hspace*{3mm} & 50.2 $\pm$ 2.9 \hspace*{3mm} \\

 hadrons$\,\pi^0\,\nu_{\tau}$ & / & hadrons$\,\pi^0\,\nu_{\tau}$ & 
 58.7 $\pm$ 1.2 \hspace*{3mm} & 79.9 $\pm$ 1.6 \hspace*{3mm} \\

 hadrons$\,\pi^0\,\nu_{\tau}$ & / & e$\nu_{\mathrm e}\nu_{\tau}$ & 
 35.2 $\pm$ 1.3 \hspace*{3mm} & 48.5 $\pm$ 2.1 \hspace*{3mm} \\

 e$\nu_{\mathrm e}\nu_{\tau}$ & / & e$\nu_{\mathrm e}\nu_{\tau}$ & 
 0.0 \hspace*{15mm} & 0.0 \hspace*{15mm} \\

\hline
\end{tabular}
\end{center}
\end{table}

\begin{table}[htbp]
\mycaption{\label{TAUSYS}{Contributions to the systematic uncertainties on the 
ditau cross section measurements, for all energies and for both inclusive and 
exclusive processes. All quoted values are in percent.}}
\begin{center}
\begin{tabular}{| c | l || l | l | l | l | l |}
\hline\strutl
 $\sqrt{s^{\prime}/s}$ & \multicolumn{1}{c||}{Description} & 
                         \multicolumn{5}{c|}{$\mathrm E_{cm}$ (GeV)} \\
\cline{3-7}
          cut          &             & 130  & 136  & 161  & 172  & 183  \\ 
\hline\hline
  0.1 & MC statistics                & 0.9  & 0.9  & 1.1  & 1.1  & 1.0  \\
      & Energy scale                 & 1.6  & 1.7  & 1.6  & 1.6  & 1.5  \\ 
      & \ggtt                        & 0.5  & 0.6  & 0.7  & 1.2  & 0.5  \\
      & \ggmm                        & 0.8  & 0.9  & 0.8  & 0.9  & 0.1  \\
      & \qq                          & 0.4  & 0.4  & 0.1  & 0.3  & 0.3  \\
      & \ee                          & 0.7  & 0.7  & 0.5  & 1.0  & 0.6  \\
      & \WW                          & ---  & ---  & ---  & 0.4  & 0.7  \\
      & Luminosity                   & 1.0  & 1.0  & 0.7  & 0.7  & 0.5  \\
\hline
  0.9 & MC statistics                & 1.4  & 1.3  & 1.4  & 1.2  & 1.2  \\
      & Energy scale                 & 2.0  & 2.0  & 2.2  & 2.1  & 2.3  \\ 
      & \ggtt                        & 0.5  & 0.8  & 0.6  & 1.0  & 0.4  \\
      & \qq                          & 0.4  & 0.6  & 0.7  & 0.3  & 0.3  \\
      & \ee                          & 1.3  & 0.9  & 1.0  & 0.8  & 0.9  \\
      & \mumu                        & 0.1  & 0.1  & 0.1  & 0.1  & 0.1  \\
      & \WW                          & ---  & ---  & ---  & 0.2  & 0.4  \\
      & Radiative background         & 1.0  & 1.4  & 2.7  & 0.5  & 0.7  \\
      & Luminosity                   & 1.0  & 1.0  & 0.7  & 0.7  & 0.5  \\
\hline
\end{tabular}
\end{center}
\end{table}

%%%%%%%%%%%%%%%%%%%%%%%%%%%%%%%%%%%%%%%%%%%%%%%%%%%%%%%%
% asymmetries
%%%%%%%%%%%%%%%%%%%%%%%%%%%%%%%%%%%%%%%%%%%%%%%%%%%%%%%%
\subsection {Measurement of the Lepton Asymmetries}
\label{sec:lepton_asym}

\begin{figure}[htbp]
\begin{center}
\mbox{\epsfig{figure=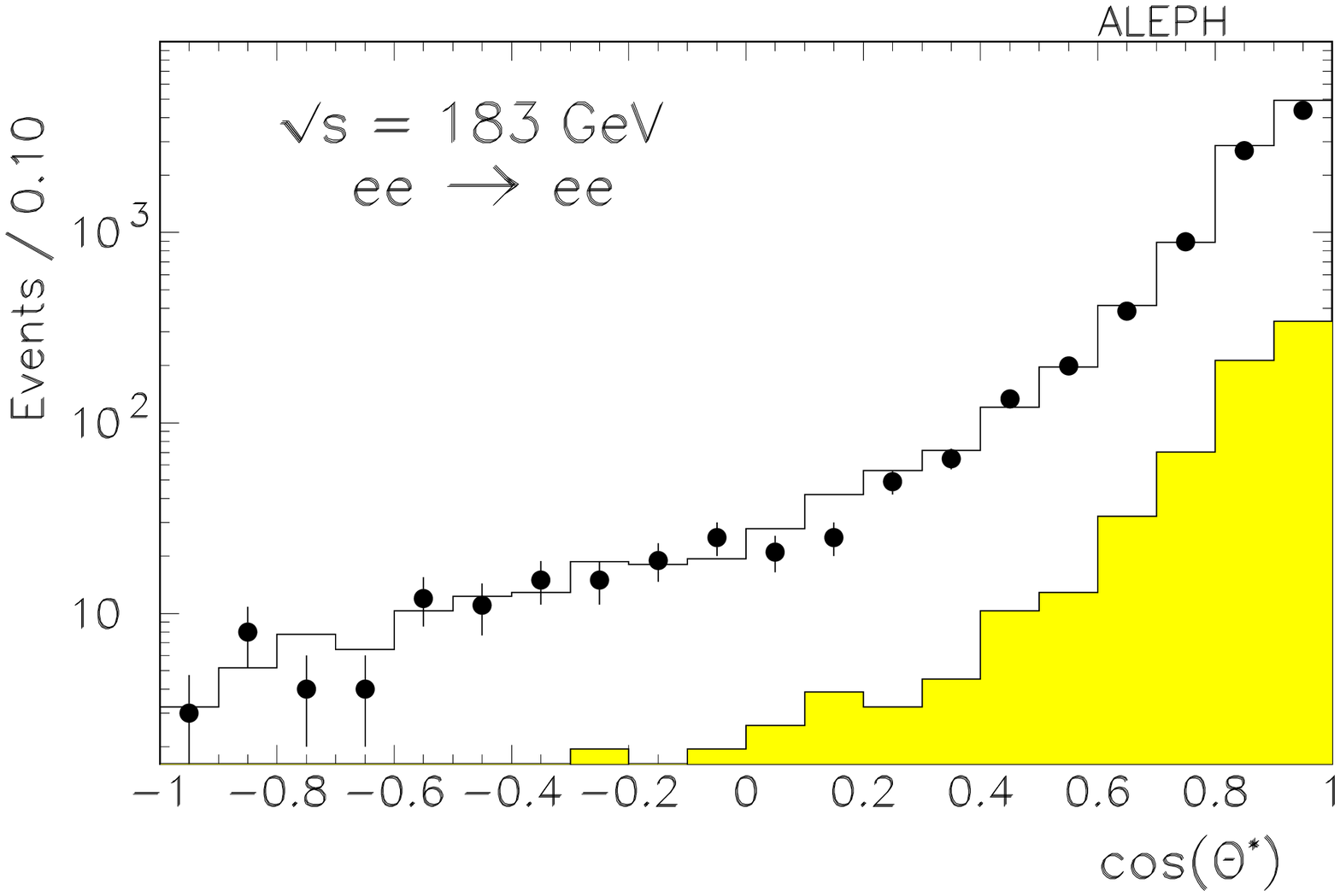,width=0.8\textwidth}} 
\mycaption{Distribution of $\cos\theta^*$ in Bhabha events at 183~GeV with
$\sqrt{s^{\prime}_m/s} > 0.9$. The data are represented by points, whilst
the Monte Carlo expectation, normalized to the same integrated luminosity, 
is shown by the white histogram. The expected background is indicated by 
the shaded histogram.
\label{cstel183}}
\end{center}
\end{figure}

\begin{figure}[htbp]
\begin{center}
\mbox{\epsfig{figure=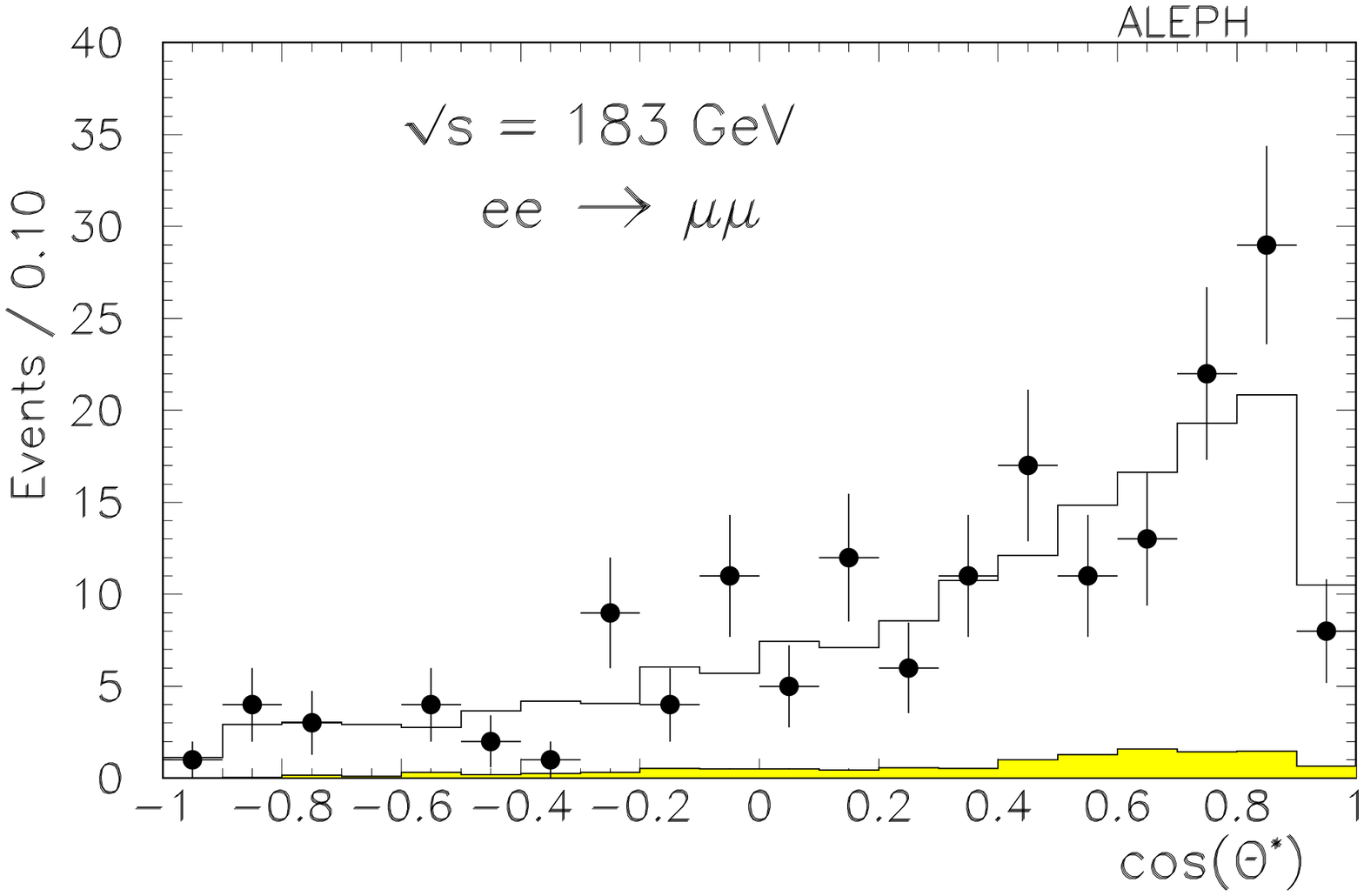,width=0.8\textwidth}} 
\mycaption{Distribution of $\cos\theta^*$ in \mumu\ events at 183~GeV with
$\sqrt{s^{\prime}_m/s} > 0.9$. The data are represented by points, whilst
the Monte Carlo expectation, normalized to the same integrated luminosity, 
is shown by the white histogram. The expected background is indicated by 
the shaded histogram.
\label{cstmu183}}
\end{center}
\end{figure}

\begin{figure}[htbp]
\begin{center}
\mbox{\epsfig{figure=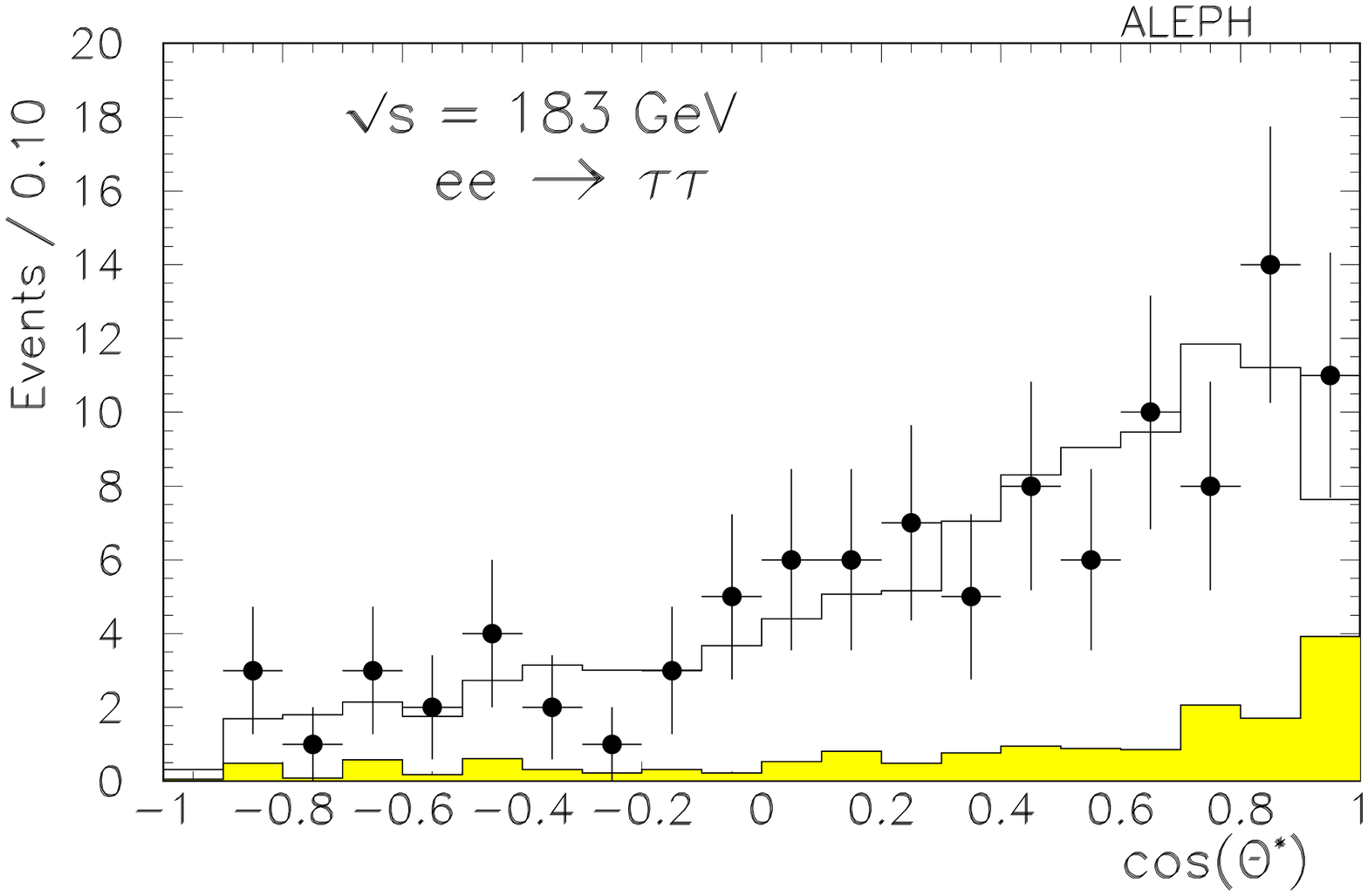,width=0.8\textwidth}} 
\mycaption{Distribution of $\cos\theta^*$ in ditau events at 183~GeV with
$\sqrt{s^{\prime}_m/s} > 0.9$. The data are represented by points, whilst
the Monte Carlo expectation, normalized to the same integrated luminosity, 
is shown by the white histogram. The expected background is indicated by 
the shaded histogram.
\label{cstau183}}
\end{center}
\end{figure}

Figures~\ref{cstel183}, \ref{cstmu183} and \ref{cstau183} show the
observed $\cos \theta^*$ distributions for electron, muon and tau pair events 
passing the exclusive selections at a centre-of-mass energy of 183~GeV. 

As discussed in Section~\ref{def}, the dilepton asymmetries are determined 
from the fraction of events in which the negatively charged lepton enters 
the forward/backward hemispheres with respect to the incoming electron.
They are measured only for the exclusive process and defined in the
range $|{\cos\theta}|<0.95$, where $\theta$ is the polar angle of the
outgoing lepton.
To ensure that the lepton charges are well measured, only events with
$|{\cos\theta^*}|<0.9$ are used. Furthermore, only unambiguously charged 
events 
are kept, i.e., the product of the charge of the two leptons must be $-1$. 
This requirement removes about 0.5\% of the muon pairs and 6\% of the tau 
pairs. The remaining charge misidentification level is estimated using 
simulated events to be 0.002\% for muon pairs and 0.05\% for tau pairs. 
The asymmetries are corrected for backgrounds and acceptance, using the 
appropriate Monte Carlo samples. This includes the effect of extrapolating
from $|{\cos\theta^*}|<0.9$ to $|{\cos\theta}|<0.95$. 

The measured \mumu\ and \tautau\ asymmetries are shown in Table~\ref{ASYM_ALL}.
For the Bhabha channel, rather than quoting an asymmetry, it is preferred to 
give the differential cross section with respect to $\theta^*$, because the 
reaction is dominated by $t$~channel photon exchange. This is given in 
Table~\ref{Bhab_dsig}. 

A major contribution to the systematic uncertainties on the asymmetries is 
the background subtraction. The ditau channel is particularly sensitive
to the Bhabha background since this is very peaked in the forward direction.
%The bias on an asymmetry \Afb{}\ due to background events can be written
%%
%\begin{equation}
%\delta\Afb{} = -\sum_{i=1}^{n_{bin}} C_i A_i +
%                \sum_{i=1}^{n_{bin}} C_i A_i^{\mathrm bkg} 
%             =  \sum_{i=1}^{n_{bin}} C_i (A_i^{\mathrm bkg} - A_i)~,
%\end{equation}
%
%where
%
%\begin{equation}
%A_{i}^{\mathrm bkg} = \frac{B_{i}^+ - B_{i}^-}
%                           {B_{i}^+ + B_{i}^-}
%\end{equation}
%%
%with $B_{i}^+$ and $B_{i}^-$ being the number of background events for two
%opposite $\pm\cos\theta$ intervals, $C_i = (B_i^+ + B_i^-)/N$ being the
%background contamination in the total number of selected events $N$, and
%%
%\begin{equation}
%A^{\mathrm bkg} = \sum_{i=1}^{n_{bin}} A_{i}^{\mathrm bkg}\frac{N_{i}^+ - N_{i}^-}{N}
%\end{equation}
%%
%is the asymmetry of the angular distribution of the background events.
%
%The first term in $\delta\Afb{}$ is due to background events with a 
%symmetrical forward-backward distribution, which is not the case here.
%The second term is due to asymmetrical backgrounds. For taus the asymmetry 
%of the background is larger than the ditau asymmetry.
%It produces a shift in the measured value of \Afb{}\ of size
%$\delta\Afb{} = C^{\mathrm bkg} (A^{\mathrm bkg} - \Afb{})$.
%From the known asymmetry of electron pair events with $|{\cos\theta}|<0.95$ of 
%0.994 and the estimated fraction $C^{\mathrm bkg}$ of this background in 
%events selected as tau pairs, the contribution of $\delta\Afb{}$ is estimated 
%for each energy point and the derived asymmetry corrected for. 
For the 
tau pair asymmetry, this gives a correction of 1.3--3.8\% for centre-of-mass 
energies of 130--183~GeV, respectively. The Monte Carlo statistical error on 
this correction enters as a part of the systematic uncertainty of the ditau 
asymmetry.
Several other sources of systematic uncertainty are considered.
The correction for event acceptance related to the extrapolation in polar
angle introduces a significant uncertainty related to the finite Monte Carlo
statistics. It also leads to a theoretical uncertainty due to ISR/FSR 
interference. This is assessed using ZFITTER.
Systematic uncertainties from charge misidentification are negligible.

%%%%%%%%
%--- tables asymmetries
%%%%%%%%
%\input ew2_conf_vanc_tabasym.tex
%%%%%%%%%%%%%%%%%%%%%%%%%%%%%%%%%%%%%%%%%%%%%%%%%%%%%%%%
%
% table: asymmetries
%
%%%%%%%%%%%%%%%%%%%%%%%%%%%%%%%%%%%%%%%%%%%%%%%%%%%%%%%%

\begin{table}[htbp]
\mycaption{\label{ASYM_ALL} Lepton forward-backward asymmetries with statistical and systematic
uncertainties, calculated for $\sqrt{s^\prime/s}\,>\,$0.9 in the range $|{\cos\theta}|<0.95$.}
\begin{center}
\begin{tabular}{| c | c || c | c |} 
\hline
 $\mathrm E_{cm}$  & Lepton & A$_{\mathrm FB}$ & SM prediction \\
        (GeV)      &  Type  &                  &                 \\
\hline\hline
  130  &   $\mu^+\mu^-$  &  0.83 $\pm$ 0.08 $\pm$ 0.03 & 0.70 \\
       &  $\tau^+\tau^-$ &  0.56 $\pm$ 0.12 $\pm$ 0.05 & 0.70 \\
\hline
  136  &   $\mu^+\mu^-$  &  0.63 $\pm$ 0.12 $\pm$ 0.03 & 0.68 \\
       &  $\tau^+\tau^-$ &  0.65 $\pm$ 0.15 $\pm$ 0.04 & 0.68 \\
\hline 
  161  & $\mu^+\mu^-$    &  0.63 $\pm$ 0.11 $\pm$ 0.03 & 0.61 \\
       & $\tau^+\tau^-$  &  0.48 $\pm$ 0.14 $\pm$ 0.04 & 0.61 \\  
\hline
  172  & $\mu^+\mu^-$    &  0.72 $\pm$ 0.13 $\pm$ 0.04 & 0.59 \\
       & $\tau^+\tau^-$  &  0.44 $\pm$ 0.20 $\pm$ 0.04 & 0.59 \\  
\hline
  183  &  $\mu^+\mu^-$   &  0.54 $\pm$ 0.06 $\pm$ 0.03 & 0.58 \\
       &  $\tau^+\tau^-$ &  0.52 $\pm$ 0.08 $\pm$ 0.04 & 0.58 \\
\hline
\end{tabular}
\end{center}
\end{table}

%%%%%%%%
%--- tables bhabha dsig
%%%%%%%%
%\input ew2_conf_vanc_tabdsig.tex
%%%%%%%%%%%%%%%%%%%%%%%%%%%%%%%%%%%%%%%%%%%%%%%%%%%%%%%%
%
% table: dsig/dtheta for bhabhas
%
%%%%%%%%%%%%%%%%%%%%%%%%%%%%%%%%%%%%%%%%%%%%%%%%%%%%%%%%
\begin{table}[p]
\vspace{-1.5cm}
\mycaption{Cross sections to produce electron pairs with 
$\sqrt{s^\prime/s}>0.9$ and $\cos\theta^*$ in the quoted ranges.
The quoted uncertainties include statistical and systematic components. 
\label{Bhab_dsig}}
\begin{center}
\begin{tabular}{|c| @{\mbox{\hspace{0.7cm}}}r r@{\mbox{\hspace{0.7cm}}} || 
r@{$\,\;\pm\;\,$}l@{\mbox{\hspace{0.5cm}}} | r@{\mbox{\hspace{1cm}}} |}
\hline
 $\mathrm E_{cm}$ (GeV) &
  \multicolumn{2}{c||}{$\cos\theta^*_{\mathrm min},\cos\theta^*_{\mathrm max}$} & 
  \multicolumn{2}{c|}{$\sigma$ (pb)} & \multicolumn{1}{c|}{SM prediction} \\ 
\hline
\hline
 130 & $-0.9$, & $-0.7$  &   0.19 &     0.34 &  0.37 \\
     & $-0.7$, & $-0.5$  &   1.41 &     0.35 &  0.55 \\
     & $-0.5$, & $-0.3$  &   1.36 &     0.45 &  1.09 \\
     & $-0.3$, & $-0.1$  &   1.23 &     0.48 &  1.19 \\
     & $-0.1$, & $ 0.1$  &   2.60 &     0.69 &  2.45 \\
     & $ 0.1$, & $ 0.3$  &   3.78 &     0.83 &  3.82 \\
     & $ 0.3$, & $ 0.5$  &   8.88 &     1.18 &  7.36 \\
     & $ 0.5$, & $ 0.7$  &  21.63 &     2.12 & 22.20 \\
     & $ 0.7$, & $ 0.9$  & 149.61 &     6.22 &148.00 \\
\hline
\hline
 136 & $-0.9$, & $-0.7$ &   0.73 &     0.20 &   0.22 \\
     & $-0.7$, & $-0.5$ &   1.16 &     0.36 &   0.62 \\
     & $-0.5$, & $-0.3$ &   0.54 &     0.35 &   0.49 \\
     & $-0.3$, & $-0.1$ &   0.52 &     0.41 &   0.89 \\
     & $-0.1$, & $ 0.1$ &   1.46 &     0.62 &   2.09 \\
     & $ 0.1$, & $ 0.3$ &   2.09 &     0.74 &   2.96 \\
     & $ 0.3$, & $ 0.5$ &   6.68 &     1.08 &   6.13 \\
     & $ 0.5$, & $ 0.7$ &  16.58 &     1.97 &  20.50 \\
     & $ 0.7$, & $ 0.9$ & 132.55 &     5.85 & 133.00 \\
\hline
\hline
 161 & $-0.9$, & $-0.7$ &   0.46 &     0.21 &   0.37 \\
     & $-0.7$, & $-0.5$ &   0.88 &     0.21 &   0.44 \\
     & $-0.5$, & $-0.3$ &   0.55 &     0.28 &   0.79 \\
     & $-0.3$, & $-0.1$ &   0.39 &     0.26 &   0.62 \\
     & $-0.1$, & $ 0.1$ &   1.24 &     0.40 &   1.43 \\
     & $ 0.1$, & $ 0.3$ &   2.37 &     0.47 &   2.07 \\
     & $ 0.3$, & $ 0.5$ &   5.35 &     0.73 &   4.95 \\
     & $ 0.5$, & $ 0.7$ &  14.38 &     1.27 &  14.10 \\
     & $ 0.7$, & $ 0.9$ &  93.76 &     3.76 &  94.20 \\
\hline
\hline
 172 & $-0.9$, & $-0.7$ &   0.32 &     0.19 &   0.28 \\
     & $-0.7$, & $-0.5$ &   0.88 &     0.19 &   0.34 \\
     & $-0.5$, & $-0.3$ &   0.66 &     0.24 &   0.58 \\
     & $-0.3$, & $-0.1$ &   0.61 &     0.23 &   0.44 \\
     & $-0.1$, & $ 0.1$ &   0.95 &     0.36 &   1.23 \\
     & $ 0.1$, & $ 0.3$ &   1.80 &     0.47 &   1.93 \\
     & $ 0.3$, & $ 0.5$ &   4.92 &     0.71 &   4.24 \\
     & $ 0.5$, & $ 0.7$ &  13.07 &     1.20 &  12.40 \\
     & $ 0.7$, & $ 0.9$ &  84.61 &     3.51 &  81.10 \\
\hline
\hline
 183 & $-0.9$, & $-0.7$ &   0.24 &     0.07 &   0.21 \\
     & $-0.7$, & $-0.5$ &   0.29 &     0.07 &   0.25 \\
     & $-0.5$, & $-0.3$ &   0.46 &     0.10 &   0.51 \\
     & $-0.3$, & $-0.1$ &   0.71 &     0.12 &   0.64 \\
     & $-0.1$, & $ 0.1$ &   0.83 &     0.14 &   0.90 \\
     & $ 0.1$, & $ 0.3$ &   1.42 &     0.20 &   1.83 \\
     & $ 0.3$, & $ 0.5$ &   3.90 &     0.29 &   3.66 \\
     & $ 0.5$, & $ 0.7$ &  12.47 &     0.56 &  11.10 \\
     & $ 0.7$, & $ 0.9$ &  71.90 &     1.86 &  71.80 \\
\hline
\end{tabular}
\end{center}
\end{table}

\clearpage

%%%%%%%%
%--- section : results
%%%%%%%%
%\input ew2_conf_vanc_results.tex
%%%%%%%%%%%%%%%%%%%%%%%%%%%%%%%%%%%%%%%%%%%%%%%%%%%%%%%%
%
% discussion of results 
%
%%%%%%%%%%%%%%%%%%%%%%%%%%%%%%%%%%%%%%%%%%%%%%%%%%%%%%%%

\section{Interpretation in Terms of New Physics}
 \label{interpretations} 

% smpred
\subsection{Comparison with Standard Model Predictions}
 \label{smpred}

\begin{figure}[htbp]
\begin{center}
\mbox{\epsfig{figure=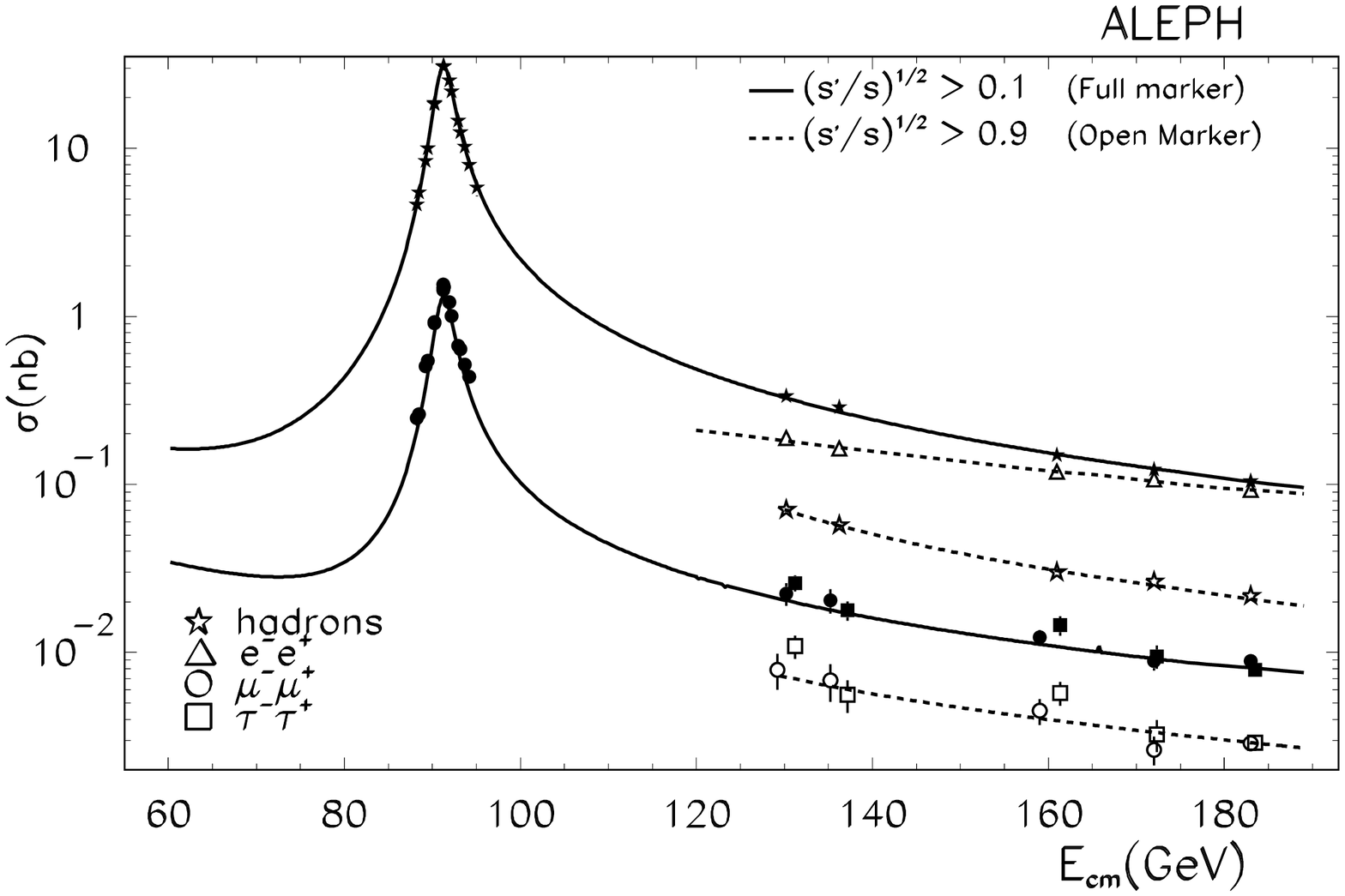,width=0.9\textwidth}} 
\mycaption{Measured cross sections for fermion pair production. The 
curves indicate the predictions obtained from BHWIDE for the Bhabha process
and from ZFITTER for the other channels. (Some of the points are shifted
slightly along the horizontal axis to prevent them overlapping).
\label{line_lep}}
\end{center}
\end{figure}

\begin{figure}[htbp]
\begin{center}
\mbox{\epsfig{figure=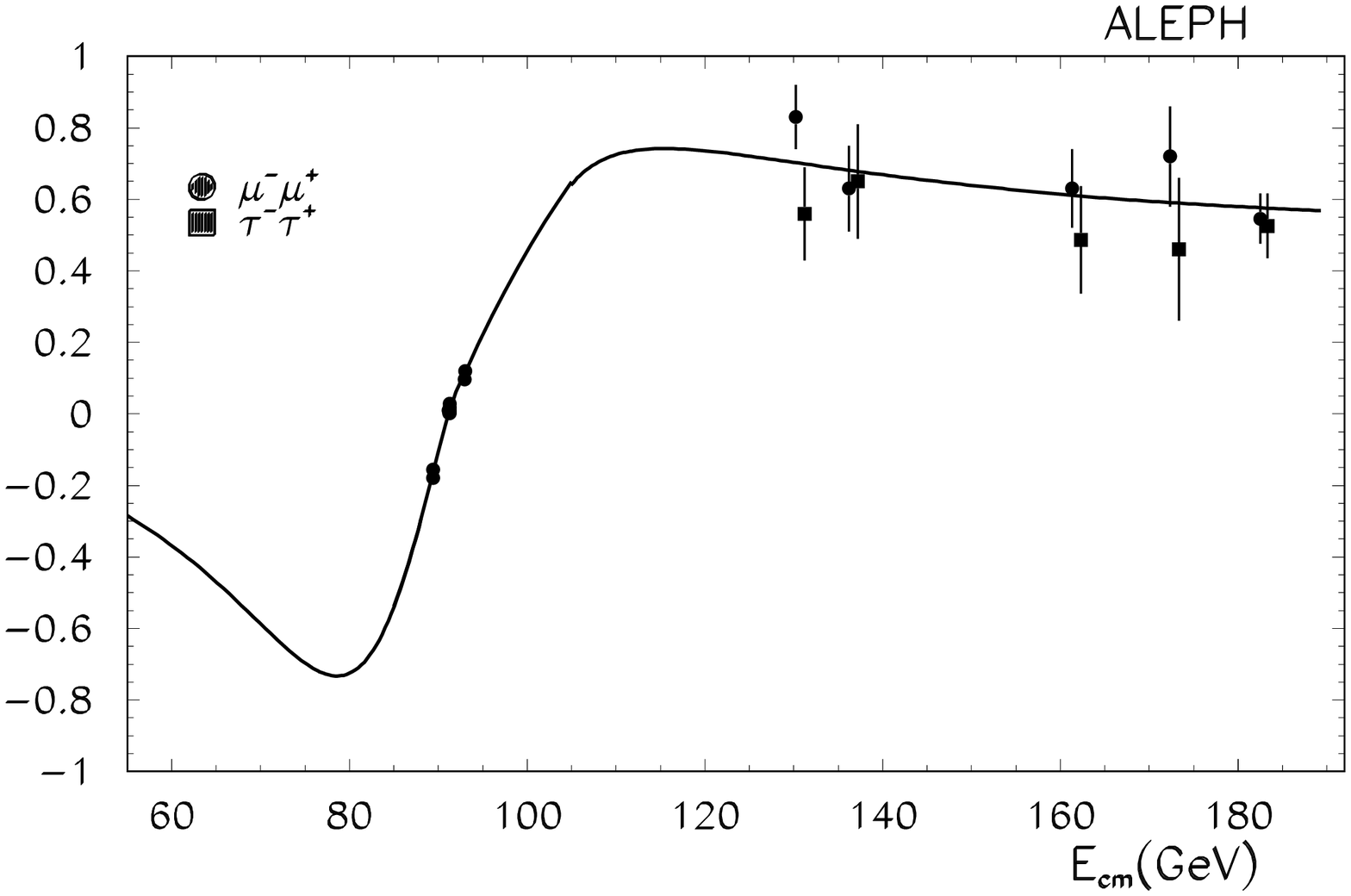,width=0.9\textwidth}} 
\mycaption{Measured asymmetries for muon and tau pair production.
The curves indicate the predictions obtained from ZFITTER. (Some of the 
points are shifted slightly along the horizontal axis to prevent them 
overlapping).
\label{asym_lep2}}
\end{center}
\end{figure}

The measured cross sections and asymmetries from Tables~\ref{CROSSALL} 
and \ref{ASYM_ALL} are plotted as a function of centre-of-mass energy
in Figs.~\ref{line_lep} and \ref{asym_lep2}, respectively.
The results are compared with SM predictions based on 
BHWIDE \cite{BHWIDE} for electron pair production and ZFITTER \cite{ZFITTER} 
for all other processes. The ZFITTER predictions
\footnote{
Default ZFITTER flags are used, except for
$\mathrm BOXD=1$, $\mathrm CONV=1$, $\mathrm INTF=1$ and $\mathrm INCL=0$.
The flag $\mathrm FINR=0$ is used for hadronic events and $\mathrm FINR=1$ for
dilepton events. 
}
are computed from the input values 
%the latest results from the LEP line-shape analysis as well as direct 
%measurements of the W and top mass at %the Tevatron~\cite{WW98}, i.e., 
$m_{\mathrm Z}=91.1867$~\GeVcc, $m_{\mathrm t}=174.1$~\GeVcc, 
$m_{\mathrm H}=127.0$~\GeVcc, $\alpha_{\mathrm em}(M_{\mathrm Z})=1/128.896$ 
and $\alpha_s(M_{\mathrm Z})=0.120$. The exclusive cross sections and asymmetries are given 
in the restricted angular range for the outgoing fermion direction, 
$|{\cos\theta}|<0.95$. The inclusive cross sections correspond to the full 
angular range.

Because of the poor knowledge of the contribution of ISR/FSR interference,
the ZFITTER predictions are assigned a systematic uncertainty 
equal to the difference in the predictions with and without interference.
This amounts to 1.5\% for the hadronic cross section and
2\% for the \mumu, \tautau, \bb\  and \cc\  cross sections. These uncertainties would 
almost double if one attempted to extrapolate the results to the full angular
range. The Bhabha cross section is assumed to be uncertain by 3\%, based on
a study of different event generators. These uncertainties are taken into 
account in the calculation of limits on physics beyond the SM given in the 
following sections.

The measured cross sections and asymmetries are consistent with 
SM predictions. Similar results have been published by 
the OPAL Collaboration~\cite{OPAL}. The L3 Collaboration has published results
covering centre-of-mass energies up to 172~GeV~\cite{L3:cross}.

\subsection{Limits on Four-Fermion Contact Interactions}
\label{contacts}

Comparing the measured exclusive difermion cross sections and 
angular distributions with SM predictions allows one to place
limits upon many possible extensions to the SM.
One convenient parametrization of such effects is given by the addition of 
four-fermion contact interactions \cite{Eichten:1983} to the known 
SM processes. Such contact interactions are characterized by 
a scale \lam, interpreted as the mass of a new heavy 
particle exchanged between the incoming and outgoing fermion pairs, and a 
coupling $g$ giving the strength of the interaction. Contact interactions 
are, for example, expected to occur if fermions are composite.

Following the notation of Ref.~\cite{Kroha:1992}, the effective 
Lagrangian for the four-fermion contact interaction in the process
$\ee\rightarrow\ff$ is given by 
\begin{equation}
\label{lagrangian}
 {\cal L}^{CI} = \frac{g^2\eta_{\mathrm sign}}{(1+\delta) \lam^2} 
                 \sum_{i,j=L,R} \eta_{ij}
                 [\bar{\mathrm e}_i \gamma^\mu {\mathrm e}_i]
                 [\bar{\mathrm f}_j \gamma_\mu {\mathrm f}_j]~,
\end{equation}
with $\delta = 1$ if $\mathrm f = e$, or 0 otherwise. The fields ${\mathrm e}_{L,R}$ 
(${\mathrm f}_{L,R}$) are the left- and right-handed chirality 
projections of electron (fermion) spinors. The coefficients
$\eta_{ij}$, which take a value between $-1$ and $+1$, indicate the relative 
contribution of the different chirality combinations to the Lagrangian. 
The sign of $\eta_{\mathrm sign}$ determines whether the contact interaction
interferes constructively or destructively with the SM amplitude.
Several different models are considered in this analysis, corresponding
to the choices of the $\eta_{\mathrm sign}$ and $\eta_{ij}$ given in 
Table~\ref{models}. 

\begin{table}[hbt]
\mycaption{\label{models} Four-fermion interaction models.}
\begin{center}
\begin{tabular}{|c||c|c|c|c|c|}
\hline
    Model    & $\eta_{\mathrm sign}$ & $\eta_{\mathrm{LL}}$ & 
 $\eta_{\mathrm{RR}}$ & $\eta_{\mathrm{LR}}$ & $\eta_{\mathrm{RL}}$ \\
\hline\hline
  LL$^{\pm}$   &  $\pm 1$  &  1  &  0  &   0  &   0   \\
  RR$^{\pm}$   &  $\pm 1$  &  0  &  1  &   0  &   0   \\
  VV$^{\pm}$   &  $\pm 1$  &  1  &  1  &   1  &   1   \\
  AA$^{\pm}$   &  $\pm 1$  &  1  &  1  & $-1$ & $-1$  \\
  LR$^{\pm}$   &  $\pm 1$  &  0  &  0  &   1  &   0   \\
  RL$^{\pm}$   &  $\pm 1$  &  0  &  0  &   0  &   1   \\
 LL+RR$^{\pm}$ &  $\pm 1$  &  1  &  1  &   0  &   0   \\
 LR+RL$^{\pm}$ &  $\pm 1$  &  0  &  0  &   1  &   1   \\
\hline
\end{tabular}
\end{center}
\end{table}

In the presence of contact interactions the differential cross section for 
$\ee\rightarrow\ff$ 
as a function of the polar angle $\theta$ of the outgoing fermion
with respect to the $\mathrm e^-$ beam line can be written as
\begin{equation}
\label{xsection}
\frac{{d}\sigma}{{d}\cos\theta} = F_{\mathrm SM}(s,t)
  \left[1 + \eps   \frac{F_{\mathrm IF}^{\mathrm Born}(s,t)}
                        {F_{\mathrm SM}^{\mathrm Born}(s,t)}
          + \eps^2 \frac{F_{\mathrm CI}^{\mathrm Born}(s,t)}
                        {F_{\mathrm SM}^{\mathrm Born}(s,t)} \right]
\end{equation}
with $s$ and $t$ being the Mandelstam variables and 
$\eps = g^2\eta_{\mathrm sign}/(4\pi\lam^2)$. The SM cross section 
$F_{\mathrm SM}$ is computed as described in Section~\ref{smpred}. 
The contributions to the cross section from the SM -- contact interaction 
interference and from the pure contact interaction are denoted by 
$F_{\mathrm IF}^{\mathrm Born}$ and $F_{\mathrm CI}^{\mathrm Born}$, 
respectively. They are calculated in the improved Born approximation. The 
Born level formulae can be found in Ref.~\cite{Kroha:1992} and these are
corrected for ISR according to Ref.~\cite{MIZA}.
Because no higher order calculations are available for the contact 
interactions, the ratios of these with the improved Born predictions for the 
SM cross sections are taken, to allow for a partial cancellation of 
higher order effects. 

The predictions of Equation~\ref{xsection} are fitted to the data using a 
binned maximum likelihood method. For contact interactions affecting the
dilepton channels, the likelihood function ${\cal L}$ is defined by
\begin{equation}
 \label{likeli}
 {\cal L} = G(\alpha^{\mathrm corr};1)\,\prod_{i}\,G(\alpha^{\mathrm uncorr}_i;1)\,
            \prod_{k}\,{\cal P}
 \left(\,
  N_{ik}^{\mathrm data} , \left[ N_{ik}^{\mathrm pred}(\eps) + \alpha^{\mathrm corr}\Delta n^{\mathrm corr}_{ik}
                                         + \alpha^{\mathrm uncorr}_i\Delta n^{\mathrm uncorr}_{ik}\right]\,
 \right)~.
\end{equation}
The indices $i$ and $k$ run over the centre-of-mass energy points and 
angular bins in $\cos\theta^*$, respectively.
The function ${\cal P}$ gives the Poisson probability to observe 
$N_{ik}^{\mathrm data}$ events in the data if $N_{ik}^{\mathrm pred}$ are
expected. The systematic uncertainties on the expected number of events which 
are (un)correlated between the centre-of-mass energy points are represented 
by ($\Delta n^{\mathrm uncorr}$) $\Delta n^{\mathrm corr}$, respectively.
These uncertainties are taken into account using the parameters 
$\alpha^{\mathrm corr}$ and $\alpha^{\mathrm uncorr}_i$, which are 
constrained using Gaussian distributions $G$ with zero 
mean and unit standard deviation. 
These parameters are fitted together with the parameter \eps.
%The uncertainties  of the theoretical predictions are of the
%order of $3\%$ for Bhabha scattering and $1\%$ for the other channels.
The fit range in the angular distribution is chosen to be
$|{\cos\theta^*}| < 0.9$ for the Bhabha channel and
$|{\cos\theta^*}|< 0.95$ for the muon and tau pair channels.

For contact interactions affecting hadronic events, the sum over angular
bins is dropped, and instead two additional terms are added to the likelihood
function to take into account the constraints from the measurement of 
the jet charge asymmetries, given by Equation~\ref{Afbeqn} and Table~\ref{Afb}.
The contact interaction can be assumed to couple to all quark flavours
with equal strength. In this case, the jet charge asymmetry measurements 
improve the limits for some models by up to 70\%, primarily because of their 
sensitivity to the relative cross sections of up- and down-type quarks.
Alternatively, one can assume that contact interactions only affect the \bb\
final state. Such interactions are strongly constrained using the measurement
of \Rb\ from Section~\ref{rb} and the jet charge asymmetry in b-enriched
events of Section~\ref{jet_charge}.

Because of the quadratic dependence of the theoretical cross
sections upon \eps, the likelihood function can have two
maxima. 
%Therefore the central value of \eps\ and its asymmetric uncertainties 
%$\Delta \eps^{\pm}$ are estimated as follows~: 
%\begin{equation}
% \label{mean}
%   \int_{-\infty}^{\eps} {\cal L}(\eps') d\eps' = 
%   0.5\, \int_{-\infty}^{\infty} {\cal L}(\eps') d\eps' \quad ,
%\end{equation}
%\begin{equation}
% \label{uncertainty}
%   \int_{\eps}^{\eps+\Delta\eps^+} {\cal L}(\eps') d\eps' = 
%   0.34\, \int_{-\infty}^{\infty} {\cal L}(\eps') d\eps' \quad , \quad
%   \int_{\eps-\Delta\eps^-}^{\eps} {\cal L}(\eps') d\eps' = 
%   0.34\, \int_{-\infty}^{\infty} {\cal L}(\eps') d\eps' \quad ,
%\end{equation}
The 68\% confidence level limits (\eps$^+$ and \eps$^-$) on \eps\ 
are therefore estimated as follows:
\begin{equation}
 \label{uncertainty}
   \int_{-\infty}^{\eps^-} {\cal L}(\eps') d\eps' = 
    \int_{\eps^+}^{\infty} {\cal L}(\eps') d\eps' = 
   0.16\, \int_{-\infty}^{\infty} {\cal L}(\eps') d\eps'~,
\end{equation}
where for each value of $\eps'$ the parameters $\alpha_c$
and $\alpha_i$ are chosen which maximize the likelihood.
The results for contact interaction affecting leptonic final states are 
listed in Table \ref{lepton_results}. Table \ref{quark_results} gives the 
results obtained for contact interactions affecting both hadronic events 
and all difermion events.

% The results are summarized in Figs.~\ref{eps_plot1} and \ref{eps_plot2}. 

\begin{table}[pht]
\mycaption{\label{lepton_results} Limits on contact interactions coupling to
dilepton final states.
The 68\% confidence level range is given for $\eps$ whilst the 95\% confidence
level limits are given for $\Lambda$.
The results presented for \myll\  assume lepton universality.}
\begin{center}
\begin{tabular}{|c||r|c|c|}
\hline
         Model                    & [$\eps^-$,$\eps^+$] (TeV$^{-2}$)\qquad & $\Lambda^-$ (TeV) & $\Lambda^+$ (TeV) \\
\hline\hline
 $\ee\rightarrow\ee$              &                   &                   &                   \\
          LL                      & [$-0.067,+0.021$] &       3.2         &      3.5          \\
          RR                      & [$-0.067,+0.022$] &       3.2         &      3.4          \\
          VV                      & [$-0.017,+0.003$] &       6.4         &      8.0          \\
          AA                      & [$-0.018,+0.019$] &       4.2         &      5.5          \\
          LR                      & [$-0.042,+0.015$] &       4.0         &      4.2          \\
        LL+RR                     & [$-0.038,+0.009$] &       4.2         &      5.0          \\
        LR+RL                     & [$-0.022,+0.006$] &       5.5         &      6.5          \\
\hline
 $\ee\rightarrow\mumu$            &                   &                   &                   \\
          LL                      & [$-0.014,+0.040$] &       4.7         &      4.0          \\
          RR                      & [$-0.016,+0.043$] &       4.4         &      3.8          \\
          VV                      & [$-0.005,+0.016$] &       7.7         &      6.3          \\
          AA                      & [$-0.009,+0.015$] &       6.8         &      6.2          \\
          LR                      & [$-0.270,+0.025$] &       1.8         &      3.8          \\
        LL+RR                     & [$-0.007,+0.022$] &       6.6         &      5.4          \\
        LR+RL                     & [$-0.260,+0.019$] &       1.9         &      5.1          \\
\hline
 $\ee\rightarrow\tau^+\tau^-$     &                   &                   &                   \\
          LL                      & [$-0.039,+0.032$] &       3.7         &      3.9          \\
          RR                      & [$-0.046,+0.034$] &       3.4         &      3.7          \\
          VV                      & [$-0.012,+0.016$] &       6.2         &      5.9          \\
          AA                      & [$-0.022,+0.013$] &       5.2         &      5.6          \\
          LR                      & [$-0.275,+0.033$] &       1.8         &      3.3          \\
        LL+RR                     & [$-0.020,+0.018$] &       5.2         &      5.2          \\
        LR+RL                     & [$-0.265,+0.025$] &       1.8         &      4.3          \\
\hline
 $\ee\rightarrow\myll$            &                   &                   &                   \\
          LL                      & [$-0.014,+0.020$] &       5.5         &      5.3          \\
          RR                      & [$-0.016,+0.021$] &       5.3         &      5.1          \\
          VV                      & [$-0.005,+0.006$] &       9.5         &      9.3          \\
          AA                      & [$-0.007,+0.010$] &       8.0         &      7.5          \\
          LR                      & [$-0.023,+0.019$] &       4.8         &      5.0          \\
         LL+RR                    & [$-0.008,+0.010$] &       7.7         &      7.3          \\
         LR+RL                    & [$-0.011,+0.009$] &       7.1         &      7.2          \\
\hline
\end{tabular}
\end{center}
\end{table}
\begin{table}[htbp]
\mycaption{\label{quark_results} Limits on contact interactions coupling to
hadronic or to all difermion final states.
The 68\% confidence level range is given for $\eps$ whilst the 95\% confidence
level limits are given for $\Lambda$.
The results presented for \ff\  assume that the contact interaction couples 
to all the outgoing fermion types equally.}
\begin{center}
\begin{tabular}{|c||r|c|c|}
\hline
         Model                    & [$\eps^-$,$\eps^+$] (TeV$^{-2}$)\qquad & $\Lambda^-$ (TeV) & $\Lambda^+$ (TeV) \\
\hline\hline\strutl
 $\ee\rightarrow\bb$              &                   &                   &                   \\
          LL                      & [$-0.024,+0.013$] &       4.9         &      5.6          \\
          RR                      & [$-0.232,-0.004$] &       1.9         &      3.9          \\
          VV                      & [$-0.029,+0.007$] &       4.6         &      6.5          \\
          AA                      & [$-0.016,+0.009$] &       5.9         &      7.0          \\
          LR                      & [$-0.143,+0.054$] &       2.3         &      3.0          \\
          RL                      & [$-0.028,+0.232$] &       3.6         &      1.9          \\
       LL+RR                      & [$-0.018,+0.009$] &       5.7         &      6.6          \\
       LR+RL                      & [$-0.036,+0.101$] &       3.6         &      2.8          \\
\hline\strutl
 $\ee\rightarrow\qq$              &                   &                   &                   \\
          LL                      & [$-0.008,+0.022$] &       6.2         &      5.4          \\
          RR                      & [$-0.025,+0.036$] &       4.4         &      3.9          \\
          VV                      & [$-0.010,+0.013$] &       7.1         &      6.4          \\
          AA                      & [$-0.004,+0.013$] &       7.9         &      7.2          \\
          LR                      & [$-0.055,+0.079$] &       3.3         &      3.0          \\
          RL                      & [$-0.045,+0.076$] &       4.0         &      2.4          \\
        LL+RR                     & [$-0.007,+0.014$] &       7.4         &      6.7          \\
        LR+RL                     & [$-0.029,+0.099$] &       4.5         &      2.9          \\
\hline\strutl
 $\ee\rightarrow\ff$              &                   &                   &                   \\
          LL                      & [$-0.006,+0.016$] &       7.2         &      6.2          \\
          RR                      & [$-0.013,+0.019$] &       5.8         &      5.4          \\
          VV                      & [$-0.005,+0.005$] &      10.1         &      9.8          \\
          AA                      & [$-0.003,+0.009$] &       9.8         &      8.4          \\
          LR                      & [$-0.024,+0.020$] &       4.8         &      4.9          \\
          RL                      & [$-0.029,+0.006$] &       4.9         &      5.7          \\
         LL+RR                    & [$-0.004,+0.009$] &       9.2         &      8.1          \\
         LR+RL                    & [$-0.014,+0.006$] &       6.8         &      7.6          \\
\hline
\end{tabular}
\end{center}
\end{table}

Although all the physics content is described by the well-defined 
parameter \eps, it is conventional to
extract limits on the energy scale \lam, assuming $g^2=4\pi$. The 
$95\%$ confidence level limits $\eps_{95}^{\pm}$ are computed according to 
\begin{equation}
   \int_{0}^{\eps_{95}^{+}} {\cal L}(\eps') d\eps' = 
   0.95\, \int_{0}^{\infty} {\cal L}(\eps') d\eps'~, \qquad
   \int_{\eps_{95}^{-}}^{0} {\cal L}(\eps') d\eps' = 
   0.95\, \int_{-\infty}^{0} {\cal L}(\eps') d\eps'~, 
\end{equation}
which are then used to obtain
\begin{equation}
 \label{lambda}
      \lam^{\pm} = 1 \left/ \sqrt{|\eps_{95}^{\pm}|}\right.~.
\end{equation}
Limits on the energy scale are listed in Tables \ref{lepton_results} and 
\ref{quark_results}. One can drop the assumption $g^2=4\pi$, in which case
these results become limits on $\sqrt{4\pi}\lam/g$.

These results are competitive with previous analyses of contact interactions
already performed at LEP \cite{OPAL,L3:CI:1998}, at the Tevatron
\cite{CDF:CI:1997} and at HERA \cite{H1:CI:1995}. However,
models of $\ee\uu$ and $\ee\dd$ 
contact interactions which violate parity (LL, RR, LR and RL) are already 
severely constrained by atomic physics parity violation experiments,
which quote limits of the order of $15$~TeV~\cite{Deandrea:1997}. 
The LEP limits for the fully leptonic couplings or those involving b~quarks 
are of particular interest since they are inaccessible 
at $\mathrm p\bar p$ or $\mathrm ep$ colliders. 

% However, they can be eluded if the new contact interactions
% obey global symmetries \cite{Nelson}. In that case the limits for the LR and
% RL models obtained here are of relevance.

%The Bhabha scattering process is the channel with the
%highest sensitivity, which is due to the interference between
%the SM t-channel and the contact interaction amplitudes. 
%The excess of events at high $Q^2$ at HERA \cite{H1:excess,ZEUS:excess} 
%has also been interpreted \cite{HERA:nonSM, HERA:contact} as being due to
%a contact interaction. However, the improved limit on 
%$\Lambda^{-d}_{\mathrm LR}$ presented here makes such an interpretation 
%less likely.

%%%%%%%%
%--- subsection : sneutrinos
%%%%%%%%
%\input ew2_conf_vanc_snu.tex
%%%%%%%%%%%%%%%%%%%%%%%%%%%%%%%%%%%%%%%%%%%%%%%%%%%%%%%%
%
% sneutrinos
%
%%%%%%%%%%%%%%%%%%%%%%%%%%%%%%%%%%%%%%%%%%%%%%%%%%%%%%%%
%
\subsection{\boldmath Limits on R-parity Violating Sneutrinos}
 \label{snu}
Supersymmetric theories with R-parity violation have terms in
the Lagrangian of the form $\lambda_{ijk} {\mathrm L}_i {\mathrm L}_j 
\bar {\mathrm E}_k$, where $\mathrm L$ denotes a lepton doublet superfield
and $\bar{\mathrm E}$ denotes a lepton singlet superfield.
The parameter $\lambda$ is a Yukawa coupling, and $i$, $j$, $k = 1$, 2, 3 are 
generation indices. The $\lambda_{ijk}$, which for the purposes of this 
analysis are assumed to be real, are non-vanishing only for $i < j$. These 
terms allow for single production of sleptons at \ee\ collider experiments.

At LEP, dilepton production cross sections could then differ from their 
SM expectations as a result of the exchange of R-parity 
violating sneutrinos in the $s$ or $t$~channels \cite{snu_theory}. 
Table~\ref{snu_table} shows the most interesting possibilities.
Those involving $s$~channel sneutrino exchange lead to a resonance.
For the results presented here, this resonance is assumed to have a width
of 1~\GeVcc, which can occur if the sneutrino also has R-parity conserving
decay modes~\cite{snu_theory}. For a sneutrino only having the R-parity
violating decay mode into lepton pairs, the width would be much less
than this, leading to slightly better limits.

\begin{table}[htbp]
\mycaption{\label{snu_table} 
For each dilepton channel, the table shows the coupling amplitude, 
the sneutrino type exchanged, and an indication of whether the exchange 
occurs in the $s$ or $t$~channel.}
\begin{center}
\begin{tabular}{|c||c|c|c|}
\hline
 $\lambda^2$       &    \ee    &  $\mu^+\mu^-$  & $\tau^+\tau^-$ \\
\hline
 $\lambda_{121}^2$ & $\tilde\nu_\mu$ (s,t)  & $\tilde\nu_e$ (t)  & --- \\
 $\lambda_{131}^2$ & $\tilde\nu_\tau$ (s,t) & --- & $\tilde\nu_e$ (t)  \\ 
 $\lambda_{121}\lambda_{233}$ & ---  & --- & $\tilde\nu_\mu$ (s) \\
 $\lambda_{131}\lambda_{232}$ & ---  & $\tilde\nu_\tau$ (s) & --- \\
\hline
\end{tabular}
\end{center}
\end{table}

Direct searches for R-parity violating sneutrinos at LEP have led to lower 
limits on their masses of 72~\GeVcc\ for $\tilde\nu_e$ and 49~\GeVcc\ for 
$\tilde\nu_\mu$ and $\tilde\nu_\tau$ \cite{ALEPH_snu}.
Indirect limits based upon lepton universality and leptonic tau decays imply 
that the couplings in Table~\ref{snu_table} must approximately satisfy 
$\lambda < 0.1 (m_{\tilde l_R}/200)$~\GeVcc, where $m_{\tilde l_R}$
is the mass of the appropriate right-handed charged slepton \cite{snu_theory}.
These limits can be improved using the dilepton 
cross section and asymmetry data presented in this paper.

Limits on the couplings are obtained by comparing the measured dilepton 
differential cross sections with respect to the polar angle with the 
theoretical cross sections in reference \cite{snu_theory}. The likelihood 
function used in the fit and the corrections for ISR, etc. follow the procedure
of Section~\ref{contacts}.
The fit is performed in terms of the parameter $\lambda^2$. Since the
likelihood function can have two minima, limits are again determined by 
integrating the likelihood function with respect to $\lambda^2$.
A one-sided limit is used when $\lambda^2$ is positive definite, which occurs
when $\lambda^2 = \lambda_{121}^2$ or $\lambda_{131}^2$, but 
not when $\lambda^2 = \lambda_{121}\lambda_{233}$ or
$\lambda_{131}\lambda_{232}$.

\begin{figure}[htbp]
\begin{center}
\mbox{\epsfig{file=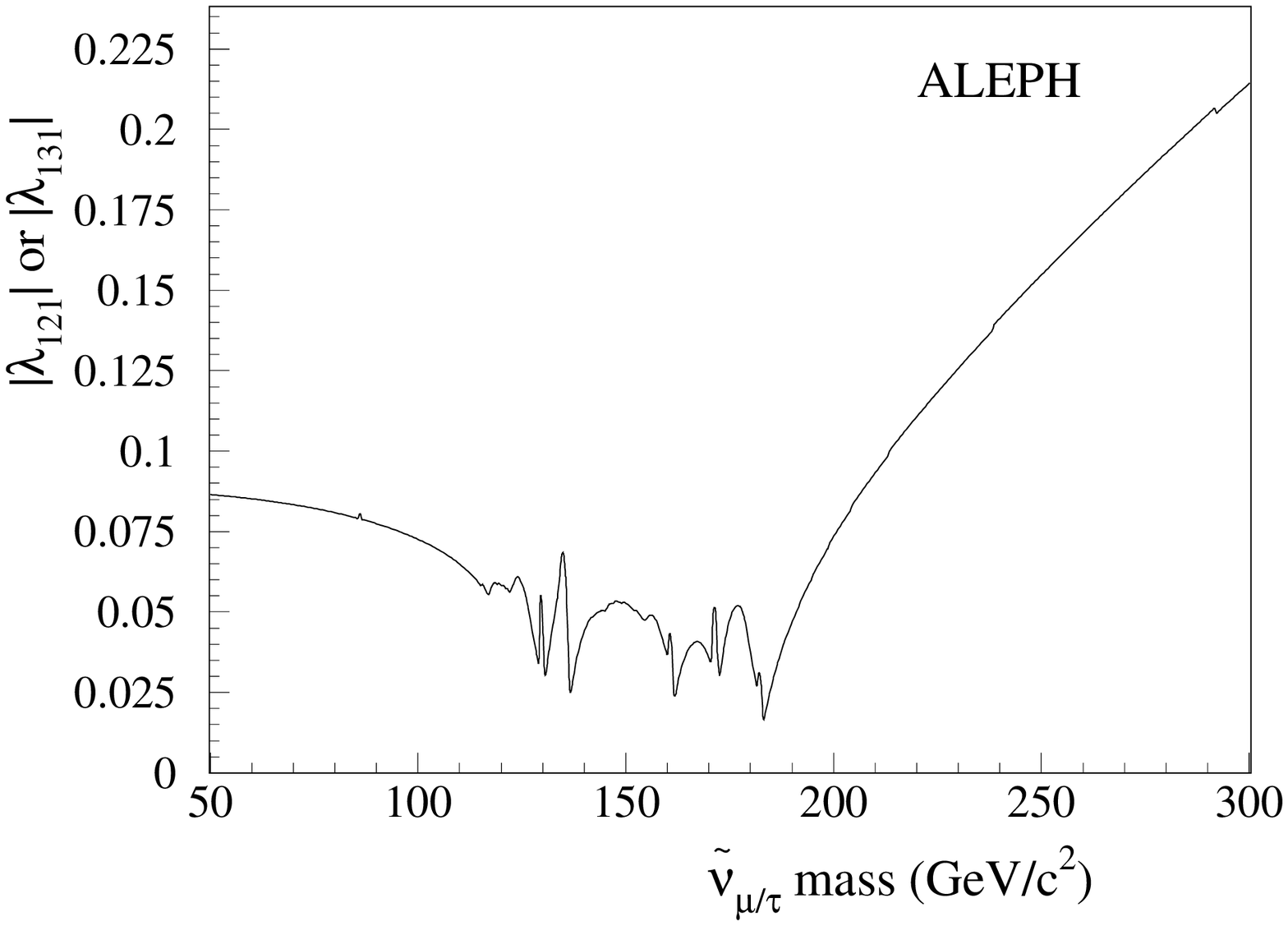,height=0.35\textheight}}
\end{center}
\mycaption{\label{e_snu} 95\% confidence level upper limits, obtained from the 
         Bhabha cross sections,  on $|\lambda_{121}|$ versus the assumed 
         $\tilde\nu_\mu$ mass and on $|\lambda_{131}|$ versus the assumed 
         $\tilde\nu_\tau$ mass.}
\end{figure}

\begin{figure}[htbp]
\begin{center}
\mbox{\epsfig{file=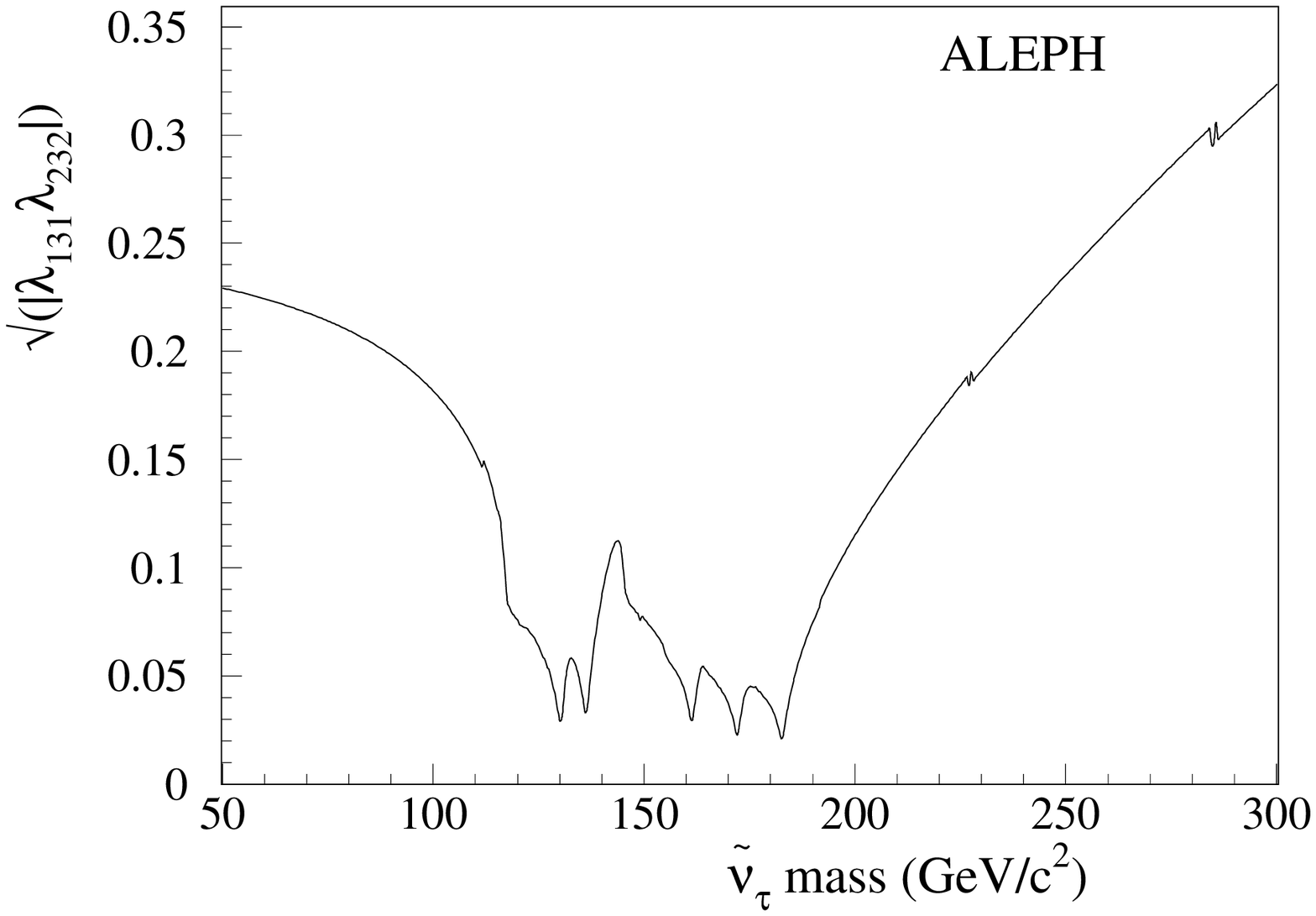,height=0.35\textheight}}
\end{center}
\mycaption{\label{mu_snu} 95\% confidence level upper limits, obtained from the 
            \mumu\ cross sections, on $\sqrt{|\lambda_{131}\lambda_{232}|}$ 
            versus the assumed $\tilde\nu_\tau$ mass.}
\end{figure}

\begin{figure}[htbp]
\begin{center}
\mbox{\epsfig{file=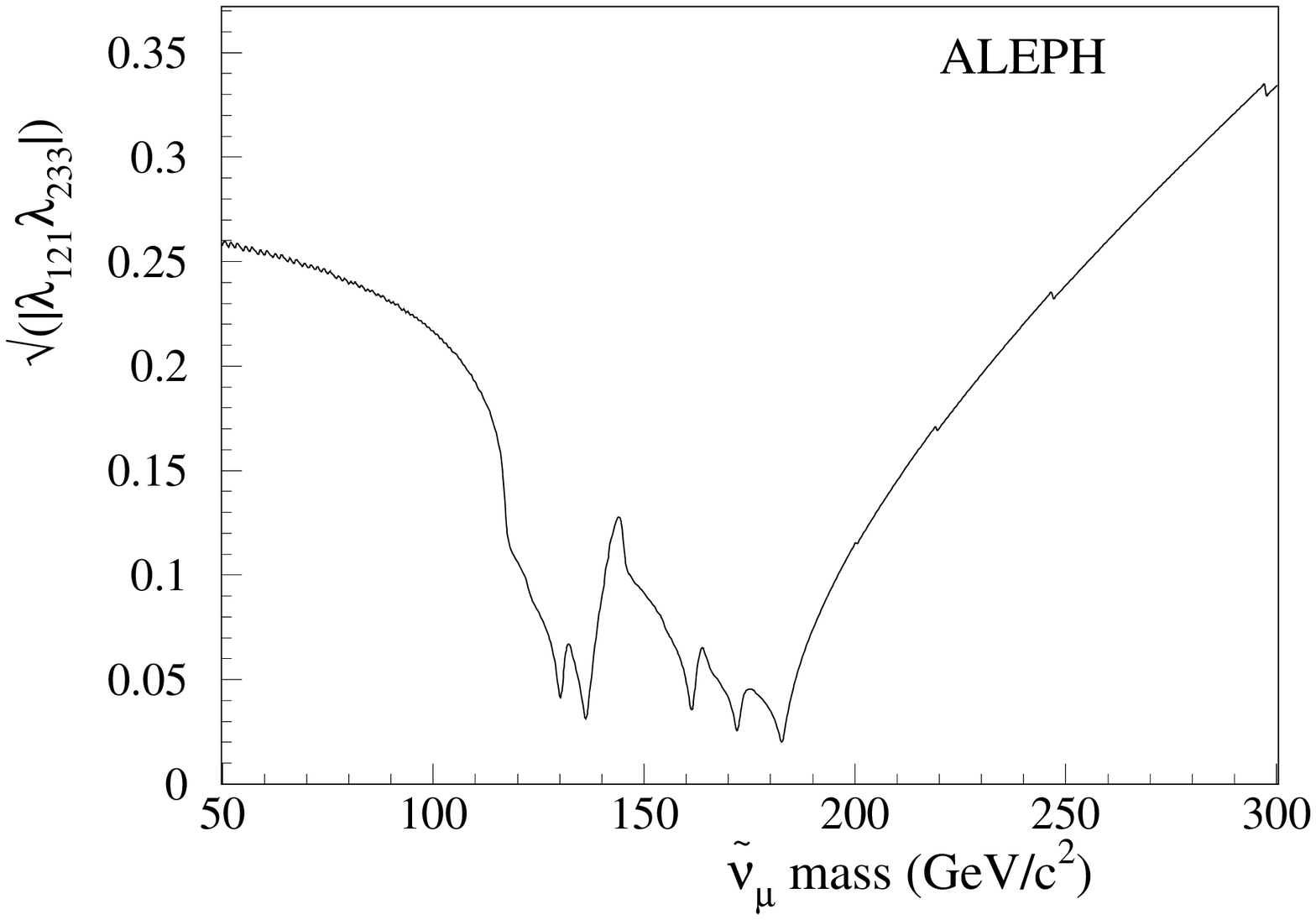,height=0.35\textheight}}
\end{center}
\mycaption{\label{tau_snu} 95\% confidence level upper limits, obtained from 
           the ditau cross sections, on $\sqrt{|\lambda_{121}\lambda_{233}|}$ 
           versus the assumed $\tilde\nu_\mu$ mass.}
\end{figure}

Figures~\ref{e_snu}, \ref{mu_snu} and \ref{tau_snu} 
show the results for those processes involving sneutrino exchange in the 
$s$~channel. Similar results have been obtained by the OPAL 
Collaboration~\cite{OPAL} and
for $\tilde\nu_\tau$ by the L3 Collaboration \cite{L3_snu}.

Limits on $\lambda_{121}$ and $\lambda_{131}$ from $t$~channel exchange of 
$\tilde\nu_e$ in muon and tau pair production, respectively, give much weaker 
limits. These rise from $|\lambda_{1j1}| < 0.5$ at $\tilde\nu_e =$ 
100~\GeVcc\ to $|\lambda_{1j1}| < 0.9$ at $\tilde\nu_e =$ 300~\GeVcc. 
%These limits are only competitive with those in Fig.~\ref{e_snu} in the 
%unlikely case where $\tilde\nu_e$ is much lighter than $\tilde\nu_\mu$ or 
%$\tilde\nu_\tau$. 

%%%%%%%%
%--- subsection : leptoquarks
%%%%%%%%
%\input ew2_conf_vanc_leptoquarks.tex
%%%%%%%%%%%%%%%%%%%%%%%%%%%%%%%%%%%%%%%%%%%%%%%%%%%%%%%%
%
% Leptoquarks
%
%%%%%%%%%%%%%%%%%%%%%%%%%%%%%%%%%%%%%%%%%%%%%%%%%%%%%%%%

\subsection{Limits on Leptoquarks and R-Parity Violating Squarks}
 \label{leptoquarks}

At LEP, the $t$~channel exchange of a leptoquark can modify the \qq\ cross section 
and jet charge asymmetry, as described by the Born level equations given in 
Ref.~\cite{KAL}. A comparison of the measurements with these equations 
allows upper limits to be set on the leptoquark's couplings $g$ as a function 
of its mass \MLQ, using the same fit technique and corrections for ISR, etc. 
as employed in Section~\ref{contacts}. 
Although the leptoquark $t$~channel exchange alters the angular distribution 
of the outgoing \qq\ system, this has negligible effect on the \qq\ selection 
efficiency.
Limits are obtained for each possible leptoquark species. 
The allowed species can be classified according to their spin, weak isospin 
$I$ and hypercharge. Scalar and vector leptoquarks are denoted by symbols 
${\mathrm S}_I$ and ${\mathrm V}_I$ respectively, and isomultiplets with 
different hypercharges are distinguished by a tilde. 
An indication ``(L)'' or ``(R)'' after the
name indicates if the leptoquark couples to left- or right-handed leptons.
Where both chirality couplings are possible, limits are set for the two
cases independently, assuming that left- and right-handed couplings are not 
present simultaneously. It is also assumed that leptoquarks within a given 
isomultiplet are mass degenerate \cite{KAL}.

The \SBL{\half} and \SL{0} leptoquarks are equivalent to up-type anti-squarks and 
down-type squarks, respectively, in supersymmetric theories with an R-parity 
breaking term $\lambda'_{1jk} {\mathrm L}_1 {\mathrm Q}_j \bar{\mathrm D}_k$ 
$(j,k=1,2,3)$. Limits in terms of the leptoquark coupling are then 
exactly equivalent to limits in terms of $\lambda'_{1jk}$.

Table~\ref{lqtab} gives for each leptoquark type, the 95\% confidence level 
lower limits on its mass \MLQ, assuming that it has a coupling 
strength equal to the electromagnetic coupling $g = e$. The limits
are given separately, assuming that (i) the leptoquark couples to only
$1^{\mathrm st}$ or only $2^{\mathrm nd}$ generation quarks, or (ii)
to only $3^{\mathrm rd}$ generation quarks. The former limits are derived
using the measured hadronic cross section and jet charge asymmetries, whilst
the latter uses the \bb\ cross section and jet charge asymmetries.
For $g\not= e$, the mass limit scales approximately in proportion to the 
coupling if it exceeds about 200~\GeVcc. (This is the contact term limit.)

\begin{table}[htbp]
\mycaption{\label{lqtab} 
95\% confidence level lower limits on the leptoquark mass for each species.
Limits are given separately according to the quark generation to which the 
leptoquark is assumed to couple. A dash indicates that no limit can be 
set and ``N.A.'' denotes leptoquarks coupling only to top quarks and hence
not visible at LEP.}
\begin{center}
\begin{tabular}{|c||c c c c c c c|}
\hline\strutl
  Quark     & \multicolumn{7}{c|}{Limit on scalar leptoquark mass (\GeVcc)} \\
\cline{2-8}\strutd
 Generation & \SL{0} & \SR{0} & \SBR{0} & \SL{1} & \SR{\half} & \SL{\half} & \SBL{\half} \\
\hline\hline
\strutl $1^{\mathrm st}$ or $2^{\mathrm nd}$ 
            &   200  &   ---  &    70   &   240  &   ---  &    20  &   ---   \\
\strutl $3^{\mathrm rd}$  
            &  N.A.  &  N.A.  &   180   &   450  &   ---  &  N.A.  &    50   \\
\hline
\multicolumn{8}{c}{\null}\\
\hline\strutl
  Quark     & \multicolumn{7}{c|}{Limit on vector leptoquark mass (\GeVcc)} \\
\cline{2-8}\strutd
 Generation & \VR{\half} & \VL{\half} & \VBL{\half} & \VL{0} & \VR{0} & \VBR{0} & \VL{1} \\
\hline\hline
\strutl $1^{\mathrm st}$ or $2^{\mathrm nd}$ 
            &   150  &   130  &    90   &   340  &   120  &   280   &   470  \\
\strutl $3^{\mathrm rd}$ 
            &   260  &   160  &  N.A.   &   400  &   140  &  N.A.   &   400  \\
\hline
\end{tabular}
\end{center}
\end{table}

Similar limits have been obtained by OPAL and L3~\cite{OPAL, L3:CI:1998}. 
Limits from the Tevatron \cite{TEVATRON, D0VEC} depend upon the assumed 
branching ratio of the charged leptonic decay mode. However, if this is 100\%,
then the Tevatron excludes leptoquark masses below 
$\sim$ 225~\GeVcc.  Other experiments can place limits on leptoquarks 
which couple to first generation quarks. In particular, low energy data such 
as atomic parity violation and rare decays give very stringent limits, usually 
as a function of the ratio $g/M_{\mathrm LQ}$. If $g=e$, they imply a 
lower limit on the leptoquark mass in the range 430--1500~\GeVcc\ 
\cite{LQlow}, depending on the leptoquark species. Preliminary results from 
HERA~\cite{HERA} exclude scalar leptoquarks with masses below $\approx 250$~\GeVcc\ if 
$g = e$. 

%%%%%%%%
%--- subsection : Zprime
%%%%%%%%
%\input Zprime.tex
%%%%%%%%%%%%%%%%%%%%%%%%%%%%%%%%%%%%%%%%%%%%%%%%%%%%%%%%
%
% Zprime
%
%%%%%%%%%%%%%%%%%%%%%%%%%%%%%%%%%%%%%%%%%%%%%%%%%%%%%%%%

\subsection{Limits on Extra Z Bosons}
 \label{Zprime}
To unify the strong and electroweak interactions, Grand Unification Theories 
(GUT) extend the SM gauge group to a group of higher rank, predicting 
therefore the presence of at least one extra neutral gauge boson \ZP. 
The theories which are considered in this section are the \ES\ \cite{ZPRIME} 
and the Left-Right (LR) models \cite{ZPRIME}. 
In the \ES\ model, the unification group \ES\ can break into the SM 
$SU_{C}(3) \otimes SU_{L}(2) \otimes U_{Y}(1)$ in different ways. Each 
symmetry breaking pattern leads to the presence of at least one extra U(1) 
symmetry and therefore one extra gauge boson. This is characterized by a 
parameter \TES\ ($-\pi/2 \leq$ \TES $\leq \pi/2$) which entirely defines its
couplings to conventional fermions. Contributions from exotic particles 
predicted by \ES\ or supersymmetric particles are ignored in 
this analysis. Four models derived from \ES\ are studied here, the \ESCHI, 
\ESPSI, \ESETA\ and \ESI\ defined by \TES $= 0$, $\pi/2$, 
$-\arctan{\sqrt{5/3}}$ and $\arctan{\sqrt{3/5}}$, respectively
\cite{ZPRIME}.
In the LR model, the SM group is extended to $SU_{C}(3) \otimes SU_{L}(2) 
\otimes SU_{R}(2) \otimes U_{B-L}(1)$, where B and L are the baryon and lepton
number, respectively. The SM $U_{Y}(1)$ symmetry is recovered by a linear 
combination of the generator of $U_{B-L}(1)$ and the third component of 
$SU_{R}(2)$. This model, which can arise from symmetry breaking of the GUT 
group SO(10), leads to the presence of one extra neutral gauge boson.
Contributions from the extra ${\mathrm W}_{R}$ of $SU(2)_{R}$ are neglected in this 
analysis. Couplings to conventional fermions depend only on one parameter 
\ALR, where $\sqrt{2/3} \leq \alpha_{LR} \leq 
\sqrt{(1-2\sin^2{\theta}_W)/\sin^2{\theta}_W}$, which is function of the 
coupling constants $g_{L,R}$ of $SU_{L,R}(2)$ and $\sin^2{\theta}_W$. 
Only the Left-Right Symmetric Model (LRS) will be studied here, which 
has $g_{L}=g_{R}$, implying that if $\sin^2{\theta}_W=0.23$ then
$\ALR=1.53$~.

Outside the context of GUT, a limit is derived on the mass of the 
non-gauge-invariant sequential SM (SSM) \ZP~boson, having the same 
couplings as the SM Z, but with a higher mass. Limits are also placed 
upon the axial and vector couplings of an arbitrary \ZP\ as a function of
its mass. 

In all the models mentioned above, the symmetry eigenstate \ZZEROP\ of the 
extra U(1) or $SU_{R}(2)$ can mix with the symmetry eigenstate \ZZERO\ of 
$SU_{L}(2) \otimes U_{Y}(1)$ with a mixing angle \TMIX. In such a case, the 
Z resonance observed at LEP I must be identified as one of the mass 
eigenstates of the \ZZEROP--\ZZERO\ system while the second mass eigenstate 
\ZP\ of mass \MZP\ is a free parameter \cite{ZPRIMEMIX, ZEFIT}.

To obtain the 95\% confidence level exclusion limits on the various free 
parameters, least squares fits are performed using the set of ALEPH 
measurements given in Table~\ref{tab:observ}, taking into account the 
correlations between them. The LEP1 measurements are taken from 
Ref.~\cite{EWLEP1}, whilst the LEP2 measurements are presented in 
Tables~\ref{CROSSALL}, \ref{rbcross} and \ref{ASYM_ALL}.

\begin{table}[htbp]
\mycaption{\label{tab:observ} Set of observables used in the \ZP\ analyses and 
the corresponding SM \CHI\ values per degree of freedom.}
\begin{center}
\begin{tabular}{|c||c|c|}
\hline
  & Observables & \CHISM/NDF. \\ 
\hline\hline
 LEP1 &
 $\sigma^{\mathrm l^+l^-}$, $A^{\mathrm l^+l^-}_{FB}$, $\sigma^{{\mathrm q} \bar{\mathrm q}}$, $l=\mu, \tau$ &
130.9/120  \\
\hline
 LEP2 &
 $\sigma^{\mathrm l^+l^-}$, $A^{\mathrm l^+l^-}_{FB}$, $\sigma^{{\mathrm q} \bar{\mathrm q}}$, \Rb &
 19.4/29 \\
 130--183 GeV &
 $l=\mu, \tau$, $\sqrt{s'/s} > 0.9$, $|{\cos\theta}| < 0.95$ & \\
\hline
\end{tabular}
\end{center}
\end{table}

Theoretical predictions for difermion cross sections and asymmetries are
obtained from the program ZEFIT 5.0 \cite{ZEFIT}, which is an extension of 
ZFITTER 5.0 \cite{ZFITTER} including the one extra neutral gauge boson of 
the \ES, LR or SSM models. Theoretical uncertainties on the SM 
predictions are taken into account as described in Section~\ref{smpred}. 
For all models, the minimum \CHI\ is found to occur when no \ZP\ boson is 
present. 

\begin{figure}[p]
  \mbox{\epsfig{file= 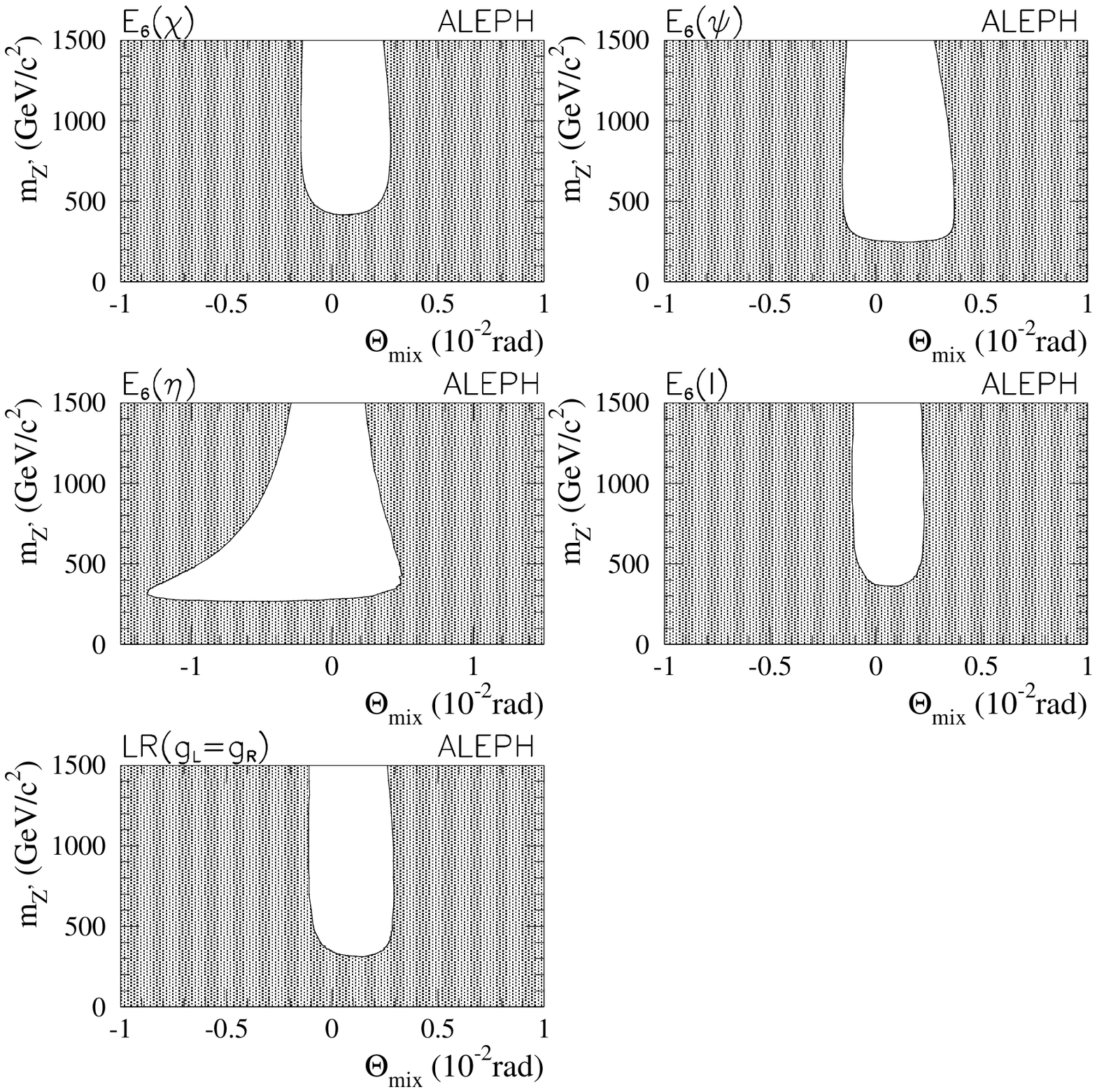,width=0.9\textwidth}}
  \mycaption{\label{limits2p} 95\% confidence level limits in the \MZP\ vs. 
           \TMIX\ plane for the  \ESCHI, \ESPSI, \ESETA, \ESI\ and \LRS\ 
           models. The shaded regions are excluded.}
\end{figure}

For the five models \ESCHI, \ESPSI, \ESETA, \ESI\ and \LRS, 
Fig.~\ref{limits2p} shows the 95\% confidence level limits obtained in the 
plane of \ZP\ mass versus mixing angle \TMIX. Both parameters are treated
as independent, so these limits correspond to a \CHI\ increase of 5.99~.
The LEP1 data mainly constrain the mixing angle, whilst the LEP2 data 
mainly constrain the \ZP\ mass at small mixing angles.

Alternatively, assuming $\TMIX=0$, lower limits on the \ZP\ mass can be 
obtained using a one-sided, one-parameter fit ($\Delta\chi^2=2.71$). The 
resulting limits are given in Table~\ref{tab:limits1D}, where they are 
compared with those from direct \ZP\ searches performed by the CDF 
Collaboration \cite{ZPCDF}. This table also gives the mass limit for the 
SSM \ZP, which is superior to the limit from CDF.

\begin{table}[htbp]
    \mycaption{\label{tab:limits1D} Comparison of 95\% confidence level lower 
     limits on \MZP\ (\GeVcc) from one parameter electroweak fits (ALEPH) and 
     direct searches (CDF) for $\TMIX=0$.}

  \begin{center}
    \begin{tabular}{|c||c|c|}
\hline
Model    & ALEPH & CDF direct \\ 
\hline\hline
 \ESCHI\ & 533 & 595 \\
 \ESPSI\ & 294 & 590 \\
 \ESETA\ & 329 & 620 \\
 \ESI\   & 472 & 565 \\
\hline
 \LRS\   & 436 & 630 \\
\hline
 Sequential SM & 898 & 690  \\
\hline
    \end{tabular}
  \end{center}
\end{table}

Limits can also be placed on the vector and axial couplings of an arbitrary 
\ZP, as a function of its mass. To simplify, such limits will only be given 
here for the leptonic couplings (assuming lepton universality) and also only
for the case $\TMIX=0$. Limits are placed on the two couplings simultaneously 
($\Delta\chi^2=5.99$). The excluded region is found to be approximately 
rectangular in shape and its size is given as a function of \ZP\
mass in Table~\ref{tab:limitsCoupl}. 

\begin{table}[htbp]
 \mycaption{ \label{tab:limitsCoupl}
 95\% confidence level limits on the axial $g_a^\prime$ and vector 
 $g_v^\prime$  couplings of a \ZP\ boson of mass $m_{\mathrm Z'}$ to a 
 lepton pair.}

  \begin{center}
    \begin{tabular}{|c|c|c|}
      \hline
   $m_{\mathrm Z'}$ (\GeVcc) & $|g_a^\prime|$  &  $|g_v^\prime|$  \\ 
   \hline
   300  & $\leq 0.36$  & $\leq 0.28$   \\
   600  & $\leq 0.81$  & $\leq 0.64$   \\
   1000 & $\leq 1.39$  & $\leq 1.11$   \\
   \hline
 \end{tabular}
\end{center}
\end{table}

%%%%%%%%
%--- bibliography
%%%%%%%%
%\input ew2_conf_vanc_bib.tex
%%%%%%%%%%%%%%%%%%%%%%%%%%%%%%%%%%%%%%%%%%%%%%%%%%%%%%%%
%
% conclusions, thanx and the bibliography
%
%%%%%%%%%%%%%%%%%%%%%%%%%%%%%%%%%%%%%%%%%%%%%%%%%%%%%%%%

\section{Conclusions}

Measurements of the hadronic and leptonic 
cross sections and asymmetries at $\sqrt{s}=\mbox{130--183}$~GeV have been 
presented. 
The ratios of the \bb\ to \qq\  production cross 
sections at $\sqrt{s}=130$--183~GeV and of the \cc\ to \qq\  production cross 
sections at $\sqrt{s}=183$~GeV have been shown, as well as jet charge 
asymmetries. 
The results agree with the predictions of the Standard Model and allow 
limits to be placed on four-fermion contact interactions, R-parity violating
sneutrinos and squarks, leptoquarks and \ZP\ bosons.
The limits on the energy scale $\Lambda$ of $\ee\ff$ contact interactions 
are typically in the range from 2--10~TeV. Those for  $\ee\myll$ and $\ee\bb$ 
interactions are of particular interest, since they are inaccessible at 
colliders using proton beams. The new ALEPH limits on R-parity violating 
sneutrinos reach masses of a few hundred \GeVcc\ for large values of 
their Yukawa couplings. 

\section*{Acknowledgements}

We thank our colleagues from the CERN accelerator divisions for the successful
operation of LEP at higher energies. We are indebted to the engineers and 
technicians in all our institutions for their contribution to the continuing
good performance of ALEPH. Those of us from non-member states thank
CERN for its hospitality.

\end{document}